\documentclass{aa}

\usepackage{txfonts}
\usepackage{epstopdf}
\usepackage{mathptmx}
\usepackage{moresize}
\usepackage[percent]{overpic}
\usepackage{url}
\usepackage{natbib}
\bibpunct{(}{)}{;}{a}{}{,}
\usepackage[normalem]{ulem}
\usepackage{graphicx}
\usepackage[colorlinks]{hyperref}
\usepackage{xspace}
\usepackage{todonotes}
\usepackage{mathtools}
\usepackage{amsmath}
\usepackage{longtable}
\usepackage{comment}
\usepackage{multirow}
\usepackage{gensymb}

\usepackage{subcaption}
\usepackage{footnote}
\setcitestyle{authoryear, round, aysep={comma}}
\hypersetup{colorlinks=blue, linkcolor=blue, urlcolor=blue, citecolor=blue}

\begin{document}
\title{X-ray counterpart detection and $\gamma$-ray analysis of the SNR G279.0+01.1 with eROSITA and Fermi-LAT}

 \author{Miltiadis Michailidis\inst{1}, Gerd P{\"u}hlhofer\inst{1},  Andrea Santangelo\inst{1}, Werner Becker\inst{2,3}, Manami Sasaki\inst{4} }

 \institute{Institut für Astronomie und Astrophysik Tübingen (IAAT), Sand 1, 72076 Tübingen, Germany \and Max-Planck Institut f{\"u}r extraterrestrische Physik, Giessenbachstraße, 85748 Garching, Germany \and Max-Planck Institut für Radioastronomie, Auf dem Hügel 69, 53121 Bonn, Germany
 \and Dr.\ Karl Remeis Observatory, Erlangen Centre for Astroparticle Physics, Friedrich-Alexander-Universit\"{a}t Erlangen-N\"{u}rnberg, Sternwartstra{\ss}e 7, 96049 Bamberg, Germany}

 \date{Received 05/11/2023; accepted ?? ??, 20??}

\abstract{
 A thorough inspection of known Galactic Supernova Remnants (SNRs) along the Galactic plane with SRG/eROSITA yielded the detection of the X-ray counterpart of the SNR G279.0+01.1. The SNR is located just $1.5\degree$ above the Galactic plane.
 Its X-ray emission emerges as an incomplete, partial shell of $\sim3\degree$ angular extension. It is strongly correlated to the fragmented shell-type morphology of its radio continuum emission. The X-ray spatial morphology of the SNR seems to be linked to the presence of dust clouds in the surroundings. The majority of its X-ray emission is soft (exhibiting strong O, Ne and Mg lines), and occurs in a narrow range of energies between 0.3 and 1.5 keV. Above 2.0 keV the remnant remains undetected. The remnant's X-ray spectrum is of purely thermal nature. 
 Constraining the X-ray absorption column to values which are consistent with optical extinction data from the remnant's location favours non-equilibrium over equilibrium models. 
 A non-equilibrium two-temperature plasma model of $\mathrm{kT}\sim0.3$~keV and $\mathrm{kT}\sim0.6$~keV, and an absorption column density of  $\mathrm{N_{H}}\sim0.3~\mathrm{cm^{-2}}$ describes the spectrum of the entire remnant well. Significant temperature variations across the remnant have been detected. Employing 14.5 years of Fermi-LAT data, we carried out a comprehensive study of the extended GeV source
 4FGL J1000.0-5312e. By refining and properly modeling the GeV excess originating from the location of the remnant, we conclude that the emission is likely related to the remnant itself rather than being co-located by chance. The derived X-ray spectra are consistent with the $\sim2.5$~kpc distance estimates from the literature, which implies a source diameter of $\sim140$~pc and old age of $>7\cdot 10^5$~yrs. However, if the source is associated with any of the pulsars previously considered to be associated with the SNR, then the updated nearby pulsar distance estimates from the YMW16 electron density model place the SNR rather at a distance of $\sim0.4$~kpc. This would correspond to a $\sim20$~pc linear size and a younger age of $10^4-<7\cdot 10^5$~yrs, which would be more in line with the non-equilibrium state of the plasma.}
 
 \keywords{supernova remnants (Individual object: G279.0+01.1) --- 
multiwavelength study}
 
\titlerunning{G279.0+01.1: X-ray counterpart detection with SRG/eROSITA}
\authorrunning{M. Michailidis}
\maketitle


\section{Introduction}\label{sec:intro}

Supernova remnants (SNR) are the residua of Supernova (SN) explosions, one of the most energetic processes in the Universe. The shock waves of those bursts can efficiently accelerate charged particles from radio to X-ray emitting energies \citep{1995Natur.378..255K}, and also up to GeV/TeV energies \citep{2004Natur.432...75A,2016ApJS..224....8A,2018A&A...612A...1H}. In contrast to Supernovae, events that occur in a short period of time observable within a few years of occurrence, their remnants can remain visible for several thousand to ten-thousands of years. Depending on their evolutionary state and distance from Earth, their angular sizes (assuming Galactic SNR) can range from a few arcmin to several degrees. Only a handful of low surface brightness Earth-adjacent remnants, a few tens of hundreds of parsecs away, which are found in their most evolved state and with sizes of several degrees, have been detected. In the X-ray band particularly, even fewer findings have been reported. The improved sensitivity of the eROSITA All-Sky Survey offers a unique chance to detect such SNRs  that XMM-Newton/Chandra/Suzaku and ROSAT could not have seen (Becker et al., 2024, in prep.).

The majority of detected SNRs fall in the Galactic plane, where massive stars are most abundant. Even though they are extended objects, particularly in their evolved states, they can be partially or totally obscured (e.g., in optical and X-ray wavebands) due to the prevalence of absorbing dust in the Galactic plane. In the radio band, the sensitivity limitation of current instruments as well as the potential confusion/contamination of the emission from brighter nearby sources is another inhibitory factor in the localization of the emission originating from supernova remnants. G279.0+01.1 is such a case of a remnant. According to current literature, it is possibly located near the tangent point to the nearby Carina Spiral arm, which would place it at a distance of $2.7\pm0.3$~kpc \citep{2019RAA....19...92S}. Its center is located just $1.5\degree$ above the Galactic plane, and it has a size of $2.3\degree$ in the radio band \citep{10.1111/j.1365-2966.2009.14476.x}. Its spatial appearance in optical is morphologically consistent with the high concentration of dust on the three sides (i.e., the Southern, Western, and Eastern sides) of the remnant as reported in \citet{10.1111/j.1365-2966.2009.14476.x}. The bright radio sources at and around the remnant's vicinity make it challenging to determine its true radio extent. A GeV source, seemingly correlated with the remnant, has recently been discovered \citep{2020MNRAS.492.5980A}, whereas no X-ray counterpart had been found up to date. In this work, we report on the first X-ray counterpart detection of G279.0+01.1 by utilizing data of the first four completed eROSITA All-Sky Surveys, i.e. eRASS:4 \citep{Merloni2023}. 

In 1988, the remnant was detected for the first time in the radio continuum band \citep{1988MNRAS.234..971W}. The SNR appears as a circular shell of $\sim1.6\degree$ angular extension, quite distinguished from the radio emission related to the nearby Carina spiral arm. The North and East limbs appear to be the brightest parts of the remnant, whereas the fainter Western limb is characterized by a region of enhanced radio emission. The latter is likely attributed to an unrelated point source, which in later studies was determined to be a powerful extragalactic point-like radio emitter (G278.0+0.8) \citep{1995MNRAS.277..319D}. The Northern, radio-bright limb of the remnant lies along the line of sight of an HI region. However, the latter is highly unlikely to be interacting with the remnant given that their kinematic distances differ significantly, by $8~\mathrm{kpc}$. Moreover, \citet{1995MNRAS.277..319D} confirmed the detection of two CO clouds likely interacting with the SNR.

There are ten pulsars in with less than a $3.0\degree$ angular separation from the remnant's redefined center (refer to sec.~\ref{eROSITAA}). Three of these pulsars - B0953-52, B0959-54, and B1014-53 - have been discussed as potential associations with the remnant. B0953-52 was initially considered the most plausible counterpart, given its $0.64\degree$ angular distance from the remnant's center and alignment with the SNR's circular morphology \citep{1988MNRAS.234..971W}. However, \citet{1995MNRAS.277..319D} suggested that the pulsar B0959-54, currently named J1001-5507, is more likely associated, despite being $1.6\degree$ away from the remnant's center and outside the radio emission region. We examine the implications of potential pulsar associations with G279.0+1.1, considering a recent update to the electron density model \citep{2017ApJ...835...29Y}. This update reduces, by about an order of magnitude, distance estimates to all pulsars potentially associated with the remnant compared to values derived using the earlier model in \citet{2002astro.ph..7156C}.

In addition, more recent radio studies showcase the detection of previously missed broad filamentary structures at the North-East and South-West parts of the SNR, and strong polarization at 1.4~GHz and 2.4~GHz frequencies \citep{1995MNRAS.277..319D,1996A&AS..118..329W}. In particular, while typical SNR do not exceed radio polarization levels of $10\%$, \citet{1995MNRAS.277..319D} detected strong polarization up to $50\%$ at 2.4 GHz. The remnant has also been classified among the barrel shape SNR as introduced in \citet{1987A&A...183..118K}. A reassessment of the remnant's radio morphology was conducted in \citet{10.1111/j.1365-2966.2009.14476.x} revealing a larger, compared to previous studies \citep{1988MNRAS.234..971W,1995MNRAS.277..319D}, radio image of the SNR at 843 MHz and 4.85 GHz. A $2.3\degree$ angular size was obtained from observations of the remnant at both frequencies \citep{1998PASA...15...64C}. 

Optical H$\alpha$ emission, originating from G279.0+01.1 was detected for the first time by \citet{10.1111/j.1365-2966.2009.14476.x}.  
The detailed optical analysis revealed 14 small-scale fragmented groups of H$\alpha$ filaments spread over a $2\degree$ area within the SNR's radio shell. Those structures are concentrated at the central and North-Eastern parts of the remnant, suggesting that the high dust concentration at the South and West of the remnant prevents optical detection. Even though the strong radio source G278.0+0.8 does not have an optical H$\alpha$ counterpart, a strong enhancement in H$\alpha$ emission is being observed just to the West of the remnant. The latter H$\alpha$ excess is consistent with diffuse radio emission of the size of 24 arcmin. The emission is concluded to be unrelated to the remnant itself, and is more likely an illuminated HII region.  More recent infrared (IR) Galactic surveys (i.e., Wide-Field Infrared Survey Explorer (WISE)) confirmed the shell-type morphology of the object and classified it as an HII region that can be found under the name G277.731+00.647 in the WISE HII catalog Ver. 2.4 \citep{2014ApJS..212....1A}.

A distance estimation based on the interaction of the SNR's blast wave with interstellar clouds resulted in a 3~kpc distance \citep{1975ApJ...195..715M,10.1111/j.1365-2966.2009.14476.x} consistent with the $\Sigma-D$ estimation. A consistent distance estimation, of $2.7\pm0.3$~kpc, was obtained using optical extinction from red clump stars \citep{2019RAA....19...92S}. 

A GeV source positionally coincident to the remnant has been detected utilizing Pass 8 Fermi-LAT data \citep{2020MNRAS.492.5980A}. The $\gamma$-ray emission region, above 5~GeV, has a $\sim2.8\degree$ angular size, seemingly surpassing the radio synchrotron \citep{10.1111/j.1365-2966.2009.14476.x} towards the North-Eastern parts of the remnant. Both a leptonic and a hadronic scenario of the $\gamma-~\mathrm{ray}$ origin are discussed in \citet{2020MNRAS.492.5980A}. However, \citet{2021ApJ...910...78Z} ruled out the leptonic scenario possibility by fitting the remnant's multiwavelength spectra with hard $\gamma-~\mathrm{ray}$ spectra, extending up to 0.5 TeV, mainly due to the remnant's evolved state. There are no signs of softening of the $\gamma-~\mathrm{ray}$ spectrum at higher energies, > 0.5~TeV, but the remnant is undetected in the TeV band \citep{2018A&A...612A...3H}.

The paper is organized as follows. In section~\ref{eROSITAanalysis} we report on the outcomes of eROSITA observations and X-ray data analysis of the remnant utilizing the first four eROSITA All-Sky Survey data, eRASS:4. We also checked archival ROSAT survey data and XMM-Newton pointings towards the South-West of the remnant and briefly report those results. In section~\ref{multianalysis} we provide a multiwavelength study of the remnant employing archival radio synchrotron and GeV $\gamma$-ray data, as well as dust tracers. In Section~\ref{eROSEspectra} we report on the X-ray spectral analysis of the remnant, utilizing both eROSITA and XMM-Newton data. An updated GeV spectrum is also provided. Closing remarks are reported in Section~\ref{concl}.

\begin{table*}

\centering
\caption{eROSITA, XMM-Newton (MOS1, MOS2, PN) and ROSAT observations analyzed in this work. The pointing column describes the position of the XMM-Newton observations with respect to the remnant's center.}
\renewcommand{\arraystretch}{1.7}
\begin{minipage}{20cm}
\begin{tabular}
{p{3.0cm} p{3.5cm} p{2.0cm} p{3.5cm} p{2.5cm} p{1.5cm}}
\hline
Instrument & ObsID& Year& Mode&Exposure (ks) &Pointing  \\ \hline
eROSITA & eRASS:4 (150144,& 2019-2021&survey&25.9\footnote{Total on-source exposure time}&- \\ 
&145144,150141,145141)&&&&\\\hline
MOS1, MOS2, PN& 0823031001& 2018& Full Frame (x3)&17.7/17.7/15.8&South West\\ 
MOS1, MOS2, PN& 0823030401& 2018 & Full Frame (x3)&16.7/16.7/14.8&South West \\ 
MOS1, MOS2, PN& 0823030301& 2018 & Full Frame (x3)&9.0/9.0/7.1&North \\ \hline
ROSAT & RASS (932618,&1990&survey&14.0\footnote{Livetime, on time}& - \\ 
&932619,932716,932717)&&&&\\
\hline
\label{TAB00}
\end{tabular}
\end{minipage}
\end{table*}

\begin{figure}[h!]
    \centering
    \begin{overpic}[scale=0.407]{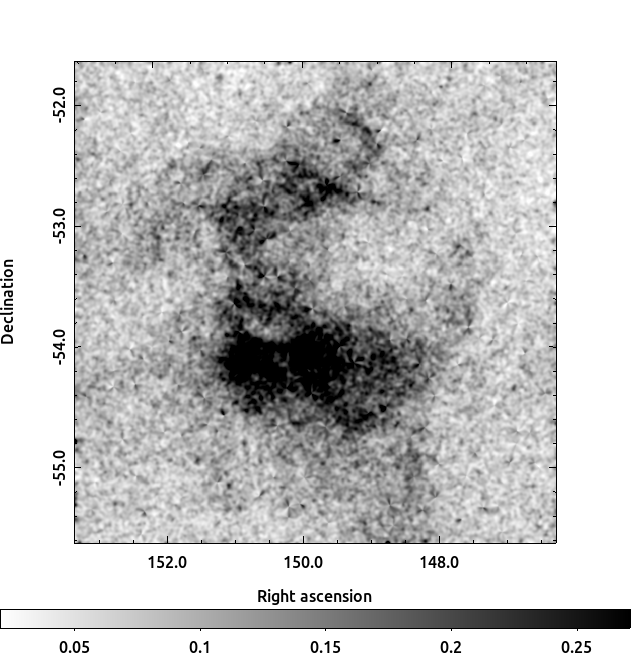}
      \put(0,7.4){\includegraphics[scale=0.546]{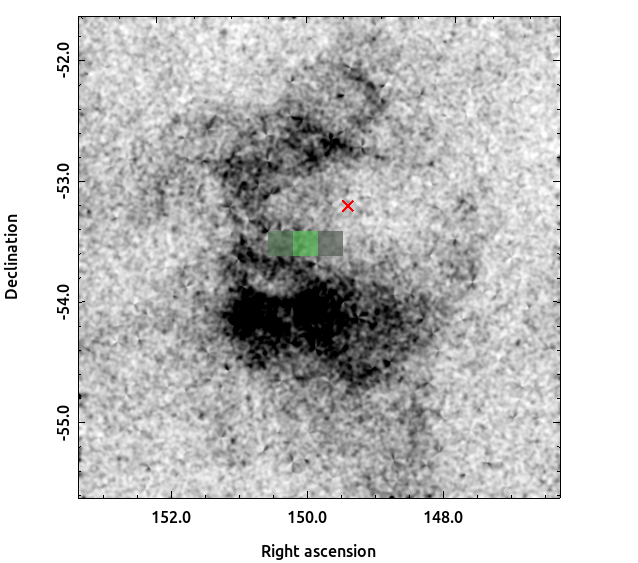}}
    \end{overpic}
    \caption{eRASS:4 exposure corrected intensity sky map in the 0.3-1.1~keV energy band, in units of counts/pixel with a pixel size of $10''$. Point sources are filtered out, and the image is convolved with a $\sigma=45''$ Gaussian to enhance the visibility of the diffuse X-ray emission originating from the source. The geometrical center of the X-ray emission from a Minkowski tensor analysis is shown by the green boxes. Brighter box means higher probability to represent the center. The red cross indicates the remnant's center based on previous radio measurements.}
    \label{net1}
\end{figure}

\section{X-ray observations and data analysis}
\label{eROSITAanalysis}

The main parameters of all X-ray observations employed in this work are summarized in Tab.~\ref{TAB00}.

\subsection{$\textit{eROSITA data}$}
\label{eROSITAA}

In this work, we use data from the eROSITA (extended ROentgen Survey Imaging Telescope Array) instrument operating in the 0.2-4.0 keV energy range \citep{2012arXiv1209.3114M,2021A&A...647A...1P}. eROSITA is one of the two scientific instruments aboard the  Russian-German Spektrum Roentgen Gamma (SRG) observatory \citep{2021A&A...656A.132S}. It hosts seven parallel-aligned X-ray telescopes (TM1-7). Each telescope has a field of view of $1\degree$. The All-Sky surveys started December 13, 2019. A (preliminary) analysis of the in-flight PSF calibration \citep{Merloni2023} showed a $\sim30"$ average spatial resolution in survey mode.

In the current analysis only data from the first four completed All-Sky Surveys (eRASS:4), were exploited, in the \texttt{c020} processing version. Data reduction and analysis was conducted utilizing the $\texttt{eSASSusers\_201009}$ version \citep{2022A&A...661A...1B} of eSASS (eROSITA Standard Analysis Software). All events that were flagged as corrupt either individually or as a whole corrupt frame were filtered out. All four legal patterns were sustained while bad patterns were identified and excluded (\texttt{pattern=15}). Disordered GTIs were recognized and repaired. In addition, eRASS:4 data were inspected for flares. The affected regions were re-processed and corrected, thus preventing possible contamination of the event files.

The eROSITA All-Sky map consists of 4700 sky tiles. Each one of them has a square morphology of $\sim3.6\degree\times3.6\degree$ size. The majority of the X-ray emission from the SNR is contained in a single sky tile. However, a total of four sky tiles were exploited in order to obtain complete coverage of the remnant and sufficient background control area. Fitting an annulus to the outermost X-ray emission ring of the remnant's fragmented shell structure, resulted in a geometrical center position of: Ra: 9:58:27.23, Dec: -53:35:46.95. We verified the above result by performing a Minkowski tensor analysis, which is an automatic bubble-recognition routine for parametrizing the shapes of bodies \citep{2021A&A...653A..16C}. The detection routine is based on the drawing of perpendicular lines to the detected structures. In our work, we perform the latter routine to the SNR fragmented shell, in the 0.3-1.1 keV energy band. Aiming to avoid contamination of our data sets and distortion of the obtained results, only X-ray diffuse structures encapsulated within the extension of the remnant's radio counterpart (see sec.~\ref{multianalysis}) were employed. Nearby structures unrelated to the remnant (e.g., the diffuse X-ray emission situated at the South of the remnant) were excluded from this analysis. All lines of the shell should meet in a small region inside the shell, thus creating high-line-density regions. The reconstructed center is shown in Fig.~\ref{net1} in green. The obtained result (central coordinates in X-rays: RA: 9:59:45.48 Dec: -53:33:11.91) is consistent with the one derived above. Consequently, to explore the remnant's X-ray spatial morphology, we construct mosaic sky maps with a size of $4\degree\times4\degree$ and a $10''$ pixel size centered on the best-fitted coordinates from the Minkowski tensor analysis.

\begin{figure}[h!]

    \includegraphics[width=0.5\textwidth]{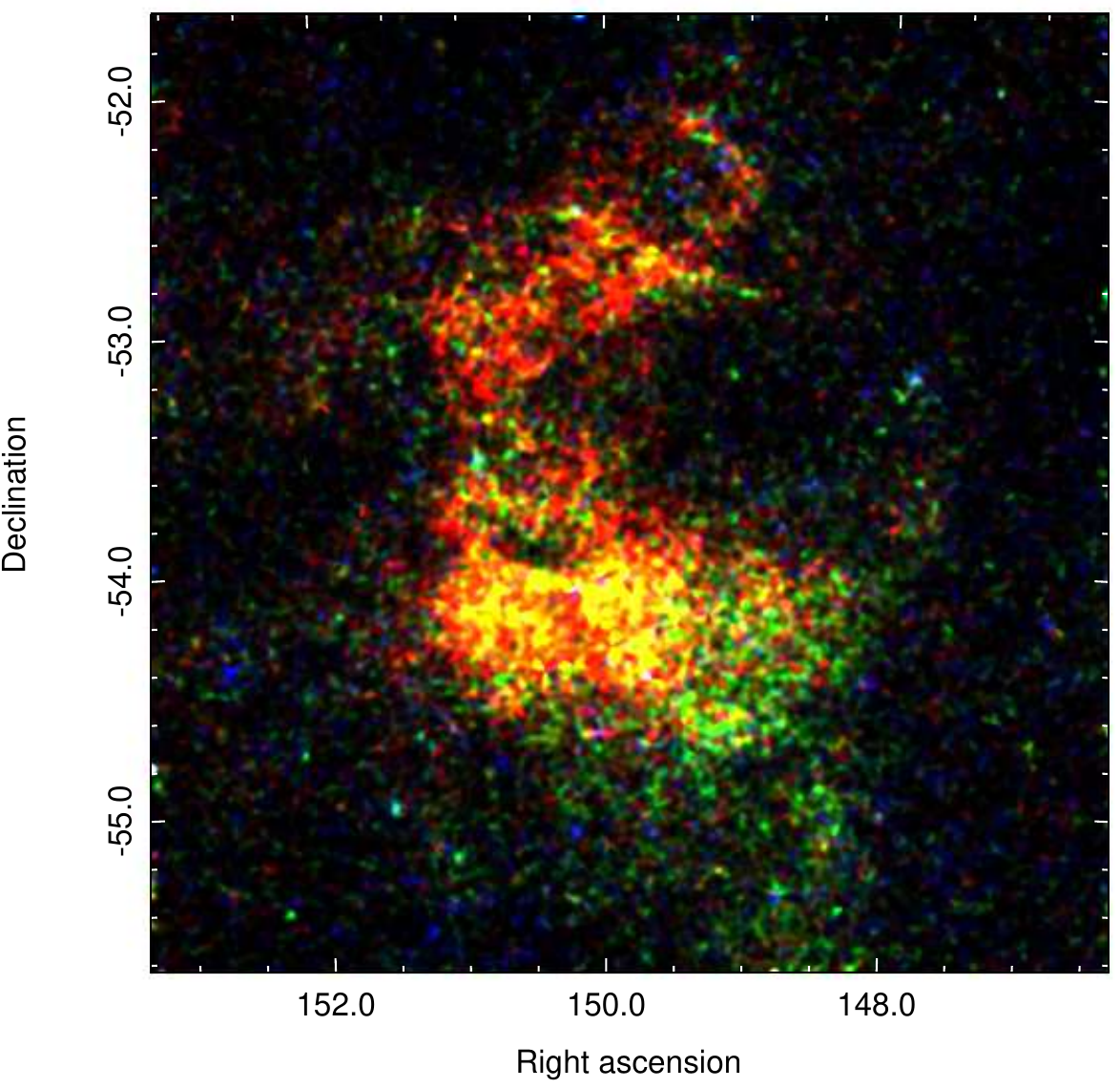}
    \caption{eRASS:4 RGB exposure-corrected intensity sky map, energy color-coded as follows, R: 0.3-0.7 keV, G: 0.7-1.1 keV, and B: 1.1-2.3 keV, in units of counts/pixel with a pixel size of $10''$. A squared-colored distribution is chosen for visual purposes. Point sources are filtered out, and the image is convolved with a $\sigma=45''$ Gaussian to enhance the visibility of the diffuse X-ray emission.}
    \label{RGB}
\end{figure}

In particular, the mosaic sky maps from the location of the remnant were produced by employing the \texttt{evtool} task of the eSASS software, combining the four aforementioned individual eROSITA sky tiles and using data from all instrument's telescopes TM1-7.
We find a strong detection of the SNR G279.0+01.1 in the narrow energy range from 0.3 to 1.1 keV, as depicted in Fig~\ref{net1}. Individual regions of the remnant, positioned mainly to the South and West, exhibit X-ray emission up to 1.5 keV. However, the remnant remains totally undetected above 2.0 keV. In addition to the soft X-ray emission that the image analysis reveals, the spatial morphology of the remnant matches with an incomplete shell (since the Western part of the shell is not observable in X-rays), or a fragmented annulus of highly asymmetric width, of $\sim3\degree$ angular size. 
The two enhanced regions of X-ray emission, in particular the two brightest X-ray "blobs" found at the South-Eastern part of the remnant (saturated blobs in Fig.~\ref{net1}), are not associated with any known astrophysical object that could account for such a type of diffuse X-ray emission. Therefore, we strongly suggest that they are part of the diffuse emission originating from the remnant itself. Further imaging analysis, color-coded RGB image (0.3-0.7 keV: red, 0.7-1.1 keV: green, 1.1-2.3 keV: blue) displayed in Fig~\ref{RGB}, indicates potential temperature variation, i.e., plasmas of different temperatures across the remnant. This is confirmed by the spectral analysis results in section~\ref{eROSEspectra}. Additionally, Fig.~\ref{RGB} confirms the lack of X-ray emission at hard X-rays by the absence of blue color, the majority of the X-ray emission is confined in the 0.3-1.1~keV energy band (red and green colors). 

\subsection{ROSAT data}\label{ROSATanalysis}
\begin{figure*}[h!]
    \centering
    
     \includegraphics[width=0.49\textwidth,clip=true, trim= 1.0cm 0.1cm 0.9cm 0.5cm]{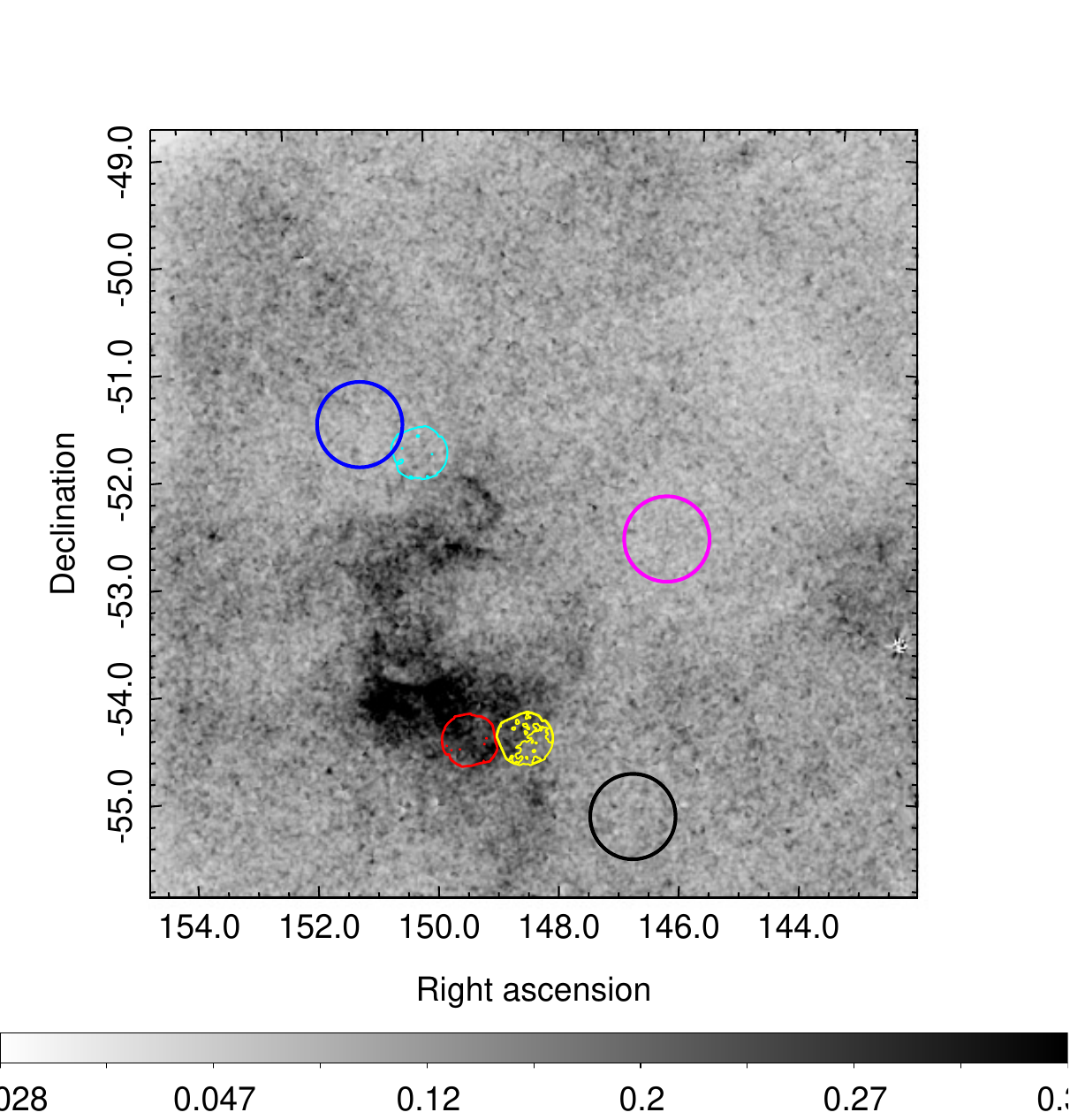}
    \includegraphics[width=0.49\textwidth,clip=true, trim= 1.0cm 0.1cm 0.9cm 0.5cm]{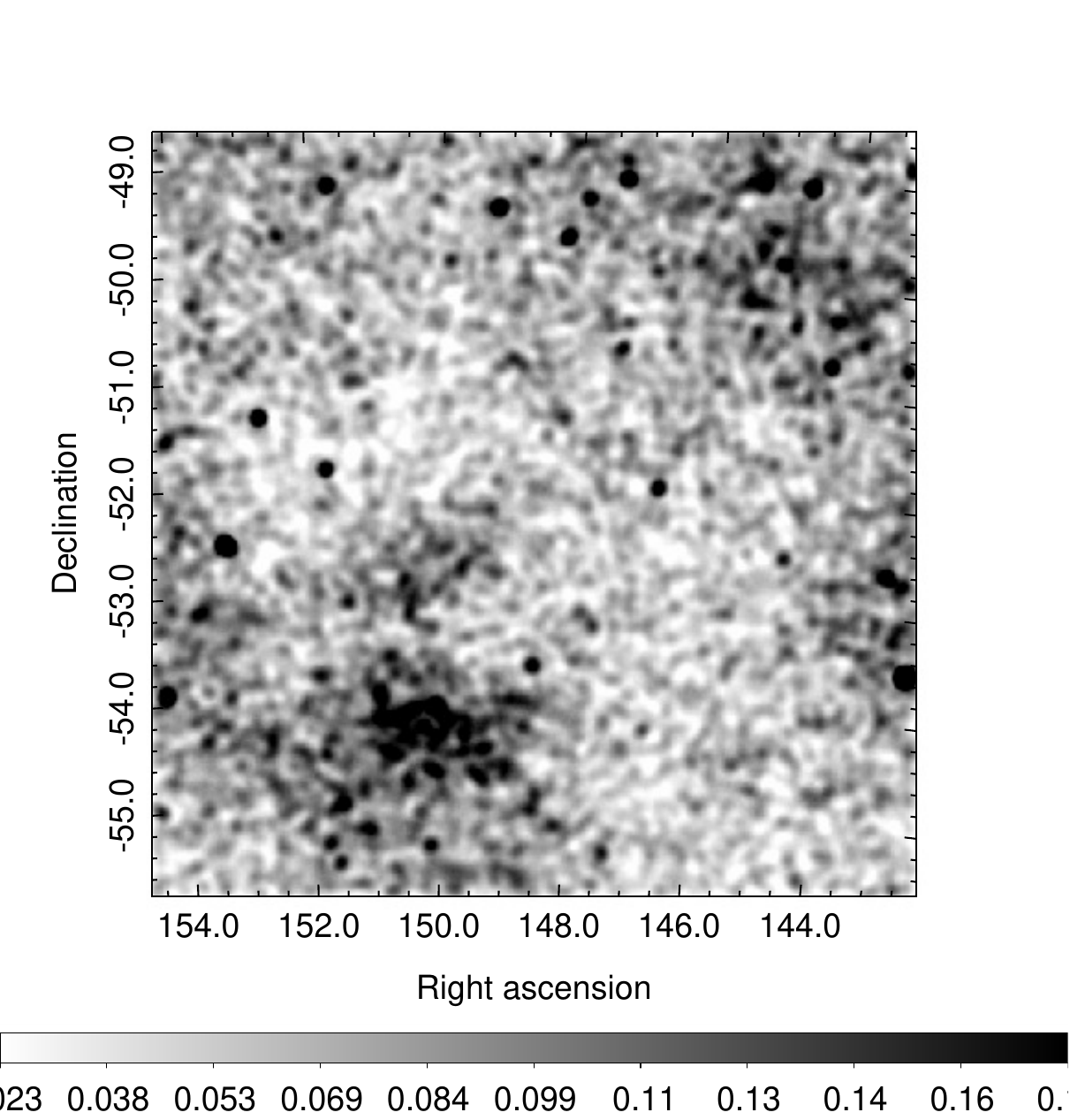}
    \caption{Left panel: eRASS:4 exposure-corrected intensity sky map in the 0.4-2.4~keV energy band, in units of counts/pixel with a pixel size of $10''$. Point sources are filtered out, and the image is convolved with a $\sigma=45''$ Gaussian. The black, magenta, and blue circles represent the three background control regions that we have selected to inspect potential background variations in the remnant's Galactic vicinity. Among those, the black circle was selected as the representative background used for the spectral analysis of the on-source regions, see section~\ref{ROSITAspec} for more details. Red, yellow, and cyan circles mark the positions of the 0823031001, 0823030401, and 0823030301 XMM-Newton pointings, respectively. Right panel: ROSAT intensity sky map in the 0.4-2.4~KeV energy band (medium RASS band). The image, with a $45''$ pixel size, is convolved with a $\sigma=3'$ Gaussian to enhance the visibility of the diffuse emission from the location of the remnant. Point sources are not removed since their proper masking requires a substantially larger extraction radius than for eROSITA, which heavily affects the faint diffuse emission originating from the remnant.}
    \label{nett1}
\end{figure*}
After the very significant detection of G279.0+01.1 in X-rays with eROSITA for the very first time we checked why the remnant has escaped detection in the ROSAT All-Sky-Survey (RASS data). We exploited publicly available data from the RASS Position Sensitive Proportional Counter detector in survey mode (PSPC) \citep{2000IAUC.7432....3V}. The medium, 0.4-2.4 keV, energy band, yielded a better signal-to-noise ratio in comparison to the narrower 0.3-1.1 keV energy range selected for eROSITA.  
As shown on the right panel of Fig~\ref{nett1},
the incomplete shell-type structure of the remnant, with much lower statistical quality in comparison to eROSITA, is visible above a strongly structured background. Both images of Fig.~\ref{nett1} are smoothed using a Gaussian function as described in the caption of the corresponding figure to enhance the visibility of the source. 5125 counts (of which 1444 are source counts) are detected with ROSAT from the location of the remnant, i.e., a circular region centered at the X-ray coordinates derived in section~\ref{eROSITAA} with a radius of $1.7\degree$, to make sure that it encircles the entire X-ray excess originating from the remnant.
The corresponding numbers for the eRASS:4 data in the same energy range are 205077 counts (of which 76651 are source counts). eRASS:4 has a $\sim53$ times higher collection area than the previous ROSAT survey (as expected), and the limited photon statistics plus the uneven background has apparently prevented a discovery with ROSAT.

\subsection{XMM-Newton data}\label{XMMspec}

The XMM-Newton data archive was inspected to see whether relevant observations exist towards the direction of G279.0+01.1 that could enhance or complement (on limited regions) the eROSITA imaging and spectral results. Indeed, two XMM-Newton observations that overlap with the SNR and one very adjacent towards the North of the SNR are found in the XMM-Newton archive (see Fig.~\ref{XMMIM} for the locations of these pointings with respect to G279.0+01.1). These observations (PI: Bettina Posselt, ObsId 0823031001, 0823030401, 0823030301) were targeted on nearby pulsars (J0957-5432, J0954-5430, J1000-5149, respectively), and no analysis on potential diffuse emission in the FoVs has been reported in the literature. We therefore analyzed these data to check for consistency with the eRASS results. Indeed, the two observations overlapping the SNR (one partially, one fully) exhibit significant diffuse emission consistent in morphology with the eRASS sky map (see Fig.~\ref{XMMIM}). We therefore extracted source spectra from the two XMM-Newton pointings from the respective on-source areas, using the source-free region and the third, off-source pointing as background control regions. eRASS spectra were extracted from the same on-source regions. Overall, there is good consistency between the spectral results of the two instruments. The XMM-Newton data were useful to verify the applicability of the spectral model ultimately chosen for the eRASS data, but did not permit to put further constraints on the ambiguities that remained in the choice of models from the eRASS spectral data analysis.

\section{G279.0+01.1 multiwavelength study}\label{multianalysis}

\subsection{Radio continuum $\&$ H$\alpha$}
\begin{figure}[h]

    \includegraphics[width=0.5\textwidth,clip=true, trim= 0.9cm 0.1cm 1.1cm 0.5cm]{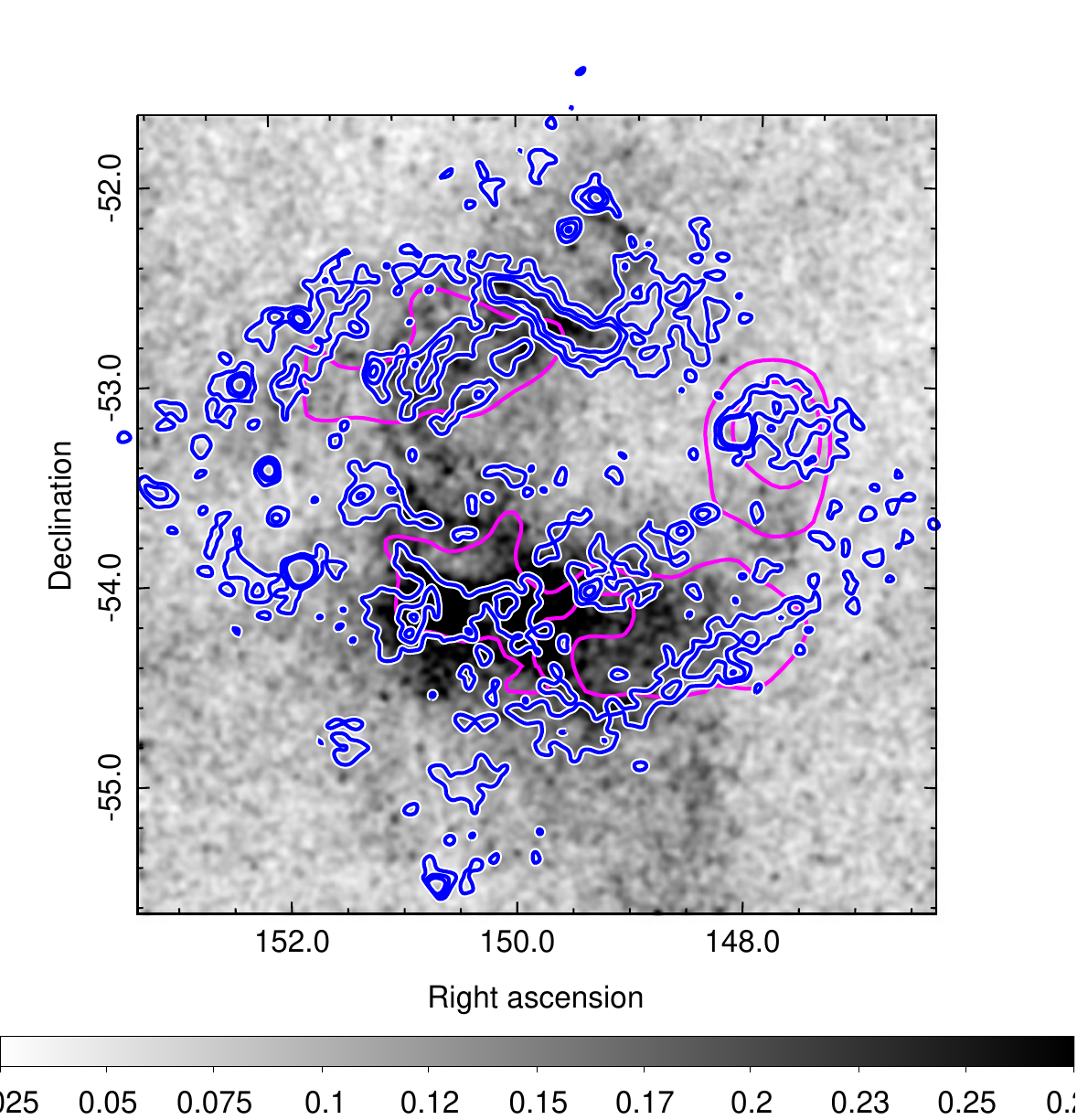}
    \caption{eRASS:4 exposure-corrected intensity sky map, with identical parameters as the one displayed on the left panel of Fig.~\ref{net1}. The blue contours mark the 4850 MHz radio data obtained from PMN \citep{1993AJ....106.1095C} Southern and tropical surveys, and GB6 \citep{1991AJ....102.2041C,1994AJ....107.1829C}. The magenta contours mark the optical H$\alpha$ data obtained from the full-sky H$\alpha$ map of \citet{2003ApJS..146..407F}, with $6'$ FWHM resolution.}
    \label{Radioandoptical}
\end{figure}

Fig.~\ref{Radioandoptical} demonstrates the spatial correlation between the X-ray emission as seen with eROSITA, using eRASS:4 data in the 0.3-1.1~keV energy range, with 4850 MHz radio data from the PMN Southern survey \citep{1993AJ....106.1095C} 
as blue contours, and full-sky H$\alpha$ data of $6'$ FWHM resolution \citep{2003ApJS..146..407F}, as magenta contours. The remnant appears as a fragmented shell of comparable radius in all three energy bands. The radio angular size seems to extend even further compared to the latest estimate of $\sim2.3\degree$ \citep{10.1111/j.1365-2966.2009.14476.x} matching its X-ray counterpart size of $\sim3\degree$, derived in this work. In particular, the bright radio limb to the North of the SNR is well complemented with a region of enhanced X-ray emission, which could possibly be associated with the presence of a CO cloud at that location of the remnant, as reported in \citet{1995MNRAS.277..319D}. An excellent visual correlation is found between the radio and X-ray data at the location of the two bright "blobs" that stand out in the eRASS:4 sky maps. The bright radio source G278.0+0.8, which is most probably of Extragalactic origin, is also detected in eRASS:4 data but masked out since this work focuses on diffuse X-ray emission from the location of the remnant. A diffuse radio emission region, of $24''$ size, observed just to the West of G278.0+0.8 is absent in the X-ray band (or too faint to be observed with eROSITA - eROSITA detects only two point sources from that area). 

No particular association between the 14 bright optical filaments detected towards G279.0+01.1  \citep{10.1111/j.1365-2966.2009.14476.x} with the eRASS:4 data has been found, whatsoever. However, collectively, they are nicely enclosed within the remnant's extension. In this work, we additionally exploited the optical H$\alpha$ data obtained from the full-sky H$\alpha$ map (of 6' FWHM resolution \citep{2003ApJS..146..407F}), which is a conglomerate of the Virginia Tech Spectral line Survey (VTSS) in the North and the Southern H$\alpha$ Sky Survey Atlas (SHASSA) in the South, to examine such an association. Two enhanced regions, in terms of H$\alpha$ emission, become clearly apparent. Both seem to be partially spatially coincident with parts of the remnant that appear bright in the eRASS:4 sky maps and well-aligned with the small-scale fragmented groups of H$\alpha$ filaments \citep{10.1111/j.1365-2966.2009.14476.x}. This association is depicted in Fig~\ref{Radioandoptical}. The H$\alpha$ contours overlaid in the aforementioned image were constructed by omitting nearby, bright optical (H$\alpha$) sources, which do not seem to be associated with the remnant. Therefore, due to the fact that the remnant falls in a highly contaminated H$\alpha$ galactic neighborhood, the available data did not allow us to perform further spectral analysis. Confirmatory spectral results are presented in \citet{10.1111/j.1365-2966.2009.14476.x} which are well-aligned with the shock excitation expected from such an old remnant and provide evidence for prominent [OII] and [OIII] lines (a potentially O-rich remnant).
\begin{figure*}[h]

    \includegraphics[width=0.49\textwidth,clip=true, trim= 1.1cm 0.1cm 1.1cm 0.5cm]{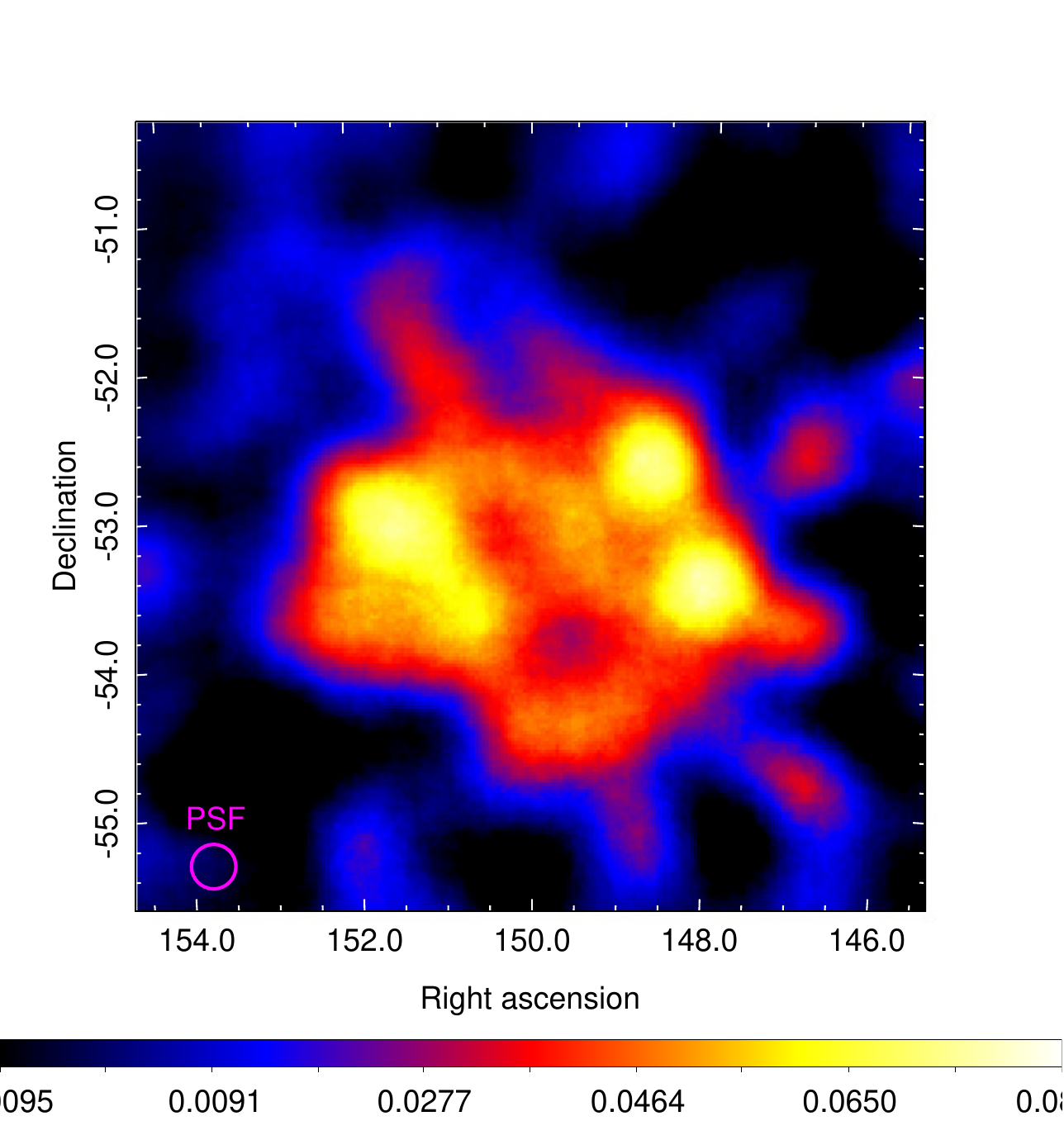}
    \includegraphics[width=0.492\textwidth,clip=true, trim= 0.88cm 0.1cm 1.1cm 0.5cm]{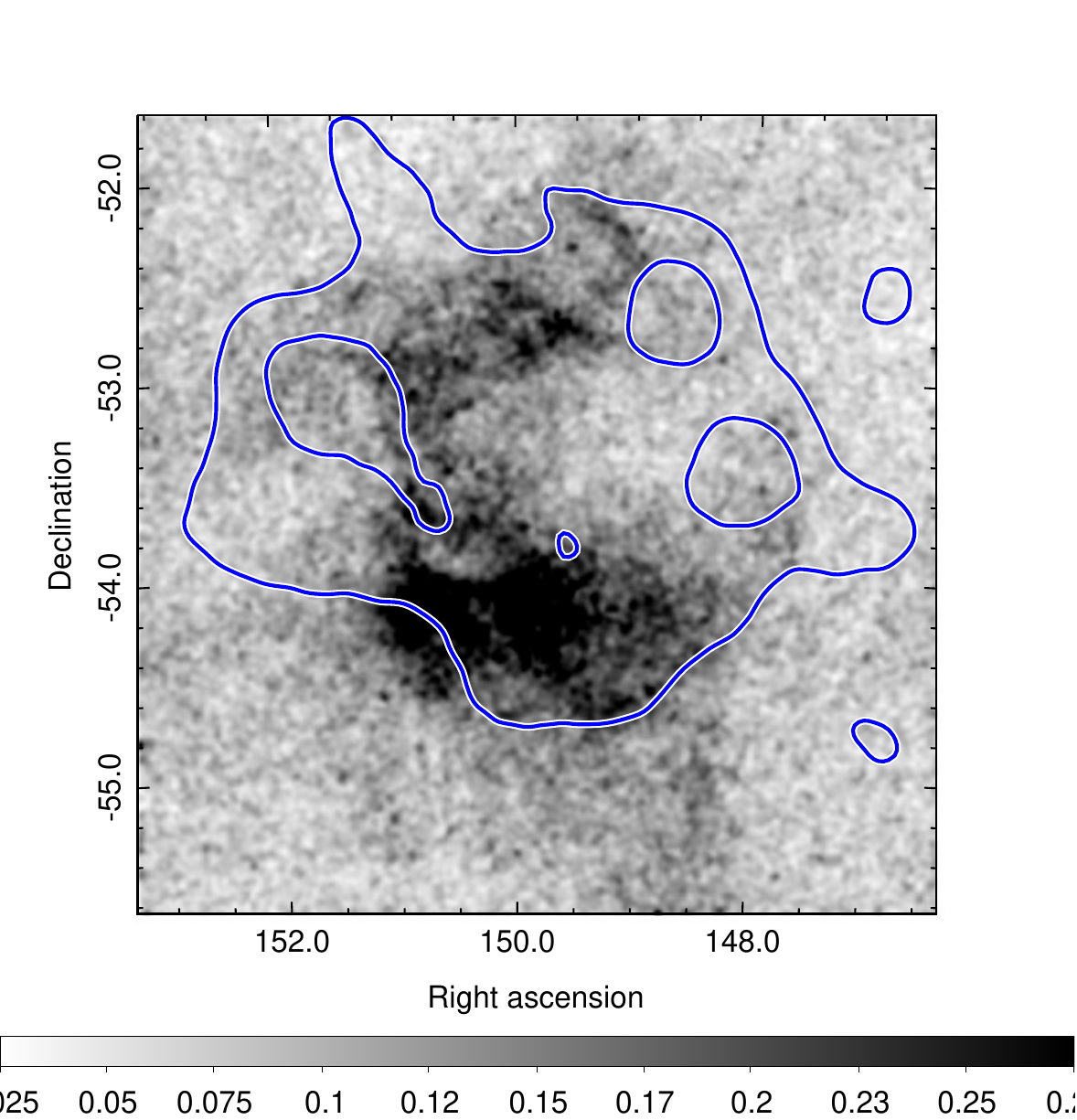}
    \caption{Left panel: $5.3\degree\times5.3\degree$ Fermi-LAT residual count map > 5~GeV centered at the coordinates used in \citet{2020MNRAS.492.5980A}, in units of counts per pixel. The image, of $90''$ pixel size, is convolved with a $\sigma=15'$ Gaussian. The magenta thick circle represents the $68\%$ containment PSF size at 5~GeV energy threshold used for the construction of the residual count map. Right panel: eRASS:4 exposure-corrected intensity sky map, with the same parameters as the one displayed on the left panel of Fig.~\ref{net1}. The blue contours mark the GeV extension of 4FGL J1000.0-5312e as displayed on the Fermi-LAT residual count map on the left panel of the figure.}
    \label{GEVRES}
\end{figure*}

\subsection{GeV $\gamma$-rays}

\citet{2020MNRAS.492.5980A} carried out a detailed Fermi-LAT data analysis from the location of the remnant, which revealed a $2.8\degree$ wide extended GeV source, currently found under the name 4FGL J1000.0-5312e. The GeV source is found to be spatially coincident with the remnant and extends slightly further to the North and East in comparison to the radio shell. $\gamma-~\mathrm{ray}$ emission, likely associated with the remnant, is detected up to $0.5$~TeV with no indication of softening at higher energies. The remnant is, however, not detected in the VHE (Very-High-Energy) band, but the available data is limited (2.7 hours of observational live time with H.E.S.S. \citep{2018A&A...612A...3H}). Later on, \citet{2021ApJ...910...78Z} attempted to fit the multiwavelength spectra of the remnant, as a part of detailed spectral modeling of a sample of 13 SNR characterized by hard GeV $\gamma-~\mathrm{ray}$. \citet{2020MNRAS.492.5980A} discusses both a leptonic and a hadronic scenario for the origin of the GeV $\gamma-~\mathrm{ray}$ emission. However, the detailed spectral modeling of G279.0+1.1 performed by \citet{2021ApJ...910...78Z}  challenges the leptonic processes, claiming that the GeV emission cannot be attributed to leptonic mechanisms due to the evolved state of the remnant, which is of age $>100$~kyr. Thus, a hadronic scenario for the production of gamma-rays is favored. 

In this work, we re-analyzed Pass 8 Fermi-LAT data (P8R3) from the location of the remnant, using $\texttt{fermitools}$ Ver. 2.0.8 standard analysis software and employing $\sim4$ additional years of data (August 2008 to March 2023) in comparison to the previous studies \citep{2020MNRAS.492.5980A}. In more detail, we performed the data reduction and analysis in a similar manner to what is reported in \citet{2020MNRAS.492.5980A}. A region of the size of $40\degree$ centered on the remnant's coordinates (identical to those employed by \citet{2020MNRAS.492.5980A}) was analyzed. Source event class data, front, and back interactions included (evclass=128, evtype=3), were exploited. The maximum data zenith angle was set to $90\degree$. An angular bin size of $0.025\degree$ was selected, in comparison to $0.1\degree$ used in \citet{2020MNRAS.492.5980A}, in order to secure a good sampling of the Fermi-LAT Point Spread Functions (PSF). Modeling of the Fermi-LAT background was performed by including the Galactic diffuse component $\texttt{(gll\_iem\_v07.fits)}$, the isotropic diffuse component $\texttt{(iso\_P8R3\_SOURCE\_V3\_v1.txt)}$ and all sources included in the Fermi-LAT 12 year source catalog (4FGL-DR3). In more detail, the normalization spectral parameter of sources within $5\degree$ of the center of the region of interest was left to vary keeping the remaining spectral parameters fixed to default catalog values. In comparison to \citet{2020MNRAS.492.5980A}, the 4FGL J1000.0-5312e extended GeV source,  seemingly associated with the remnant, appears in the 4FGL-DR3 catalog with different spectral features. In particular, a LogParabola spectrum instead of a simple powerlaw appears to provide the best fit model for the GeV source.

\begin{figure*}[h]
    \centering
    \includegraphics[width=0.415\textwidth,clip=true, trim= 0.9cm 0.1cm 1.1cm 0.5cm]{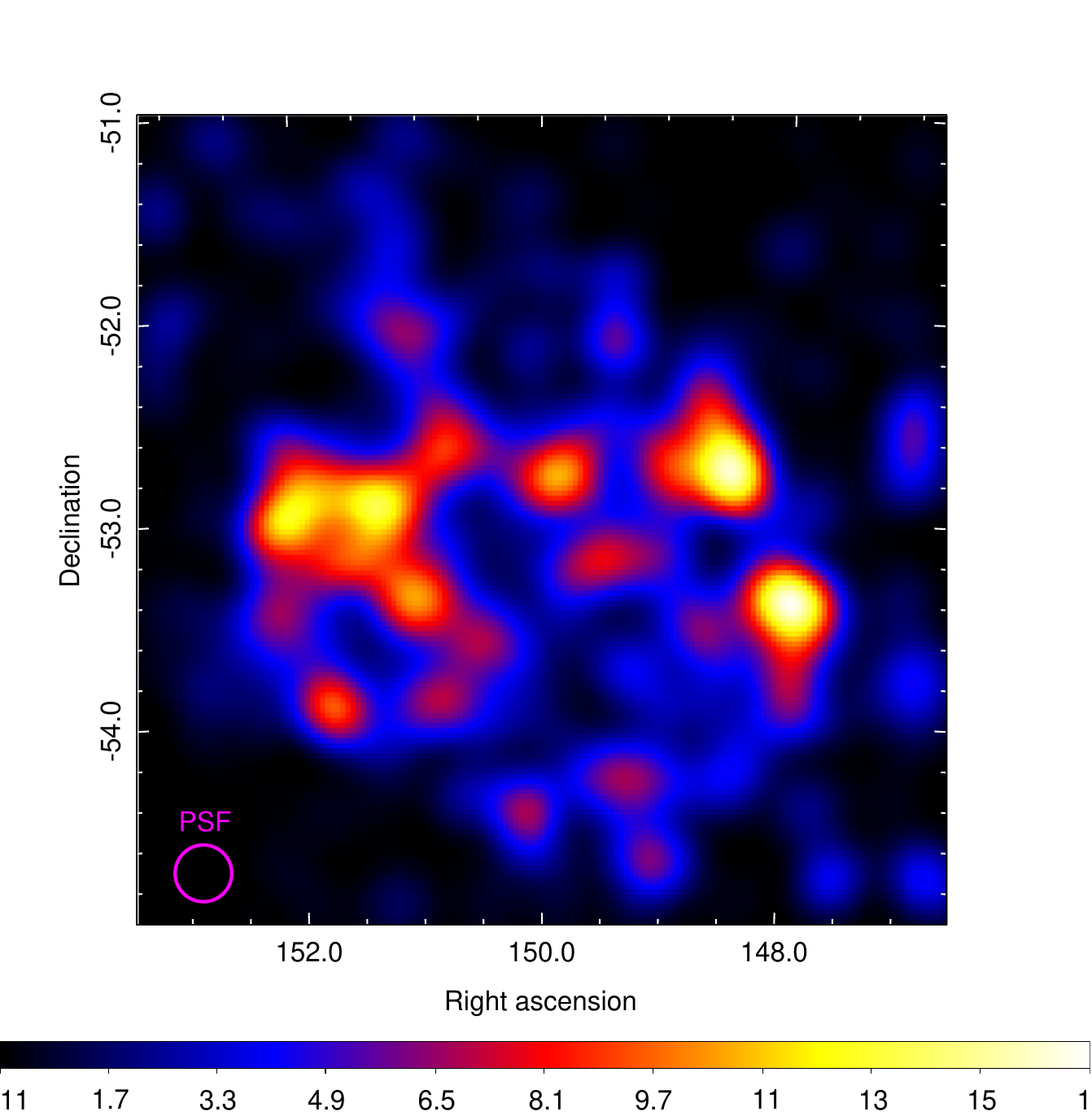}
    \includegraphics[width=0.57\textwidth]{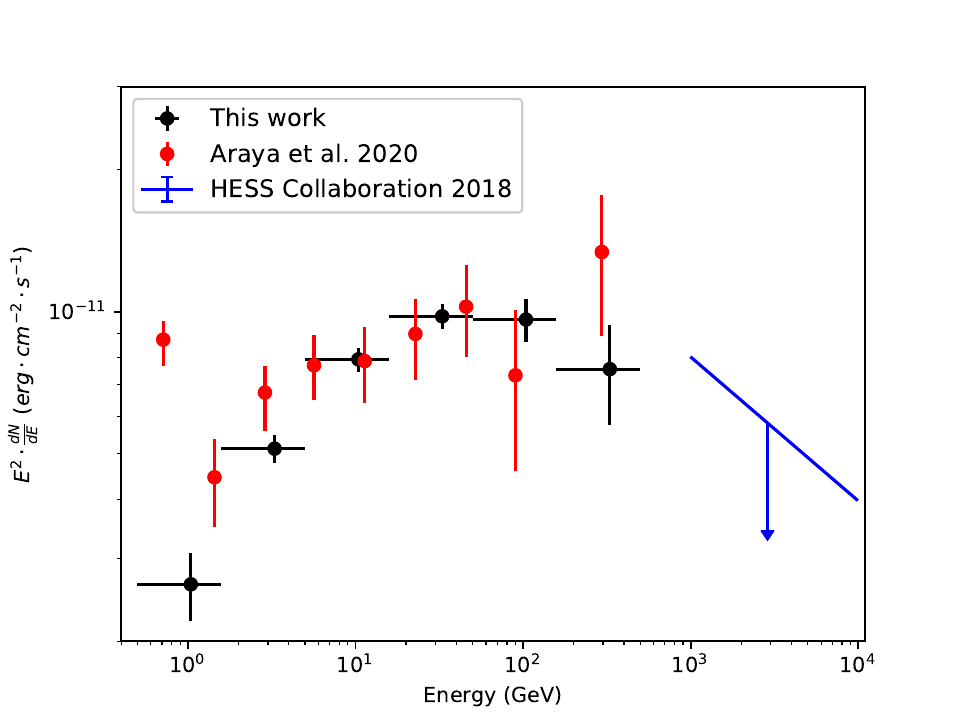}
    
    \caption{Left panel: $4\degree\times4\degree$ Fermi-LAT TS map > 5~GeV centered at the coordinates used in \citet{2020MNRAS.492.5980A}. The image, of $90''$ pixel size, is convolved with a $\sigma=6.75'$ Gaussian. The magenta thick circle represents the $68\%$ containment PSF size, applied at the 5~GeV energy threshold used for the construction of the TS map. Right panel: 4FGL J1000.0-5312e  Fermi-LAT SED. Black dots correspond to the Fermi-LAT spectrum in the 0.5-500~GeV band, obtained in this work. Red and blue dots correspond to GeV Fermi-LAT data reported in \citet{2020MNRAS.492.5980A} and TeV-H.E.S.S. upper limits reported in \citet{2018A&A...612A...3H}, respectively.}
    \label{GEVSPECONLY}
\end{figure*}
\begin{figure*}[h!]
    \centering
    \includegraphics[width=0.49\textwidth]{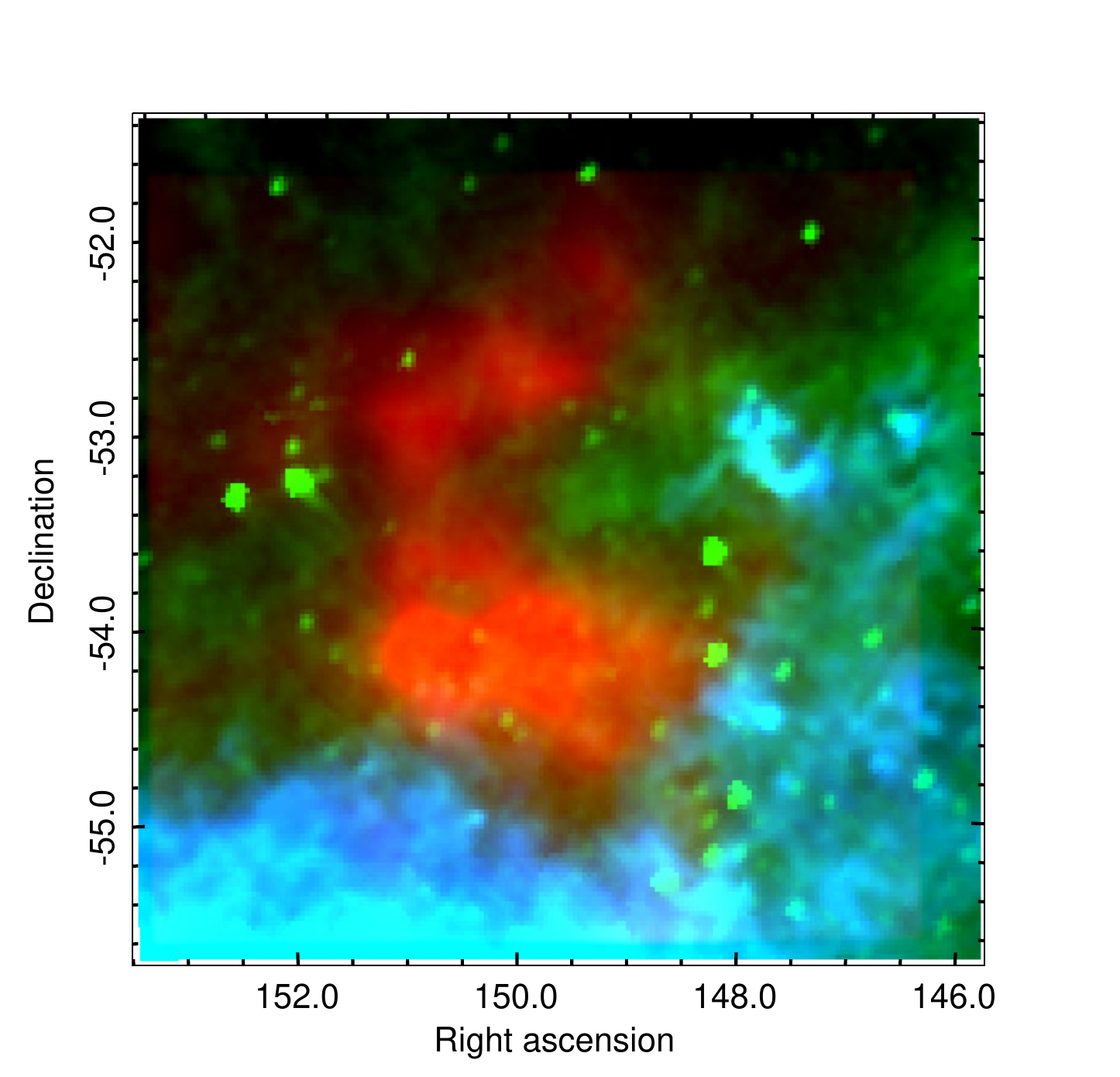}
    \includegraphics[width=0.49\textwidth]{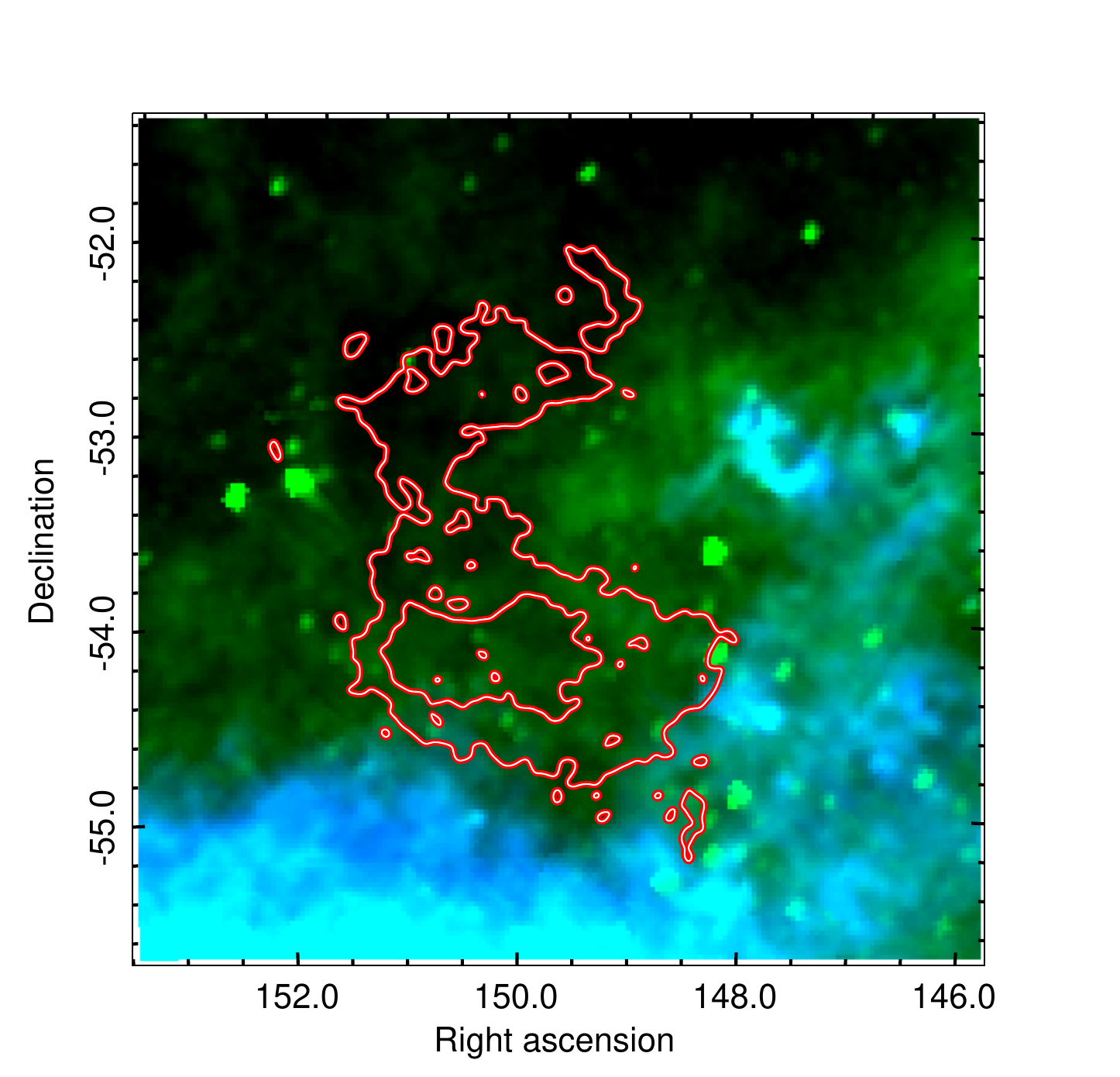}
    \caption{Left panel: RGB image, displaying combined eRASS:4 X-ray data in the 0.3-1.1~keV energy band (red), IRAS 25 $\mu$m data (green), and IRAS 100 $\mu$m data (blue) from the location of the remnant. Right panel: combined IRAS 25 $\mu$m data (green) and IRAS 100 $\mu$m data (blue) from the location of the remnant. The red contours represent two levels of eRASS:4 X-ray data in the 0.3-1.1~keV energy band which we overlaid to IRAS data sets, aiming at inspecting the IR emission at the location of the X-ray excess, as observed with eROSITA, and enhancing the apparent anti-correlation in the two distinct energy bands, i.e., how the IR emission "respects" the X-ray excess emanating from the remnant in the South and West.}
    \label{IRASS}
\end{figure*}
A series of binned analysis procedures for extended Fermi-LAT sources was carried out. Both residual count map and Test Statistic (TS) maps were produced in different energy ranges to thoroughly inspect and refine the gamma-ray emission originating from the remnant's location. Below $5$~GeV,  $\gamma-~\mathrm{ray}$ emission is barely distinguished from nearby GeV emission originating from the Galactic plane. Therefore, for the construction of both types of sky maps, we restricted ourselves to $>5$~GeV to make use of the improved spatial resolution that the Fermi-LAT PSF provides at higher energies. On the left panel of Fig~\ref{GEVRES}, the residual count map above 5 GeV is depicted, which is in good agreement with the corresponding image obtained by \citet{2020MNRAS.492.5980A}. For the TS map production, the detection significance calculation was carried out based on the maximum likelihood test statistic. In particular, the TS maps were obtained by moving an ostensible point source through the grid and obtaining the maximum likelihood fit at each position of the grid. 
From the inspection of the 5-500 GeV TS map at the location of the remnant, which is shown on the left panel of Fig.~\ref{GEVSPECONLY}, we identified 3 main regions of significant GeV emission with $4.6\sigma$, $5.5\sigma$, and $5.8\sigma$ detection significance, respectively. Moreover, $3\sigma$ significance detections are obtained at multiple regions where the remnant extends over. As displayed on the right panel of Fig.~\ref{GEVRES}, the GeV emission
encircles well both the radio and X-ray fragmented shells. It also exhibits an angular extension of $\sim3\degree$, a result obtained by fitting an annulus to the outermost part of the emission. The above value is in agreement with the findings by \citet{2020MNRAS.492.5980A}. 

Comparing eRASS:4 to Fermi-LAT data strongly suggests that the emission at the remnant site spatially anticorrelates in those two energy bands. One obtains such a result by overlaying the TS map contours to the eRASS:4 data, as seen on the right panel of Fig.~\ref{GEVRES}. However, the absence of X-ray emission accompanying the detection of significant GeV emission to the West of the remnant can be easily interpreted when taking a look at the combined eRASS1-4 (red), IRAS 25 $\mu\mathrm{m}$ (green), and IRAS 100 $\mu\mathrm{m}$ (blue) data depicted as an RGB image that is displayed in Fig.~\ref{IRASS}. Here, it becomes apparent that the SNR's structure is a fragmented shell in X-rays because it is partially occluded by dust. The surrounding dust clouds encircle the remnant, absorbing the majority of the X-ray emission to its Western part, and thus forcing this fragmented shell-type appearance of the remnant in the X-ray band. The above claim is confirmed by the X-ray spectral analysis of the remnant, performed in section~\ref{eROSEspectra}. When fitting an appropriate model to the data, a significantly increased absorption column density is obtained to the West of the remnant as shown in Fig~\ref{ABSORB}. The infrared emission to the East of the remnant appears weakened in comparison to the Southern and Western regions, and spectral analysis results do not reveal strong absorption in comparison to its neighboring regions, that are found to be bright in X-rays. Concluding, the prevalence of dust clouds in the surroundings of the remnant seems to be responsible for the morphological anti-colleration of the remnant in the two wavebands. Unlike gamma-ray emission, X-ray emission is subject to absorption.  

The origin of the gamma-ray emission is complex to derive. While there is strong absorption of the X-rays to the Western part of the SNR, Fermi-LAT data reveal two regions of enhanced GeV emission of unclear origin. The GeV blob at the South-West of the remnant overlaps only partially with the G277.731+00.647 HII region and the strong radio source G278.0+0.8, making a possible association unlikely. The GeV blob at the North-West of the remnant spatially coincides with the Western faint CO cloud reported in \citet{1995MNRAS.277..319D}, but it is not consistent with the overall spatial morphology of the cloud. Similarly, the region of bright GeV emission at the North-East of the remnant partially overlaps with the Eastern CO cloud reported in \citet{1995MNRAS.277..319D}. Therefore, it could be the case that the two faint CO clouds, reported in \citet{1995MNRAS.277..319D}, account for two of the three aforementioned, significantly GeV-emitting, regions by interacting with the remnant at the West and thus yielding GeV emission. Overall, at first glance, it is unclear whether the gamma-ray emission there originates from the remnant itself or if it occurs randomly (point source related). A combination of all of the above scenarios may apply. 

To inspect in detail whether the three regions of enhanced GeV emission belong to the diffuse GeV emission originating from the remnant or if they can be attributed to three distinct unknown point sources, we added to the spectral model three new point sources. The new point sources were centered at the location of the three regions which are bright in GeV. By adopting a simple powerlaw spectra for all three sources, we performed the same fitting process and extracted the spectrum from the location of the remnant. The computed spectral shape does not change while the derived flux is only marginally lower. Thus, we are strongly convinced that the three bright blobs are part of the diffuse GeV emission emanating from the remnant and not the result of several point source emission regions.
Regardless, the hard GeV spectral component, detected up to 0.5~TeV, as reported in \citet{2020MNRAS.492.5980A} and confirmed in this work with a slightly modified spectral shape (see section~\ref{Gevspec} for details on the GeV spectral analysis), supports the hypothesis that the extended GeV emission, or at least a good fraction of it, is the result of particle acceleration in the remnant. 
We note that the age of the remnant of $10^6$~yrs (when adopting a distance of 2.7 kpc) implies that particles at TeV energies should already have escaped the SNR. A possible solution to this apparent contradiction with the above findings might come from a revised age estimate as discussed later in the paper.

\section{\textit{X-ray spectral analysis}}\label{eROSEspectra}

\subsection{\textit{eROSITA spectra}}
\label{ROSITAspec}
\begin{figure*}[h!]
    \centering

    \includegraphics[width=0.5\textwidth,clip=true, trim= 1.2cm 0.1cm 1.3cm 0.5cm]{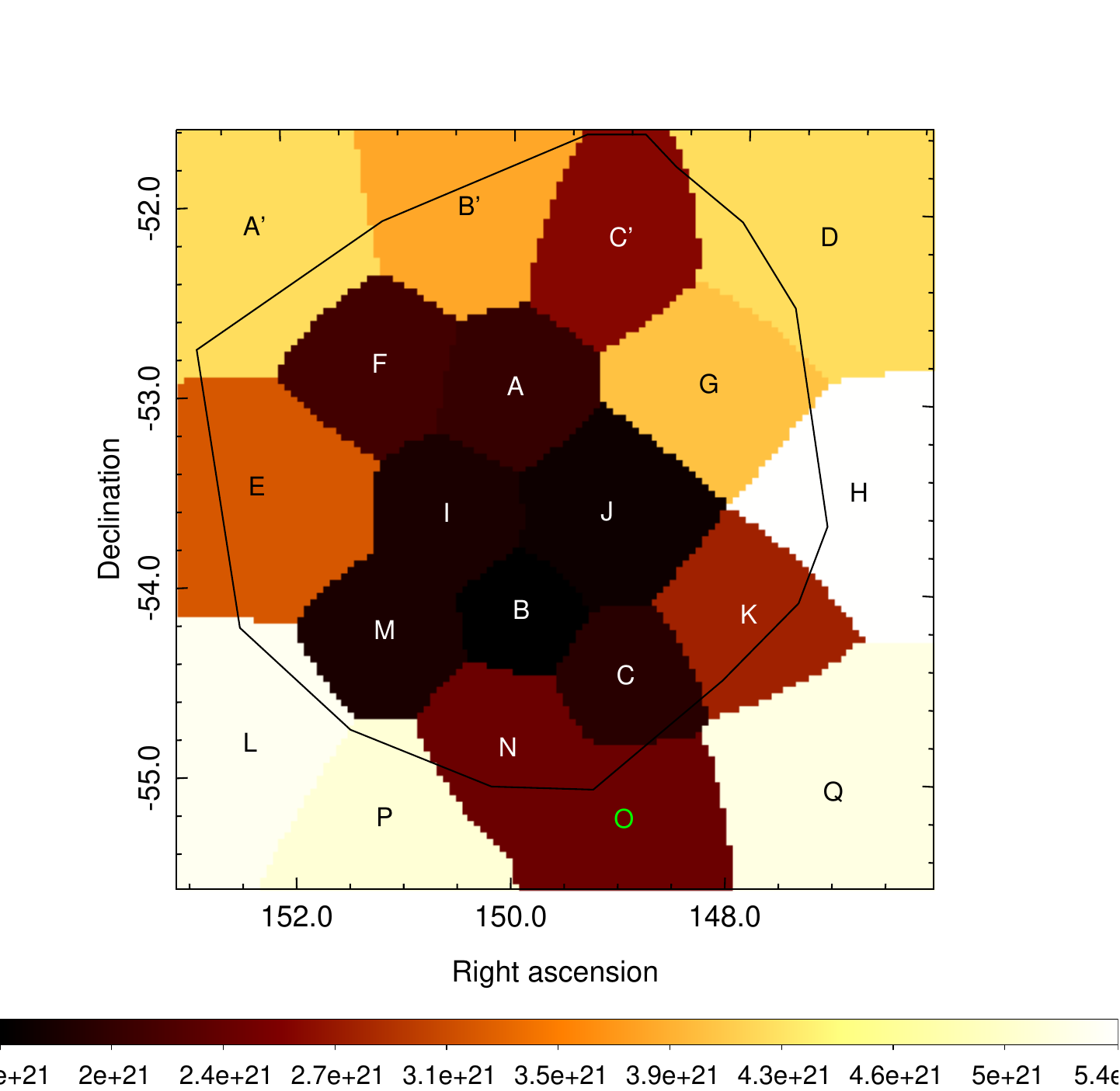}\includegraphics[width=0.52\textwidth]{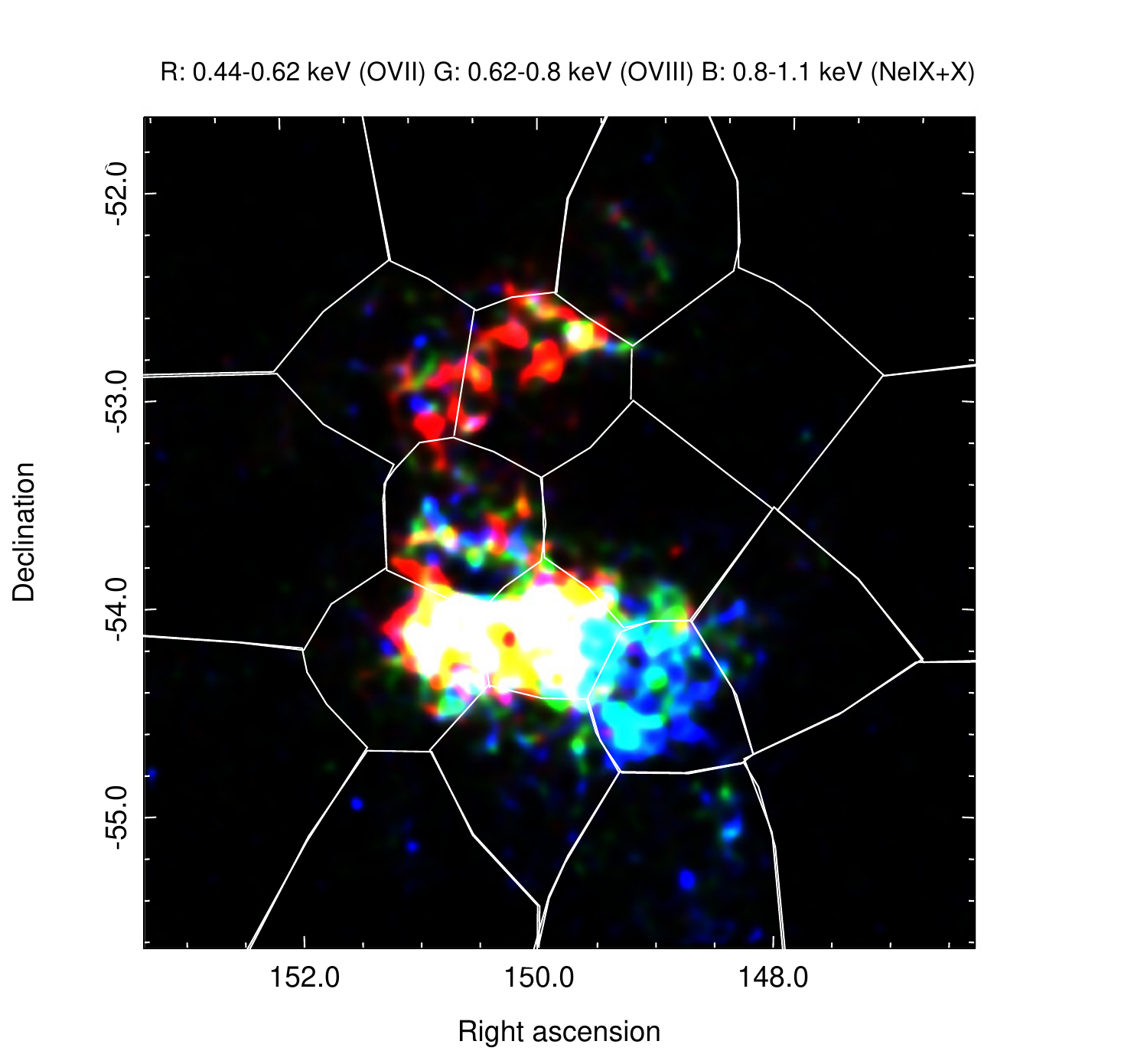}
    \caption{Left panel: Absorption column density map (in units of $\mathrm{cm^-2}$) from the location of the remnant, as computed by the best-fit absorption column density. Values are obtained from each distinct sub-region defined by the Voronoi binning algorithm. The region selected for spectral analysis of the entire remnant is shown as a black line. Regions of significant diffuse X-ray emission from the remnant are displayed in black letters whereas those surrounding regions containing faint diffuse X-ray emission from the remnant are displayed in white letters. Finally, in green the region which contains diffuse X-ray emission unrelated to the remnant is shown. Right panel: eRASS:4 RGB image (R: 0.44-0.62~keV (OVII), G: 0.62-0.80~keV (OVIII), B: 0.80-1.10~keV (NeIX+X)), identical to the lower right panel of Fig.~\ref{NARROW} but in power scale aiming to reveal the strongest elemental abundance at each sub-region where we perform spectral analysis. It displays the distribution of the different elemental abundances detected across the remnant. White contours represent the 20 distinct regions obtained from the Voronoi binning analysis to be used for further spectral analysis.} 
    \label{ABSORB}
\end{figure*}

\begin{figure*}[]
    \centering
    \includegraphics[width=0.496\textwidth]{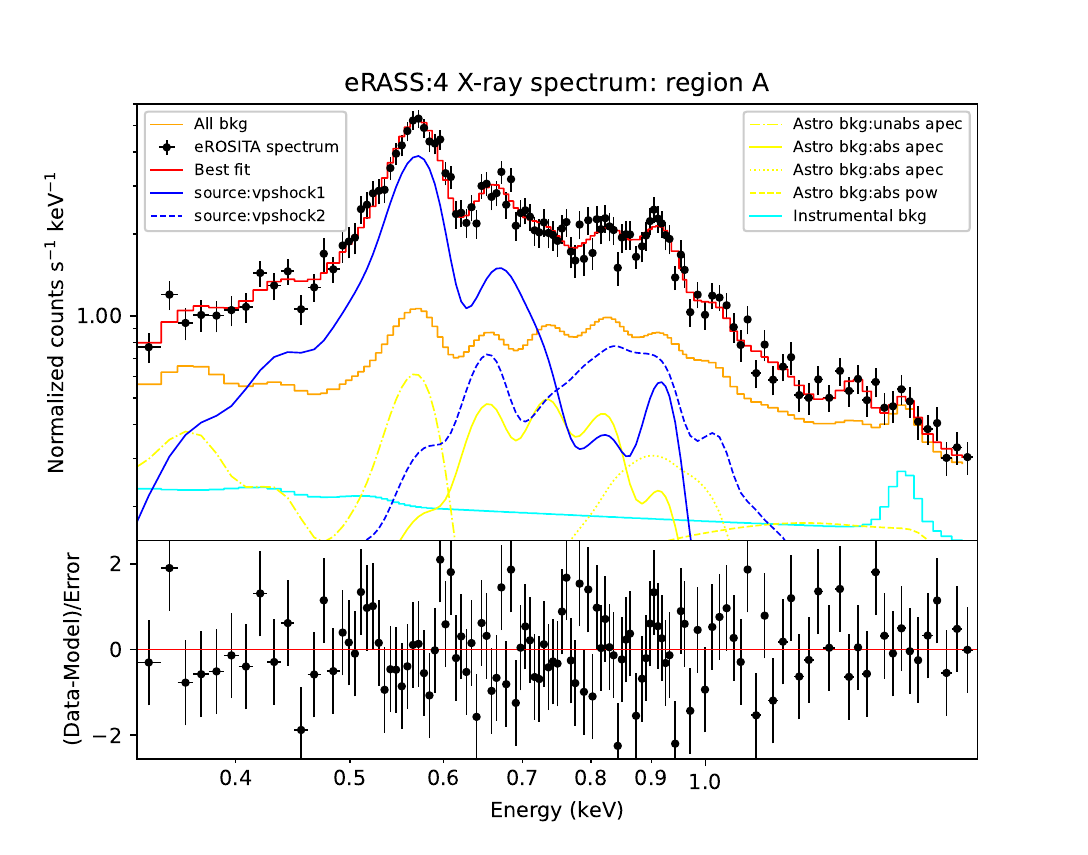}
    \includegraphics[width=0.498\textwidth]{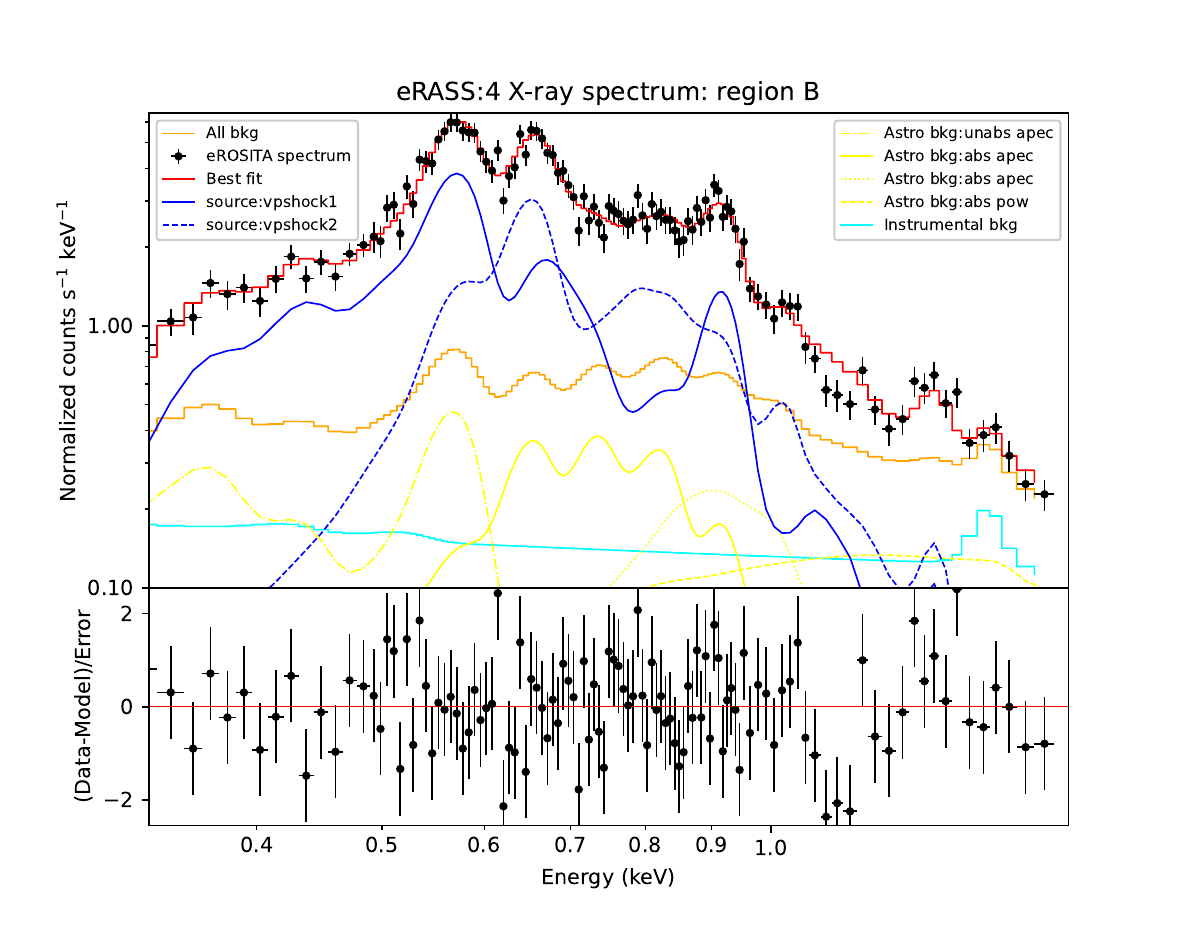} 
    \includegraphics[width=0.501\textwidth]{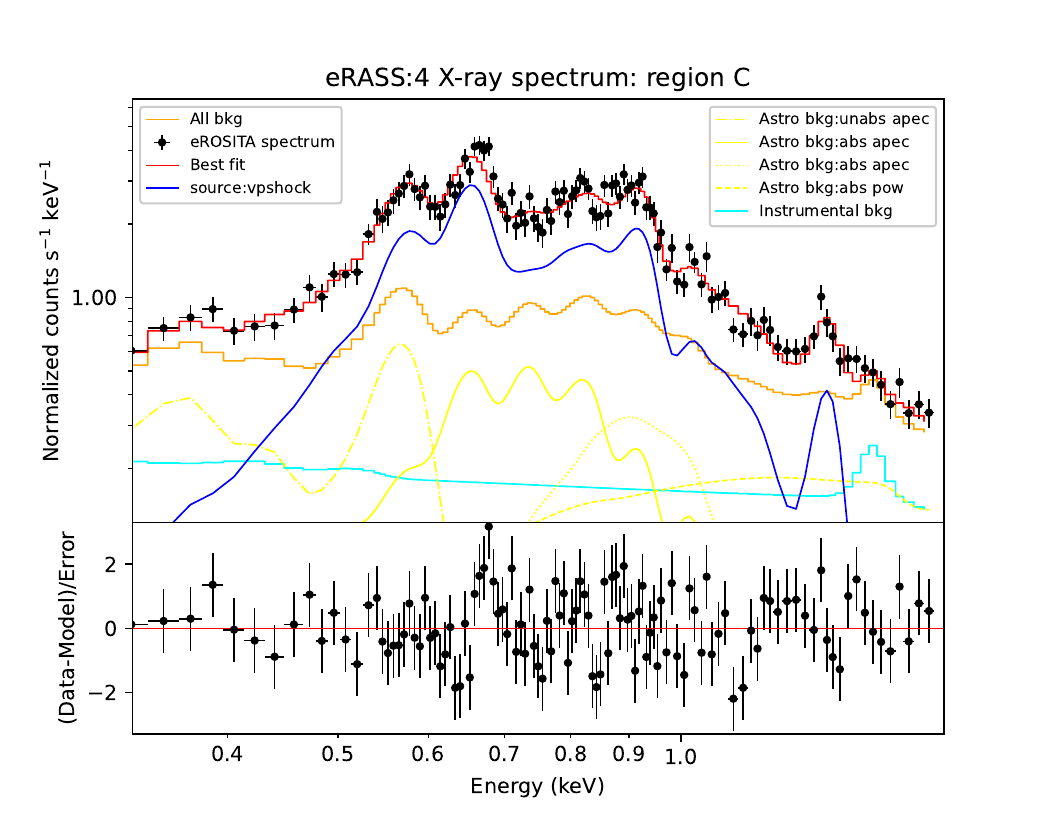}
    \includegraphics[width=0.493\textwidth]{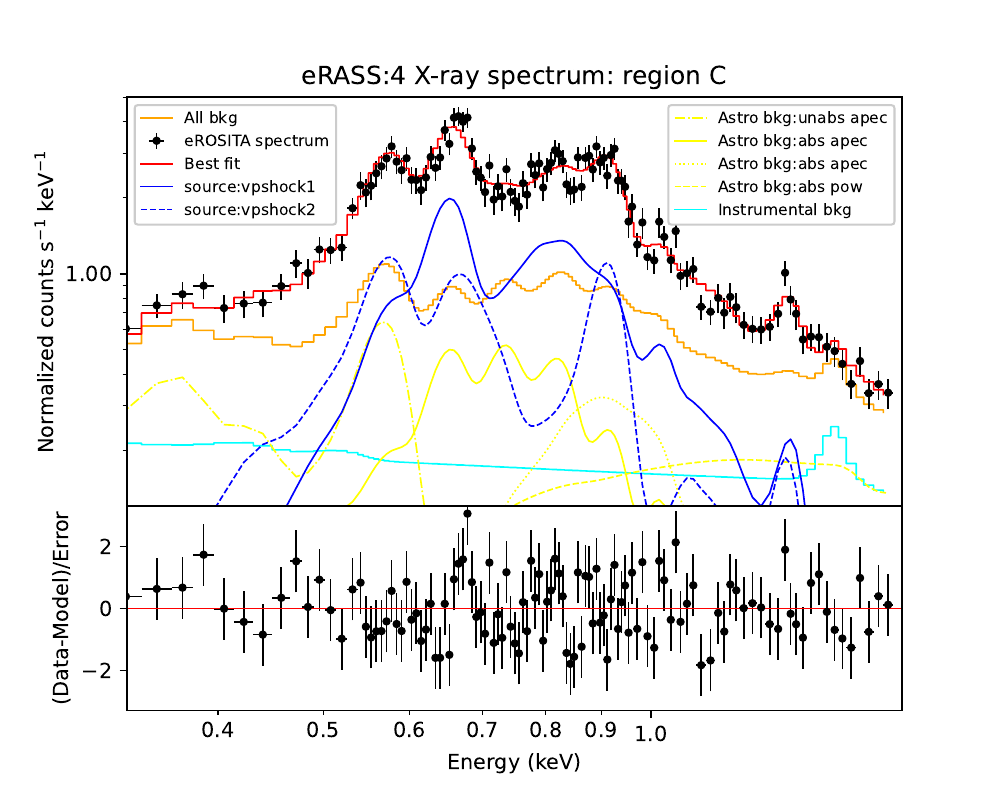}

    \caption{X-ray spectrum, eRASS:4 data in the 0.3-1.7 keV energy band, from the selected representatives regions of the remnant, which are demonstrating the spectral shape change detected across the remnant. Upper left: region A, \texttt{tbabs(vpshock+vpshock)}. Upper right: region B, \texttt{tbabs(vpshock+vpshock)}. Lower left: region C, \texttt{tbabs(vpshock)}. Lower right: region C, \texttt{tbabs(vpshock+vpshock)}. }
    \label{SPEC1}
\end{figure*}

Fig~\ref{RGB} indicates potential temperature variation across the remnant, as described in section~\ref{eROSITAA}. Therefore, aiming at assessing in detail the nature of the X-ray emission emanating from the remnant, a spectral extraction process was performed from 20 distinctive regions, which are shown on the right panel of Fig.~\ref{ABSORB} by using \texttt{SAOIMAGE DS9} \citep{joye2003new}. The selection of the regions was optimized based on the \citet{2003MNRAS.342..345C} Voronoi binning algorithm that we run on the 0.3-1.1~keV intensity map of the remnant, depicted in Fig.~\ref{net1}. Here, the image was rebinned to $2'$ pixel size so that a single pixel contains sufficient number of counts. A signal-to-noise ratio of $\mathrm{S/N=110}$ was set given the relatively faint appearance of the SNR in the X-ray energy band. The obtained regions are shown on both panels of Fig.~\ref{ABSORB}. 
We also extracted the spectrum from the entire remnant. The selected polygonal region that combines the emission from the whole remnant is overlaid on the left panel of Fig.~\ref{ABSORB} with black contours. Additionally, we extracted the spectrum from a region identical to the 0823031401 XMM-Newton observation, to be able to directly compare the X-ray spectra obtained from the two distinct instruments from the exact same location of the remnant. The X-ray spectra comparison between the two instruments is presented in sec.~\ref{XMMSPECI}.
eRASS:4 data were utilized in the spectral analysis procedure, excluding data recorded by TM5 and TM7 since the light peak suffering of those cameras \citep{2021A&A...647A...1P} strongly affects the lower energy regime where the SNR is observable. In addition, the spectra were grouped, using the grppha FTOOLS\footnote{http://heasarc.gsfc.nasa.gov/ftools/} task, to achieve a minimum of 50 counts per single bin. Spectral extraction was also performed from three additional regions representative of the background, aiming at inspecting potential background variations. 
The regions are shown on the left panel of Fig.~\ref{nett1}. Their selection was optimized based on the contamination of the surrounding regions. In particular, regions located towards the South of the remnant were excluded since they exhibit strong X-ray emission of unknown origin. However, we speculate that the nature of the emission could potentially be associated with another remnant since there is an apparent radio arc in the 
PMN data that seems to encapsulate nicely the X-ray excess. Despite the differences, one obtains compatible results when modeling the spectrum obtained from each of those off-source regions, mainly discrepancies in the normalization value of the astrophysical background components. The obtained spectral source parameters for the best-fit models are consistent within $1\sigma$ errors for all background regions when applied to the simultaneous fitting of the on-source regions, as discussed below. Therefore, the black circled region was chosen to represent the background X-ray emission from the whole remnant. For the background model, the best-fit model spectral parameters used in the simultaneous fitting of the source and background emission are fixed to the best-fit values. The normalization values are rescaled according to the area of the corresponding on-source regions.
The fitting of the background regions is performed by adopting the following model in Xspec notation: \texttt{apec+tbabs(apec+apec+pow) + gaussian + expfac(bkn2pow + powerlaw + powerlaw) + powerlaw + gaussian + gaussian + gaussian+ gaussian + gaussian + gaussian}. This can be broken down to the contribution from the astrophysical background (\texttt{apec+tbabs(apec+apec+pow)}) that includes the Local Hot Bubble (LHB) low temperature plasma, the Galactic Halo (GH) plasma, and the Cosmic X-ray Background (CXB) originating from the combined emission of unresolved Active Galactic Nuclei (AGN), and to the particle or instrumental background of eROSITA which can be best described by a combination of power law and Gaussian model components in the particular energy range that the X-ray fitting is performed: \texttt{gaussian + expfac(bkn2pow + powerlaw + powerlaw) + powerlaw + gaussian + gaussian + gaussian+ gaussian + gaussian + gaussian}. 
To avoid likely spectral contamination from point sources that fall within the extension of the remnant, we masked out with a 110 arcsec mask radius all point sources detected with a 3$\sigma$ significance level or higher, based on the latest eROSITA point source catalog. The \texttt{srctool} eSASS task was employed for the spectral extraction procedure, while Xspec (X-ray spectral fitting package; Ver. 12.12.1) was utilized to perform the spectral fitting. Given the relatively faint appearance of the remnant in the X-ray band, C-statistics \citep{1979ApJ...228..939C} were selected in the fitting procedure. A simultaneous fit of the on-source and background emission is favored over the subtraction of the background emission from the on-source spectra. 
As a cross-check, we also performed a spectral analysis on the background-subtracted spectra. Given that the normalization of the background models in the simultaneous fitting procedure was not left to vary, both methods yield, as expected, consistent results. The methodology that we employed for the fitting process of each individual region is as follows. We started by obtaining the best fit of the background subtracted spectra which give us a rough estimate of the emission nature originating from the remnant. We then proceeded to the simultaneous fitting process using as initial input parameters, of the source emission, the ones obtained from the background subtraction strategy. Since the emission emanating from the remnant is purely thermal, one can describe the optically thin plasma either in a non-equilibrium ionization state (NEI), which usually applies to young and middle-age remnants, or in collisionally ionization equilibrium (CIE) which is representative of older remnants. In our analysis, even if the remnant is considered to be old, we tested both types of models. In particular, the VAPEC model of collisionally ionized diffuse gas as a CIE model, and the non-equilibrium ionization collisional plasma (VNEI) as well as the constant temperature plane parallel shock plasma (VPSHOCK \citet{Borkowski_2001}), as NEI model representatives in Xspec notation, were employed in the fitting process. The Galactic absorption towards the source was modeled with the TBABS absorption model by \citet{2000ApJ...542..914W}. Finally, the spectral fitting is performed in the 0.3-1.7~keV energy range since above 1.7~keV the background becomes strongly dominant.
\begin{figure*}[]
    \centering
   
    \includegraphics[width=0.419\textwidth,clip=true, trim= 1.2cm 0.1cm 1.3cm 0.5cm]{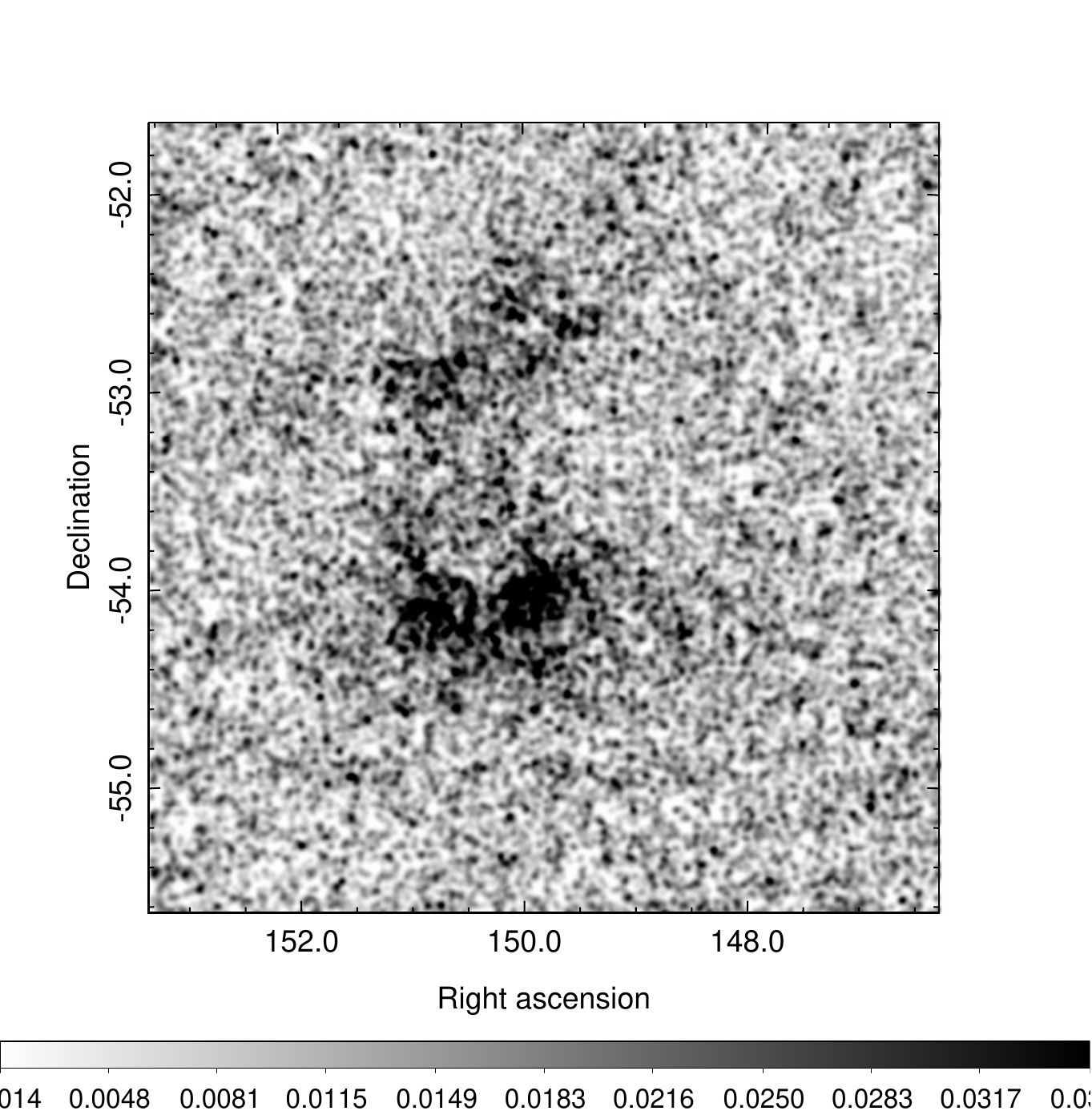}
    \includegraphics[width=0.419\textwidth,clip=true, trim= 1.2cm 0.1cm 1.3cm 0.5cm]{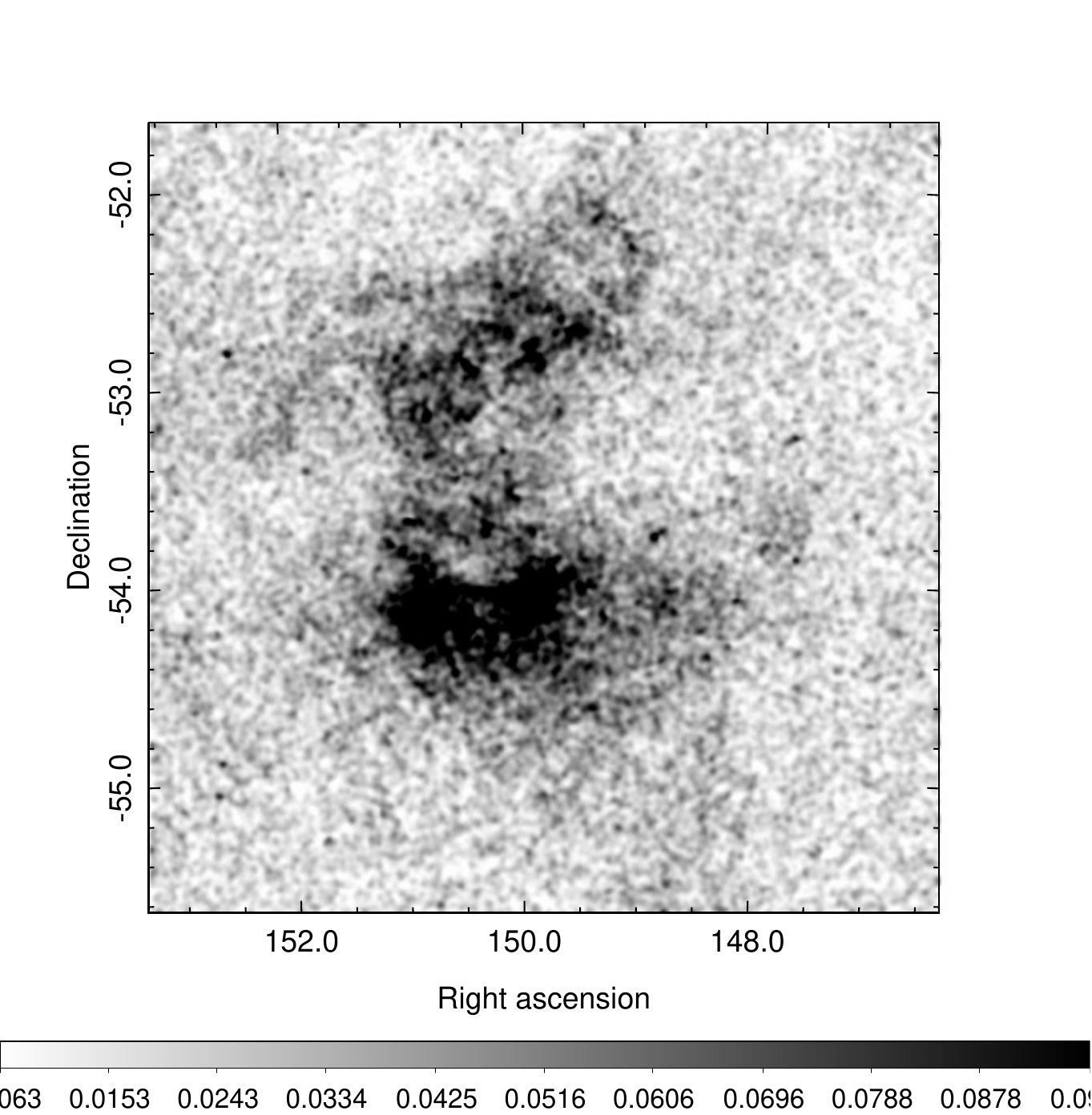} 
    \includegraphics[width=0.419\textwidth,clip=true, trim= 1.2cm 0.1cm 1.3cm 0.5cm]{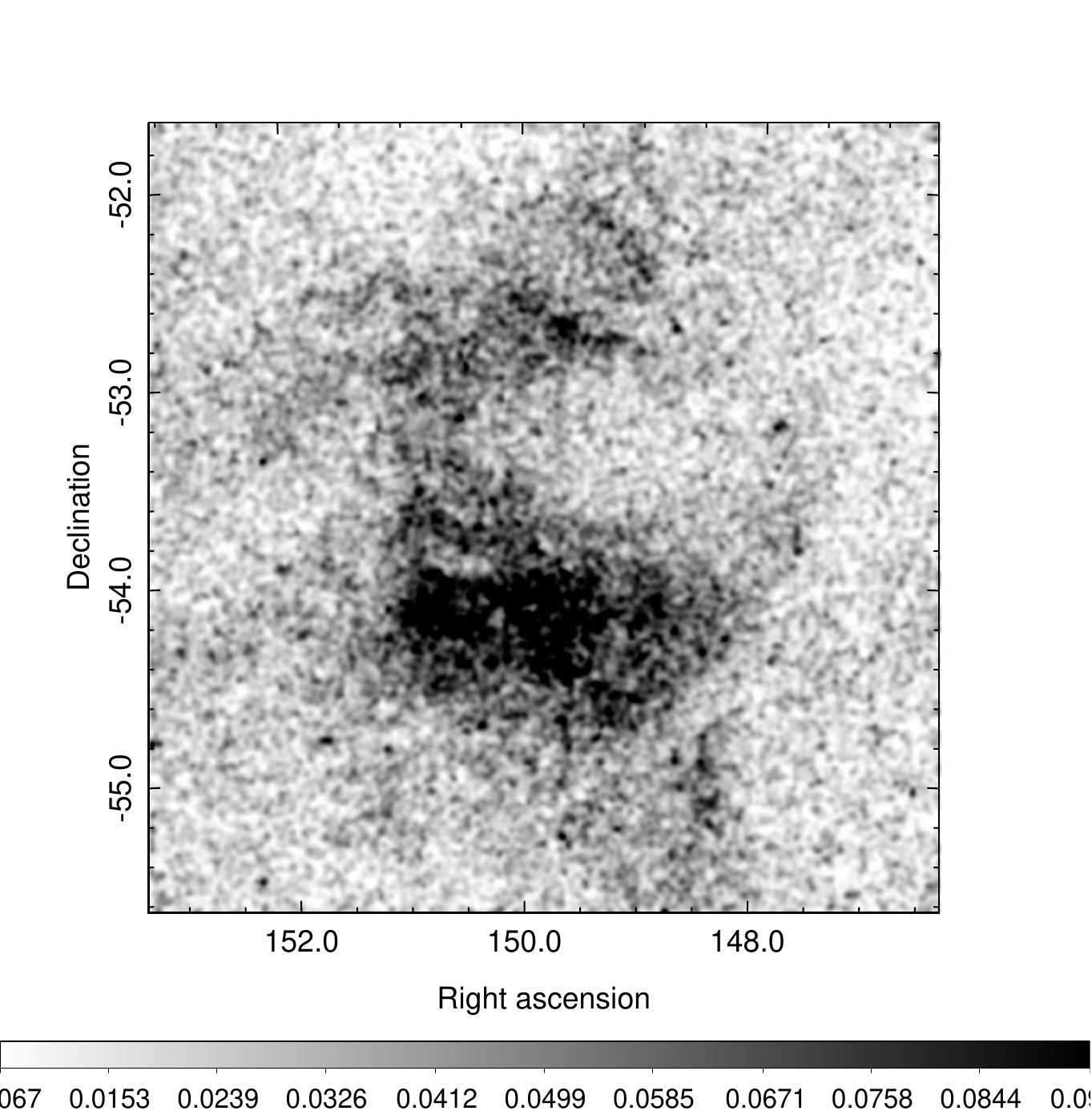}
    \includegraphics[width=0.419\textwidth,clip=true, trim= 1.2cm 0.1cm 1.3cm 0.5cm]{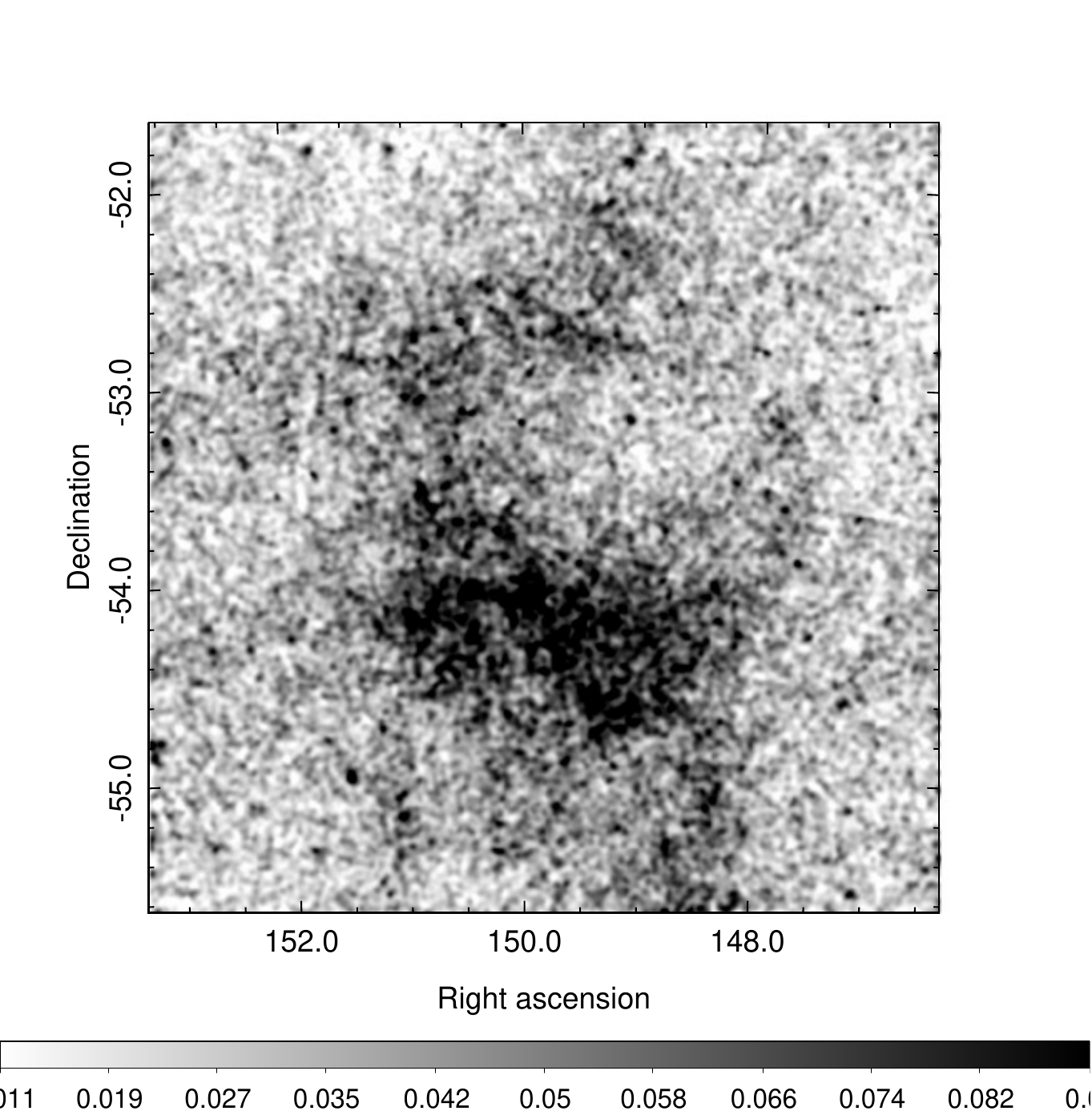}
    \includegraphics[width=0.419\textwidth,clip=true, trim= 1.2cm 0.1cm 1.3cm 0.5cm]{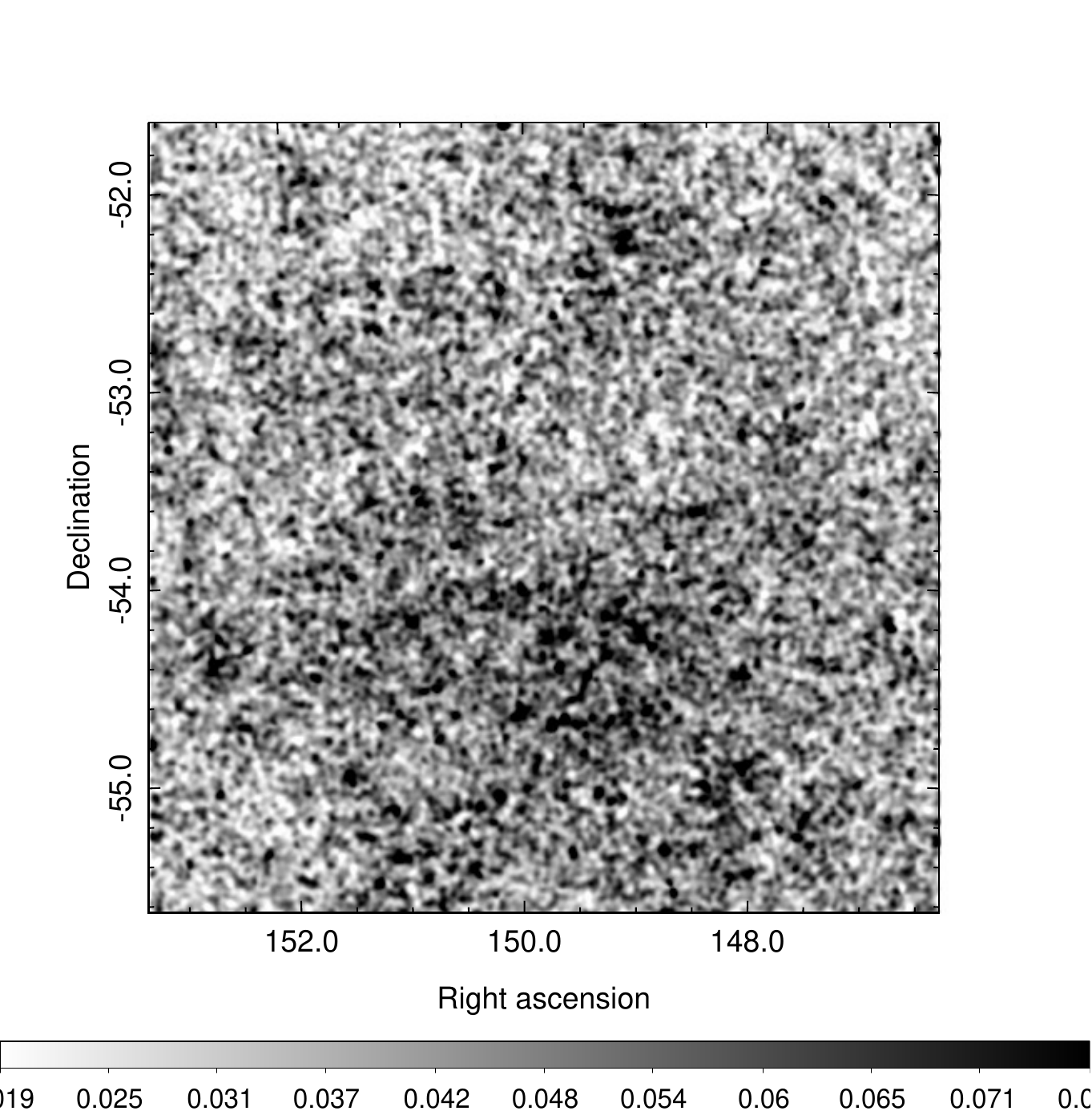}
    \includegraphics[width=0.419\textwidth,clip=true, trim= 1.2cm 0.1cm 1.3cm 0.5cm]{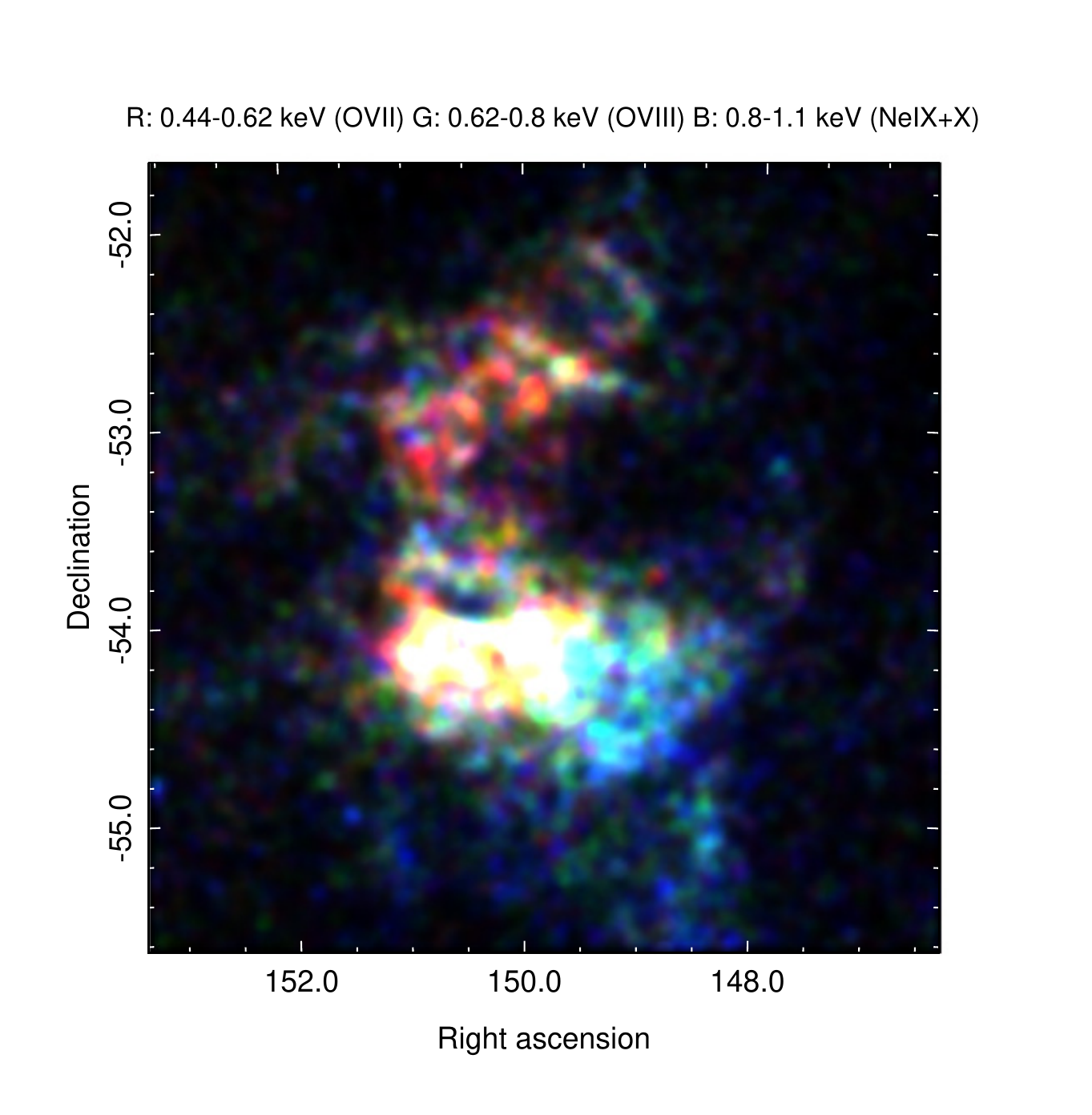}
    \caption{eRASS:4 exposure corrected intensity sky maps in units of counts per pixel. The five distinct panels depict narrow, spectrally-motivated energy bands: 0.3-0.44~keV (N, upper left), 0.44-0.62~keV (OVII, upper right), 0.62-0.80~keV (OVIII, middle left), 0.80-1.10~keV (NeIX+X, middle right), and 1.10-2.10~keV (Mg, lower left). The sixth one (RGB, lower right) displays the distribution of the different elemental abundances detected across the remnant.}
    \label{NARROW}
\end{figure*}
For the majority of the selected regions, when one tries to fit the emission spectrum with one of the aforementioned single-component models by keeping the elemental abundances fixed to solar values results in a poor fit. On the contrary, varying Oxygen (O), Neon (Ne), and Magnesium (Mg) significantly improves the fitting results. The latest assertion is confirmed by the clear identification of O VII ($\sim$0.56 keV), O VIII Ly$\alpha$ ($\sim$0.65 keV), and Ne IX ($\sim$0.905 keV) emission line features in the source spectrum. The Ne X Ly$\alpha$ line ($\sim$1.02 keV), as well as the He-like Mg XI unresolved triplet ($\sim$1.35 keV), and the Mg XII Ly$\alpha$ line ($\sim$1.47 keV) are also present for some of the regions. Finally, some Fe L-transitions are prominent in a number of regions selected for spectral analysis, in the 0.7-1.2 keV energy band, but a definitive identification of the latter is beyond the scope of this work. Since those Fe-L lines do not leave strong residuals, as is the case in e.g., \citet{2015PASJ...67...16K} due to the errors in Fe-L modeling in current version of Xspec code \citep{Borkowski_2006,10.1093/pasj/63.sp3.S837}, we did not add any additional Gaussian lines around those energies. On Fig.~\ref{NARROW} we display eRASS:4 intensity sky maps in narrower, spectrally-motivated energy bands: 0.3-0.44~keV (N), 0.44-0.62~keV (OVII), 0.62-0.80~keV (OVIII), 0.80-1.10~keV (NeIX+X), and 1.10-2.10~keV (Mg). As can be seen in Fig.~\ref{NARROW}, the North-Eastern parts of the remnant contain high abundances of OVII and OVIII, whereas the South-Western parts contain high abundances of OVIII, Ne, and Mg. It is interesting to note that the two bright blobs contain high abundances of all three elements. A similar pattern is evident in the spectra as displayed in Fig.~\ref{SPEC1} which displays significant changes in the spectral shape. In particular, the OVII abundance declines whereas the Ne and Mg abundances increase, as one moves across the remnant from the North-East (region A) to the South-West (region C).

\begin{figure*}[h!]
    \centering
    \includegraphics[width=0.501\textwidth]{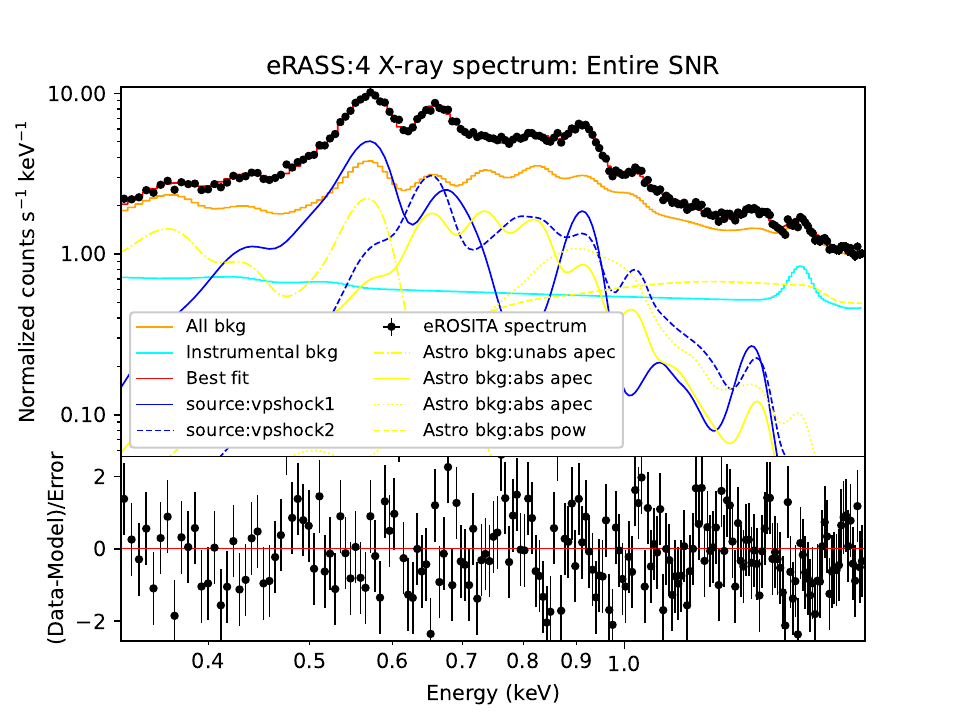}
    \includegraphics[width=0.494\textwidth]{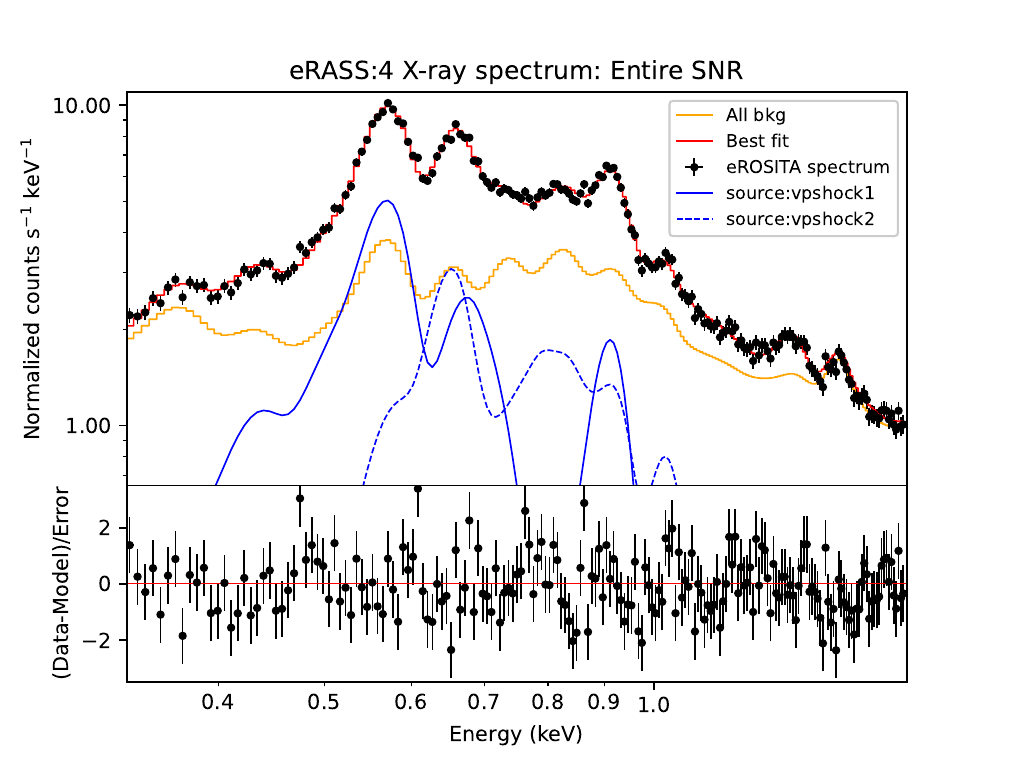}

    \caption{X-ray spectrum, eRASS:4 data in the 0.3-1.7 keV energy band, from the entire remnant. Left panel: All distinct components contributing to the spectrum (yellow: astrophysical background, cyan: instrumental background, and blue: source. Right panel: Source components, in blue, and total background contribution to the spectrum, in orange.}
    \label{SPEC2}
\end{figure*}

For the on-source regions with significant diffuse X-ray excess, the obtained reduced chi-squared values of the single-component model, after letting the above elemental abundances vary, might indicate that a single-component model does not sufficiently describe the source's emission. In particular, for the majority of the regions significant residuals are apparent at the Ne X Ly$\alpha$ line energy (at $\sim$1.02 keV) which cannot be improved neither when switching from an equilibrium to a non-equilibrium model nor when varying the elemental abundances of the corresponding model. Strong residuals are also present in the 0.7-1.0 keV energy range. Therefore, multiple-component models were also employed for those regions, aiming at improving the quality of the fit and effectively describing the source emission. In Fig.~\ref{SPEC1} we show the X-ray spectral fitting results obtained from three representative regions of the remnant (in a simultaneous fitting of the source and background emission). The regions were chosen to depict the substantial X-ray spectral variation found across the remnant, in particular, how the X-ray spectral shape changes as one moves from the North-East to the South-West of the SNR. Fig.~\ref{SPEC2} depicts the X-ray spectral fitting results that one obtains when extracting the spectrum from the entire SNR. The letters A, B, and C are used to identify these three regions in Fig.~\ref{ABSORB}. For the rest of the regions, the X-ray spectral fitting results are summarized in appendices (sec.~\ref{sec:append2}), and displayed in Fig.~\ref{ALLSUBSPEC}. For each individual region, we started the fitting procedure attempting to fit the data with the simplest possible model for an evolved SNR i.e., the VAPEC model (CIE). We then switched to non-equilibrium models (NEI), which provide a much higher temperature plasma and a significantly lower absorption column density compared to CIE models, as shown in Tab.~\ref{TABIS1}. Therefore, we conclude that NEI models are necessary at least in some regions since the derived absorption column density based on the known distance of the remnant is well-aligned with the spectral fit results of NEI models whereas it falls short of the CIE model (see sec.~\ref{dist} for a detailed discussion). This is even more true when a revised distance estimate to the SNR is considered (refer to sec.~\ref{dist}). We stress, however, that a \texttt{tbabs(vpshock)} model for regions A and C and a \texttt{tbabs(vapec)} model (ignoring $N_\mathrm{H}$ constraints) for region B seem to describe the remnant's spectral data relatively well. Even if in both cases (single CIE or single NEI model) acceptable fits were derived for specific sub-regions (under certain adjustments which are described in detail in sec.~\ref{sec:append1}), the obtained results point toward the fact that multi-temperature models further improve the fitting process.

Therefore, as a next step, we considered multiple component models, and in particular two temperature plasmas, aiming at improving our fitting results.  
We note that the best-fit spectral results for each sub-region are reported in the appendices (refer to Tab.~\ref{TABISOO}, sec.~\ref{sec:append2}). Here, we give an overview of the most important findings. In this work, we attempted to model the remnant's spectrum with all possible combinations of two temperature plasma models (i.e., equilibrium models (\texttt{tbabs(vapec+vapec)}), non-equilibrium models (\texttt{tbabs(vnei+vnei)}, \texttt{tbabs(vpshock+vpshock)}), and mixed models (\texttt{tbabs(vapec+vnei)}, \texttt{tbabs(vapec+vpshock)}).
As shown in Tab.~\ref{TABIS1}, no strong preference among the aforementioned models was obtained. However, it is worth to note that a significantly improved fit is obtained for regions A and B compared to single-component models, whereas region C can be sufficiently described by single-component models. All best-fit results are described in detail in appendices (sec.~\ref{sec:append1}) and reported in Tab.~\ref{TABIS1}. 
Finally, when fitting the spectrum of the entire remnant, two temperature plasma components provide by far better fit quality compared to single-temperature models (either CIE or NEI). In particular, among all models mentioned above, a two-temperature plasma component in non-equilibrium (\texttt{tbabs(vphock+vpshock)}), letting O and Ne to vary, provides a fit of $\chi^2/dof=1.19$ (with a total flux of $F_{total}=1.48^{+0.61}_{-0.42}\cdot 10^{-9}~\mathrm{erg/cm^2/s}$). The best fit parameters of this model are reported in Tab.~\ref{TABIS1}.

An identical spectral fitting approach was employed for the rest of the regions obtained from the Voronoi binning algorithm, when the source and background emission were simultaneously fitted with independent models. 
The obtained spectral fits for the rest of the sub-regions are summarized in appendices (sec.~\ref{sec:append2}).
The main parameters of the best-fit for the three representative regions, as well as those obtained from the entire remnant, are summarized in Tab~\ref{TABIS1}. The best-fit models from each distinct region used for the spectral fitting process confirm that the X-ray emission originating from the remnant is purely thermal. Fig.~\ref{ABSORB}, left panel, shows the absorption column density variation across the remnant as derived from the spectral analysis of individual sub-regions. We note that this map is not obtained by employing a consistent model for all sub-regions but rather by individual "preferred" models. A similar figure illustrating the temperature variation across the remnant would not be that instructive  since some regions of the remnant are sufficiently fitted with one plasma temperature model while others require two distinct temperature plasma components. Therefore, we did not include such an image in this work. The South-Western part of the remnant appears to be the hottest, and well-described by a single component model. The region of enhanced X-ray emission positioned at the South-Eastern part of the remnant, which contains the two bright "blobs", appears to be moderately cooler. The Northern part of the remnant is the coolest of all. The presence of O- and Ne-enriched ejecta material, along with the presence of strong [OII] and [OIII] lines in its optical spectrum \citep{1981ApJ...248L.105D,1979MNRAS.188..357G,4fbea282-ad61-35c8-8776-164566ece9b7,1980ApJ...242L..73M}, forces us to propose the identification of G279.0+1.1, as a new O-rich SNR. However, the dominance of oxygen over hydrogen, [OIII]/H$\beta>3$, is not that strong compared to typical O-rich remnants. Further studies of the remnant's optical spectrum are required to explore in detail the nature of its optical counterpart. Until now, only a small number of the detected optical filaments have been spectrally studied. If confirmed, G279.0+1.1 will be the first known evolved Galactic SNR that exhibits such features. It is worth noting that even if ejecta detection was expected only from young SNR, such a feature has been detected in a small number of middle-aged SNR (e.g., G292.0+1.8 \citep{1979MNRAS.189..501M,2009ApJ...692.1489W}, Puppis A \citep{Hwang_2008}). Such a finding would make G279.0+1.1 the first evolved O-rich SNR and the fifth O-rich SNR in the Milky Way (MW): Cassiopeia A \citep{1976PASP...88..587K,2001AJ....122..297T,2006ApJ...645..283F,2008ApJS..179..195H}, Puppis A \citep{1985ApJ...299..981W}, and G292.0+1.8 \citep{1979MNRAS.189..501M,2009ApJ...692.1489W} are the remaining three O-rich remnants. None of the aforementioned remnants are found in evolved states. In fact, only four such (young) remnants are known outside the MW: 0102.2-272.9, and 0103-72.6 \citep{2003ApJ...598L..95P} \citep{2006ApJ...641..919F,Banovetz_2021} in the Small Magellanic Clouds (SMC) and N132D \citep{1987ApJ...314..103H}, and 0540-69.3 \citep{1980ApJ...242L..73M,Park_2010} in the Large Magellanic Clouds (LMC). A second evolved Galactic SNR, S 147 or Spaghetti nebula, has recently been discovered to exhibit similar characteristics (i.e., ejecta material in the X-ray spectrum \citep{G279MILTOS,spaghettichugai}). However, the latter lacks the presence of [OIII] lines in its optical spectrum. Thus it cannot be classified as O-rich as of now.

\begin{table*}
\centering
\caption{Best-fit parameters, with $1\sigma$ errors, of the regions (defined via the Voronoi analysis process) that have been selected as representatives to exhibit best the spectral variation detected across the remnant. Where not defined; elemental abundances are set to solar values.}
\renewcommand{\arraystretch}{1.7}
\setlength{\tabcolsep}{9pt}
\begin{tabular}
{p{2.5cm} p{5.5cm} p{5.5cm} p{5.5cm} p{5.5cm} p{5.5cm} p{5.5cm} p{5.5cm} p{5.5cm} p{5.5cm} p{5.5cm} p{5.5cm} p{5.5cm}}
\hline
Region & \multicolumn{3}{c}{A region}& \multicolumn{3}{c}{B region} &\multicolumn{3}{c}{C region} &\multicolumn{3}{c}{Entire remnant}\\ \hline
Area ($10^{6}~\mathrm{arcs^2}$) &\multicolumn{3}{c}{7.13}& \multicolumn{3}{c}{4.23} &\multicolumn{3}{c}{5.23} &\multicolumn{3}{c}{104.36}  \\ \hline
$\mathrm{Surf\_bri}$ ($10^{-3}~\mathrm{c/arcs^2}$) &\multicolumn{3}{c}{1.92}& \multicolumn{3}{c}{3.36} &\multicolumn{3}{c}{2.25} &\multicolumn{3}{c}{1.66}  \\ \hline
Model & \multicolumn{12}{c}{vapec+vapec}  \\ \hline
kT{\ssmall (keV)}& \multicolumn{3}{c}{$0.62_{-0.08}^{+0.05}$ $0.16_{-0.01}^{+0.01}$}& \multicolumn{3}{c}{$0.54_{-0.09}^{+0.08}$ $0.19_{-0.01}^{+0.01}$} & \multicolumn{3}{c}{$0.16_{-0.01}^{+0.01}$\space NaN}& \multicolumn{3}{c}{-} \\ 
$\mathrm{N_{H}}${\ssmall ($\mathrm{10^{22}cm^{-2}}$)}&\multicolumn{3}{c}{$0.20_{-0.03}^{+0.03}$}& \multicolumn{3}{c}{$0.13_{-0.01}^{+0.02}$}  & \multicolumn{3}{c}{$0.63_{-0.04}^{+0.03}$}&\multicolumn{3}{c}{-}\\
O&\multicolumn{3}{c}{1.0 \space  $0.83_{-0.08}^{+0.09}$}& \multicolumn{3}{c}{1.0 \space  $0.71_{-0.06}^{+0.07}$} & \multicolumn{3}{c}{$1.38_{-0.30}^{+0.41}$}&\multicolumn{3}{c}{-} \\
Ne&\multicolumn{3}{c}{$3.08_{-0.92}^{+1.02}$ \space1.0}& \multicolumn{3}{c}{$8.52_{-3.44}^{+9.42}$ $1.69_{-0.21}^{+0.24}$}& \multicolumn{3}{c}{$1.17_{-0.23}^{+0.32}$}&\multicolumn{3}{c}{-}\\ 
Mg&\multicolumn{3}{c}{-}& \multicolumn{3}{c}{$9.32_{-3.25}^{+9.40}$ \space1.0}& \multicolumn{3}{c}{$2.46_{-0.67}^{+0.92}$}&\multicolumn{3}{c}{-}\\
Fe&\multicolumn{3}{c}{-}& \multicolumn{3}{c}{-}& \multicolumn{3}{c}{$3.30_{-1.47}^{+2.78}$}&\multicolumn{3}{c}{-} \\
 \hline
 $\mathrm{\chi^2/dof}$ &\multicolumn{3}{c}{1.03}& \multicolumn{3}{c}{1.01} &  \multicolumn{3}{c}{1.19}&\multicolumn{3}{c}{-} \\ 
\hline
Model & \multicolumn{12}{c}{vnei+vnei}  \\ \hline
kT{\ssmall (keV)}& \multicolumn{3}{c}{$0.68_{-0.08}^{+0.32}$ $0.29_{-0.06}^{+0.10}$}& \multicolumn{3}{c}{$0.66_{-0.12}^{+0.14}$ $0.58_{-0.17}^{+0.26}$} & \multicolumn{3}{c}{$15.30_{-4.53}^{+7.96}$ \space NaN}&\multicolumn{3}{c}{-}\\ 
$\mathrm{N_{H}}${\ssmall ($\mathrm{10^{22}cm^{-2}}$)}&\multicolumn{3}{c}{$0.20_{-0.04}^{+0.04}$}& \multicolumn{3}{c}{$0.16_{-0.02}^{+0.02}$} & \multicolumn{3}{c}{$0.20_{-0.03}^{+0.03}$} &\multicolumn{3}{c}{-} \\
O&\multicolumn{3}{c}{-}& \multicolumn{3}{c}{$3.51_{-1.36}^{+2.48}$ $0.59_{-0.07}^{+0.07}$} & \multicolumn{3}{c}{$1.15_{-0.11}^{+0.13}$} &\multicolumn{3}{c}{-}\\
Ne&\multicolumn{3}{c}{-}& \multicolumn{3}{c}{$1.55_{-0.48}^{+0.67}$ $0.94_{-0.22}^{+0.29}$}& \multicolumn{3}{c}{$1.43_{-0.15}^{+0.17}$}&\multicolumn{3}{c}{-}\\ 
Mg&\multicolumn{3}{c}{-}& \multicolumn{3}{c}{-}& \multicolumn{3}{c}{-}&\multicolumn{3}{c}{-}\\
Ionization time (\ssmall{$\mathrm{10^{10}s/cm^{3}}$)} &\multicolumn{3}{c}{$9.78_{-0.81}^{+4.00}$ $1.53_{-5.03}^{+3.02}$}& \multicolumn{3}{c}{$10.23_{-3.82}^{+5.86}$ $0.48_{-0.10}^{+0.45}$} & \multicolumn{3}{c}{$0.78_{-0.07}^{+0.10}$}&\multicolumn{3}{c}{-}\\
\hline
$\mathrm{\chi^2/dof}$ &\multicolumn{3}{c}{0.88}& \multicolumn{3}{c}{1.06} & \multicolumn{3}{c}{1.26}&\multicolumn{3}{c}{-}\\ 
 \hline
Model & \multicolumn{12}{c}{vpshock+vpshock}  \\ \hline
kT{\ssmall (keV)}& \multicolumn{3}{c}{$0.85_{-0.16}^{+0.23}$ $0.26_{-0.05}^{+0.11}$}& \multicolumn{3}{c}{$0.70$ \space $0.54_{-0.24}^{+0.27}$} & \multicolumn{3}{c}{$1.00_{-0.21}^{+0.34}$ \space NaN} & \multicolumn{3}{c}{$0.60_{-0.05}^{+0.07}$ $0.34_{-0.07}^{+0.03}$} \\ 
$\mathrm{N_{H}}${\ssmall ($\mathrm{10^{22}cm^{-2}}$)}&\multicolumn{3}{c}{$0.20_{-0.03}^{+0.03}$}& \multicolumn{3}{c}{$0.16_{-0.02}^{+0.02}$} & \multicolumn{3}{c}{$0.19_{-0.03}^{+0.04}$} & \multicolumn{3}{c}{$0.31_{-0.02}^{+0.04}$} \\
O&\multicolumn{3}{c}{-}& \multicolumn{3}{c}{$3.93_{-1.75}^{+5.91}$ $0.51_{-0.05}^{+0.06}$} & \multicolumn{3}{c}{$1.30_{-0.14}^{+0.16}$} & \multicolumn{3}{c}{$4.47_{-0.84}^{+1.30}$ $0.66_{-0.05}^{+0.04}$} \\
Ne&\multicolumn{3}{c}{-}& \multicolumn{3}{c}{$1.65_{-0.51}^{+0.90}$ $1.1_{-0.23}^{+0.47}$} & \multicolumn{3}{c}{$1.82_{-0.34}^{+0.35}$} & \multicolumn{3}{c}{$2.52_{-0.39}^{+0.38}$ $1.48_{-0.22}^{+0.26}$}\\ 
Mg&\multicolumn{3}{c}{-}& \multicolumn{3}{c}{-}& \multicolumn{3}{c}{$1.88_{-0.46}^{+0.50}$} & \multicolumn{3}{c}{1.0 \space $5.85_{-1.43}^{+4.27}$} \\ 
Ionization time (\ssmall{$\mathrm{10^{11}s/cm^{3}}$)} &\multicolumn{3}{c}{$1.57_{-0.69}^{+1.16}$ $0.52_{-0.36}^{+0.79}$}& \multicolumn{3}{c}{$2.31_{-0.95}^{+2.14}$ $0.09_{-0.04}^{+0.34}$}& \multicolumn{3}{c}{$0.41_{-0.14}^{+0.19}$} & \multicolumn{3}{c}{$2.34_{-0.69}^{+0.97}$ $0.06_{-0.01}^{+0.02}$}\\
 \hline
$\mathrm{\chi^2/dof}$ &\multicolumn{3}{c}{0.87}& \multicolumn{3}{c}{1.06} & \multicolumn{3}{c}{1.16} &\multicolumn{3}{c}{1.19} \\ 
\hline
Model & \multicolumn{12}{c}{vnei+vapec}  \\ \hline
kT{\ssmall (keV)}& \multicolumn{3}{c}{$0.65_{-0.05}^{+0.04}$ $0.16_{-0.01}^{+0.01}$}& \multicolumn{3}{c}{$0.63_{-0.12}^{+0.15}$ $0.19_{-0.01}^{+0.01}$} &\multicolumn{3}{c}{-} &\multicolumn{3}{c}{-}\\ 
$\mathrm{N_{H}}${\ssmall ($\mathrm{10^{22}cm^{-2}}$)}&\multicolumn{3}{c}{$0.18_{-0.03}^{+0.04}$}& \multicolumn{3}{c}{$0.13_{-0.02}^{+0.01}$}& \multicolumn{3}{c}{-} &\multicolumn{3}{c}{-}  \\
O&\multicolumn{3}{c}{-}& \multicolumn{3}{c}{1.0 \space $0.71_{-0.06}^{+0.07}$} &\multicolumn{3}{c}{-}&\multicolumn{3}{c}{-} \\
Ne&\multicolumn{3}{c}{-}& \multicolumn{3}{c}{$4.12_{-1.88}^{+6.26}$ $1.59_{-0.24}^{+0.25}$}&\multicolumn{3}{c}{-}&\multicolumn{3}{c}{-}\\ 
Mg&\multicolumn{3}{c}{-}& \multicolumn{3}{c}{$6.50_{-2.72}^{+6.48}$ \space 1.0}&\multicolumn{3}{c}{-}& \multicolumn{3}{c}{-}\\ 
Ionization time (\ssmall{$\mathrm{10^{11}s/cm^{3}}$)} &\multicolumn{3}{c}{$2.16_{-0.92}^{+1.22}$ \space NaN}& \multicolumn{3}{c}{$2.43$ \space NaN}&\multicolumn{3}{c}{-}& \multicolumn{3}{c}{-}\\
 \hline
$\mathrm{\chi^2/dof}$ &\multicolumn{3}{c}{1.03}& \multicolumn{3}{c}{0.99} &\multicolumn{3}{c}{-}&\multicolumn{3}{c}{-}\\ 
\hline
\label{TABIS1}
\end{tabular}

\end{table*}

\subsection{\textit{Fermi-LAT spectra $\&$ multiwavelenght SED}}
\label{Gevspec}

As a final part of the binned likelihood analysis of the extended GeV source 4FGL J1000.0-5312e, which is spatially coincident to the remnant, we report on the obtained spectral energy distribution (SED) computed in the 0.5-500 GeV energy range. Data are divided into 6 equally spaced logarithmic energy bins. The best-fit spatial template reported in \citet{2020MNRAS.492.5980A} was used. The spectral fitting procedure reveals that a LogParabola model emerges as the best fit to the data, instead of a simple power law as reported in \citet{2020MNRAS.492.5980A}. During the fitting process, the normalization of all 4FGL-DR3 sources falling within $5\degree$ distance from the source of interest was let to vary. The same approach was applied to the normalization values of the Galactic diffuse and isotropic background. In addition, the normalization 
of the LogParabola model of G279.0+1.1 was left free, with the goal of obtaining the best fit. Our results are found to be in good agreement with the updated spectral plot of the remnant\footnote{\url{https://fermi.gsfc.nasa.gov/ssc/data/access/lat/12yr_catalog/}} as illustrated in Fig.~\ref{GEVSPECONLY} and there is a discrepancy to the GeV spectrum derived by \citet{2020MNRAS.492.5980A} towards the low-energy end (refer to Fig.~\ref{GEVSPECONLY}, right panel). We stress that the SED results are largely independent of the adopted spectral model (i.e., a Log-parabola or a power-law) used to construct the SED. 
This discrepancy is likely caused by the updated model used in 4FGL-DR3 to model the Galactic diffuse component ($\texttt{gll\_iem\_v07.fits}$) and the isotropic diffuse component ($\texttt{iso\_P8R3\_SOURCE\_V3\_v1.txt}$).

The interpretation of the gamma-ray data remains challenging. Our spectral results are less compatible with the expected gamma-ray emission from a hadronic hard-spectrum particle component given the flux decrease towards 1 GeV (e.g. \citet{2018A&A...615A.108Y}, for the expected spectral shape). However, given the age of the SNR, a leptonic, Inverse Compton (IC) emission scenario continues to be questionable as well \citep{2021ApJ...910...78Z}. Additionally, no non-thermal (electron synchrotron) component has been detected in the X-ray data from the remnant, data from both eROSITA and XMM-Newton are fully consistent with pure thermal emission. A relic electron scenario (i.e. emission from GeV-TeV electrons outside of high magnetic-field areas) might explain GeV IC emission without detectable (with current sensitivity) non-thermal X-ray emission, but a quantitative exploration is beyond the scope of this work.

\subsection{Distance, age, and plasma density estimation}
\label{dist}

A kinematic distance calculation of the remnant is difficult to be implemented due to the uncertainty of the spatial distribution of the radiative shock. However, three distinct distance estimation methodologies have been carried out so far. \citet{1975ApJ...195..715M} introduced a distance estimation technique on the basis of the blast wave energy and associated cloud parameters, which when employed for G279.0+1.1 resulted in a $\sim3$~kpc distance. The $\Sigma-D$ relation for distance calculation has also been applied in the case of the remnant, converging towards a similar distance of $\sim3~\mathrm{kpc}$ \citep{1988MNRAS.234..971W}. The distance estimation of CO-emitting clouds possibly associated with the remnant, reported in \citet{1988MNRAS.234..971W}, is also placing the remnant at the same distance. More recently, \citet{2019RAA....19...92S} utilized Red Clump (RC) stars to build an extinction-to-distance relation in the direction of remnants from the fourth Galactic quadrant, reporting a $2.7\pm0.3$~kpc distance for G279.0+1.1, based on the computed optical extinction.

\begin{figure*}[h]

    \includegraphics[width=0.53\textwidth]{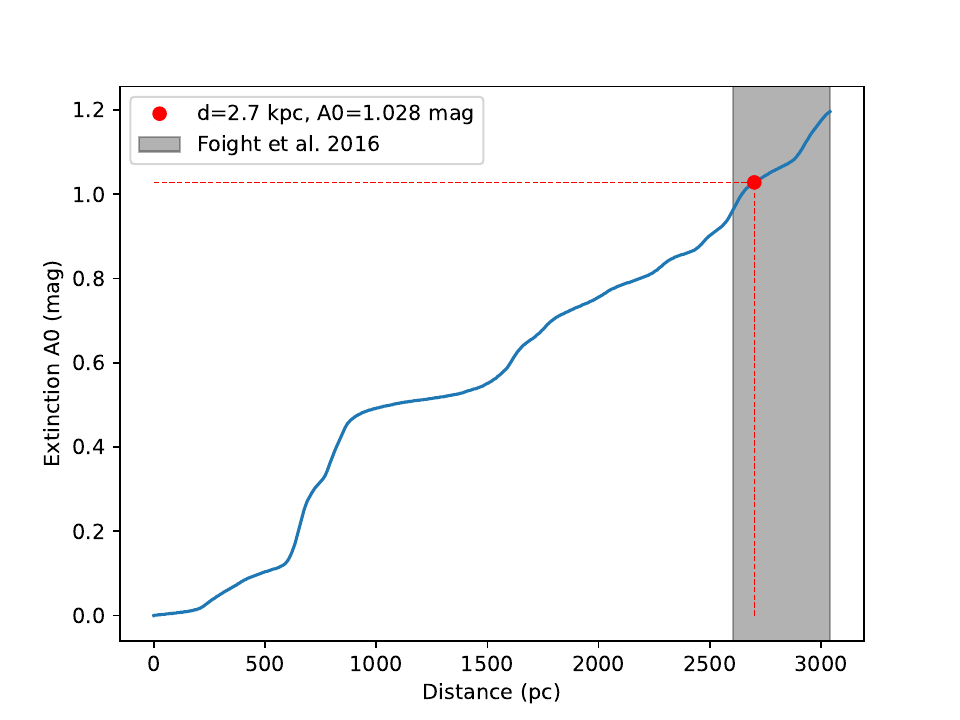}\includegraphics[width=0.53\textwidth]{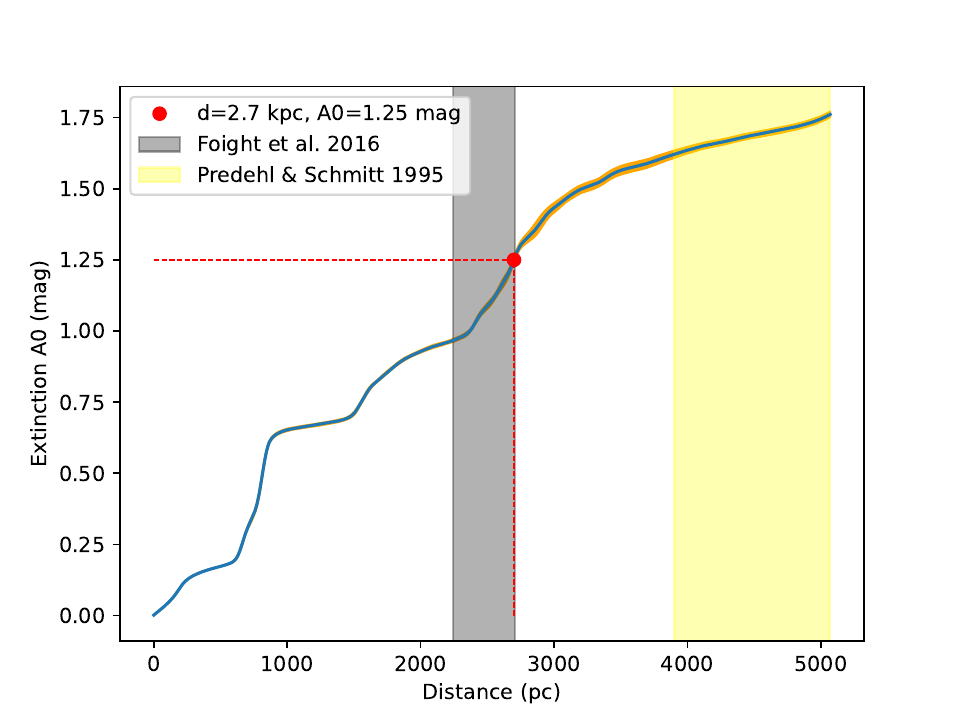}
    \caption{Let panel: One-dimensional cumulative extinction graph as a function of the distance up to $\sim3$kpc (\citet{2019A&A...625A.135L} data sets) towards G279.0+01.1 SNR, obtained by using the GAIA/2MASS tool for one-dimensional extinction computation \url{https://astro.acri-st.fr/gaia_dev/}. Right panel: One-dimensional cumulative extinction graph as a function of the distance up to $\sim5$kpc (updated \citet{2022A&A...661A.147L} data sets) towards G279.0+01.1 SNR, obtained by using the EXPLORE G-Tomo tool for one-dimensional extinction computation \url{https://explore-platform.eu/}. In both panels, the grey-shaded areas correspond to the distance uncertainty estimation when employing Eq.~\ref{math1} and the obtained best-fit value of the absorption column density derived from the spectral analysis. The yellow area indicates the distance uncertainty range when employing Eq.~\ref{math}. The red point represents the obtained extinction when assuming that the remnant is located at a distance of 2.7 kpc.}
    \label{Extinction3}
\end{figure*}
We perform an additional distance consistency check based on the estimated absorption column density parameter obtained from the best fit of the eRASS:4 X-ray spectra. In particular, we made use of the Galactic mean colour excess spatial distribution, established in \citet{1978A&A....64..367L}, and the 3D extinction maps obtained from the combination of GAIA and 2MASS photometric data reported in \citet{2019A&A...625A.135L,2022A&A...661A.147L}. 
Using the statistical relation between the observed absorption in X-rays with the mean colour excess \citep{1995A&A...293..889P}:

\begin{equation}
  \begin{array}{lcl}
       N_\mathrm{H}/E_{\mathrm{B-V}}&=&5.3\times10^{21}~\mathrm{cm^{-2}\cdot mag^{-1}}\\N_\mathrm{H}\mathrm{[cm^{-2}}/A_{\mathrm{\nu}}]&=&1.79\times10^{21}
 \label{math}
 \end{array}
\end{equation}

we derived $A_{\nu}=1.73_{-0.11}^{+0.23}$ in the direction of the remnant. The error in the latter estimate is computed based on the 1$\sigma$ error of the absorption column density as obtained by the X-ray spectral fitting of the entire remnant. Compared to the GAIA data set \citep{2019A&A...625A.135L}, this places G279.0+1.1 at a distance larger than $3$~kpc ($A0\equiv A(550$~nm). Making use of the updated \citep{2022A&A...661A.147L} data sets, which extend up to $\sim5$~kpc, a distance constraint of $4.9_{-1.0}^{>+0.17}$~kpc is obtained, as shown on the right panel of Fig.~\ref{Extinction3} in yellow.

However, more recent observations of SNRs
employing Chandra data have resulted in a modified statistical relation between X-ray absorption and mean color excess \citep{2016ApJ...826...66F}. A significantly higher proportionality factor compared to previous reports is derived:

\begin{equation}
  \begin{array}{lcl}
       N_\mathrm{H}/E_{\mathrm{B-V}}&=&8.9\times10^{21}~\mathrm{cm^{-2}\cdot mag^{-1}}\\N_\mathrm{H}\mathrm{[cm^{-2}}/A_{\mathrm{\nu}}]&=&2.87(\pm0.12)\times10^{21}
 \label{math1}
 \end{array}
\end{equation}

Using this relation, the obtained X-ray absorption column density yields an expected extinction of $A_{\nu}=1.08_{-0.12}^{+0.19}$. Employing the \citet{2019A&A...625A.135L} data sets the latter estimate results in a $2.9_{-0.30}^{>+0.14}$~kpc distance for G279.0+1.1, as shown on the left panel of Fig.~\ref{Extinction3} in black. Compared to the most recent \citet{2022A&A...661A.147L} data sets, a $2.49_{-0.25}^{+0.22}$~kpc distance is derived, as shown on the right panel of Fig.~\ref{Extinction3} in black. These distance values (using the \citet{2016ApJ...826...66F} values) are consistent with earlier estimations of the remnant's distance.

Using the data sets by \citet{2019A&A...625A.135L,2022A&A...661A.147L}, one can also perform the inverse, cross-check, procedure considering as known the remnant's distance, $2.7\pm0.3$ kpc. In this case, as shown in Fig.~\ref{Extinction3} in red, an extinction of A0=1.03 mag or A0=1.25 mag is obtained, depending on the GAIA/2MASS data sets utilized. Employing Eq.~\ref{math1} results in $N_\mathrm{H}=0.29_{-0.06}^{+0.06}\cdot10^{22}\mathrm{cm^{-2}}$ or $N_\mathrm{H}=0.36_{-0.08}^{+0.07}\cdot10^{22}\mathrm{cm^{-2}}$, respectively, for the two different data sets. These values are consistent with the best-fit values obtained from the X-ray spectral fitting. One derives even smaller $N_\mathrm{H}$ values when employing the empirical relation implemented in \citet{1995A&A...293..889P}. In particular, when employing Eq.~\ref{math} for the two distinct data sets by \citet{2019A&A...625A.135L,2022A&A...661A.147L}, one derives $N_\mathrm{H}=0.18_{-0.03}^{+0.03}\cdot10^{22}\mathrm{cm^{-2}}$ and $N_\mathrm{H}=0.22_{-0.04}^{+0.04}\cdot10^{22}\mathrm{cm^{-2}}$, respectively. To summarize, based on all the above measurements one expects a maximum $N_\mathrm{H}\sim0.3-0.35\cdot10^{22}\mathrm{cm^{-2}}$ towards the remnant, except for the parts of the remnant which are spatially coincident with dense dust clouds. In the latter case, the aforementioned value could easily be exceeded. 
 
Adopting now a distance to the remnant of 2.7 kpc, and taking into account the remnant's angular size of $\sim3\degree$, one derives a 140~pc diameter (or 70~pc radius). Hence, assuming a spherical distribution, one derives a total plasma volume of  $V=4.32~10^{61}~\mathrm{cm^{-3}}$. The emission measure of the lower and higher temperature components, using the corresponding normalization values from the two-temperature spectrum best-fit and assuming a uniform density distribution across a spherical volume of fully ionized plasma($n_\mathrm{e}=1.2n_\mathrm{H}$), was estimated to:

\begin{equation}
    \boldsymbol{EM}=\int n_\mathrm{e}\cdot n_\mathrm{H} dV=\eta\cdot1.2\cdot n_\mathrm{H}^{2}\cdot V
    \label{likelihood44}
\end{equation}

Here, $n_\mathrm{e}$ is the electron density, $n_\mathrm{H}$ is the proton density, $\eta$ is a filling factor and $V$ the volume occupied by both plasmas, as computed above. 

 At the same time, one can employ the following equation for the emission measure calculation:

 \begin{equation}
    EM=\frac{\boldsymbol{norm}\times4\pi D^{2}}{10^{-14}}
    \label{likelihood45}
\end{equation}

where D is the remnant's distance in cm. 
Combining Eq.~\ref{likelihood44} and \ref{likelihood45} one obtains the following formula for the plasma density computation:
\begin{equation}
    n_\mathrm{H}=\sqrt{\frac{norm\times4\pi\cdot D^{2}}{1.2\cdot\eta\cdot10^{-14}\cdot V}}
    \label{likelihood455}
\end{equation}

From the best-fit model of the spectrum from the entire remnant, we obtained: $norm_{\mathrm{Hot}}=0.013_{-0.001}^{+0.002}$ for the hot temperature plasma component.
 The filling factor of the two components is unknown. Assuming uniform diffuse emission covering the entire spherical volume, the filling factor is of order unity. However, given that a portion of the computed volume is free of emission (i.e., the central part of the remnant where the spectral analysis results does not suggest X-ray absorption), we estimated the fraction of the area free of plasma emission to be $\eta=0.92$. Taking into account the error on the distance of the remnant ($2.7\pm0.3$~kpc, \citet{2019RAA....19...92S} utilizing Red Clump (RC) stars), 
the resulting local density is: $n_\mathrm{H}=4.9_{-0.54}^{+0.54}~10^{-3}~\mathrm{cm^{-3}}$ (or $n_\mathrm{e}=5.9_{-0.65}^{+0.65}~10^{-3}~\mathrm{cm^{-3}}$). The normalization error is negligible here, and the error on the filling factor is unknown.

The remnant is considered to be amongst the oldest Galactic SNR. An estimated age of the order of $10^6$ years has been reported by \citet{1988MNRAS.234..971W}. One can estimate the age of the remnant by employing the same relation as in \citet{2009A&A...507..841G}: $t\sim\frac{\tau}{n_e}$, where $\tau$ is the ionization timescale of the emission plasma. Making use of the derived $n_\mathrm{e}$ value and the ionization timescale of the hot plasma component for the entire remnant (as shown in Tab.~\ref{TABIS1}), we compute the remnant's age to be $1.26_{-0.51}^{+0.66}$~Myrs. We additionally applied the evolutionary models of SNR as provided in \citet{Leahy_2017} to perform an updated age estimation. For the obtained absorption column density of the entire remnant, $N_\mathrm{H}=0.31_{-0.02}^{+0.04}~\mathrm{10^{22}cm^{-2}}$ (as shown in Tab.~\ref{TABIS1}) and a distance of $2.7\pm0.3$~kpc, one derives a local\protect\footnotemark interstellar medium (ISM) number density of $n_\mathrm{H}=0.37_{-0.07}^{+0.09}~\mathrm{cm^{-3}}$. Considering as inputs the derived local plasma density as calculated above, a typical explosion energy of the order of $10^{51}~\mathrm{erg}$ and keeping the remaining parameters to default input values, we derived an age of $9.25_{-1.55}^{+1.85}~10^5$~yrs. This value is again in great agreement with previous reports.

\footnotetext{local in the sense that the average density towards the SNR is representative of the density at the SNR}

All the estimates reported above yield a consistent picture for the SNR, using an adopted distance of $\sim$ 2.7\,kpc.  Nevertheless, all distance estimates are based on empirical relations with substantial scatter, and a critical assessment may be justified. Specifically, it is noteworthy that an update of the electron density model from NE2001 \citep{2002astro.ph..7156C} to YMW16 \citep{2017ApJ...835...29Y} has reduced the dispersion-measure based distances of all pulsars in projected vicinity to G279.0+1.1 by typically a factor of $\sim 8$. While this withdraws the basis of all proposed pulsar associations in the literature which are based on the compatibility of the SNR's distance estimate with the respective pulsar's distances estimate, it is a useful exercise to consider whether a typical (as of today) distance that would be derived from a pulsar's dispersion measure together with a proposed association with the SNR would also yield a consistent picture. To this end, we list in Tab.~\ref{psr} all pulsars that are in reasonable (within $3\degree$ of the remnant's center) angular distance to G279.0+1.1, together with their properties from the ATNF pulsar catalog\footnote{\url{https://www.atnf.csiro.au/research/pulsar/psrcat/}} \citep{2005AJ....129.1993M}. Three pulsars fall within the remnant's extension (J0955-5304 (B0953-52), J0957-5432, J0954-5430), whereas seven pulsars lie outside of the remnant's structure (J1001-5507 (B0959-54), J1000-5149, J1001-5559, J1002-5559, J1016-5345 (B1014-53), J0941-5244, J0940-5438). Amongst the above pulsars, J0940-5428 is the only one categorized as a GeV emitter, namely 4FGL J0941.1-5429 as a GeV point source \citep{Abdollahi_2020}, and it exhibits the highest spin-down power ($1.9\cdot10^{36}~\mathrm{erg\cdot s^{-1}}$). A potential association with any of the aforementioned pulsars (except for J1002.5559) would place the remnant at a much closer distance of $0.12-0.45$~kpc, based on the YMW16 electron density model \citep{2017ApJ...835...29Y}. In particular, assuming that the pulsar J1001-5507 (0959-54) is associated with the remnant (as suggested by \citet{1995MNRAS.277..319D}), the dispersion-measure based distance is 0.41\,kpc and the pulsar's spin-down age (which would then be a measure for the SNR's age) is $~0.4$~Myr.

Assuming thus a $0.4$~kpc remnant distance one can perform the same computation series as for the case of an assumed remnant's distance of $2.7\pm0.3$~kpc. Such a closer distance results in a 21~pc remnant diameter. A $V=1.42\cdot10^{59}~\mathrm{cm^{-3}}$ volume can then be derived assuming spherical geometry. For identical values of $\eta$ and $norm_\mathrm{Hot}$ one then derives (using eq.~\ref{likelihood455}) $n_\mathrm{H}=1.26_{-0.05}^{+0.09}~10^{-2}~\mathrm{cm^{-3}}$ (or $n_\mathrm{e}=1.52_{-0.06}^{+0.11}~10^{-2}~\mathrm{cm^{-3}}$) local density. Employing the \citet{2009A&A...507..841G} relation: $t\sim\frac{\tau}{n_e}$, the remnant is found to have an age of $4.9^{+2.4}_{-1.7}\cdot10^{5}$~yrs. Similarly, one can also employ the evolutionary models of SNR as provided in \citet{Leahy_2017} to derive an age estimate. With default inputs (identical to those employed for an age estimate assuming a $2.7\pm0.3$~kpc remnant's distance) except for the local ISM number density which is computed to be $n_\mathrm{H}=2.51_{-0.16}^{+0.33}~\mathrm{cm^{-3}}$  (for $N_\mathrm{H}=0.31_{-0.02}^{+0.04}~\mathrm{10^{22}cm^{-2}}$ and a distance of $0.4$~kpc), one obtains a remnant's age of $1.01_{-0.03}^{+0.09}~10^4$~yrs. Such an age would actually favor an association with the pulsar J0940-5428, the pulsar J0954-5430, or the pulsar J1001-5507, since the rest of the pulsars appear to be considerably older.
It is noteworthy that whereas a 0.38~kpc pulsar's J0940-5428 distance makes the latter object a compelling candidate to be associated with the remnant given its computed transverse velocity, its young age (0.04~Myrs), its large spin-down power, and its GeV counterpart (4FGL J0941.1-5429). A 2.7 kpc distance forbids such an association with the remnant given the extremely unrealistic transverse velocity of $\sim3236$~km/s which would be required to reach its present position.

\begin{table*}[!]
    \centering
    \begin{tabular}{|c|c|c|c|c|c|c|}
    \hline 
        \textbf{Pulsar} & Ang. sep.  & DM&\textbf{$D_1$} &$D_2$& Age & $v_{\mathrm{transv}}$  \\
         & ($\degree$) & \textbf{$\mathrm{pc\cdot cm^{-3}}$} & kpc&kpc& Myr &$\mathrm{km\cdot s^{-1}}$\\
        \hline
        J0955-5304 (B0953-52)& $0.83$ & $156.9$ & $0.40 (3.31)$ &- &$3.87$ &1.5 \\
        J0957-5432 & $0.88$ & $226.1$ & $0.45 (4.33)$ &- &$1.66$ &4.1 \\
        J0954-5430 & $1.13$ & $201.57$ & $0.43 (3.96)$  &- &$0.17$&48.8 \\\hline
        J1001-5507 (B0959-54) & $1.48$ & $130.32$ & $0.41 (2.78)$ & $0.30^{+1.1}_{-0.3}$ &$0.44$&23.5  \\
        J1000-5149 & $1.85$ & $72.8$ & $0.13 (1.93)$ &-& $4.22$ &1.0 \\
        J1001-5559 & $2.32$ & $159.3$ & $0.43 (3.32)$&- &$30.6$ &0.6\\
        J1002-5559& $2.37$ & $426.0$ & $3.27 (9.83)$&- &$7.84$ &16.9 \\
        J1016-5345 (B1014-53) & $2.55$ & $66.8$ & $0.12 (1.94)$&- &$6.33$ &0.8 \\
        J0941-5244 & $2.80$ & $157.94$ & $0.40 (3.14)$&-& $9.17$ &2.1 \\
        J0940-5428 & $2.81$ & $134.55$ & $0.38 (2.95)$&-&$0.04$ &455.5 \\ \hline
        \end{tabular}
        \caption{Pulsars within $3\degree$ of the remnant's center. The table is split into two halves: the upper half contains the first three pulsars which lie within the remnant's extension. The rest of the pulsars that lie well outside the remnant's structure are displayed in the lower half of the table. The first and second columns give the pulsar's name and angular separation from the remnant's center (as re-defined in this work). The third column gives the Dispersion Measure. The fourth and fifth columns give the pulsar's distance from Earth based on DM measurements and potential associations, respectively. The values within parenthesis correspond to older distance estimates based on the NE2001 electron density model \citep{2002astro.ph..7156C}. Since 2017, YMW16 is considered the default model for DM-based distance calculations \citep{2017ApJ...835...29Y}. The sixth column corresponds to the pulsar's spin-down age. The seventh column displays the transverse velocity required for each pulsar to move from the remnant's center to its present location.}
        \label{psr}
\end{table*}

\section{Summary}\label{concl}

We report on the discovery of the X-ray counterpart of the SNR G279.0+01.1, which is found to be of $\sim3\degree$ size, using eRASS data from the first four completed SRG/eROSITA All-Sky Surveys, eRASS:4. We perform a comprehensive X-ray imaging and spectral analysis, using eRASS:4 data, and complement the findings with archival data from ROSAT and XMM-Newton. 
The obtained results from all X-ray datasets were found to be in excellent agreement, taking into account the restrictions  that the ROSAT and XMM-Newton data are subject to. 

The majority of the remnant's X-ray emission is restricted to the 0.3-1.1 keV energy band. In the 1.1-1.5 keV band, only portions of the remnant are observable, while above 2 keV no emission from the remnant is detected at all. The emission from the remnant can be described with thermal, thin-plasma emission. No sign for non-thermal emission was detected. The data from the entire remnant can be described by a two-temperature plane-parallel shocked plasma in non-equilibrium, with temperatures of kT$\sim0.6$~keV and $\sim0.3$~ keV, respectively, and an average absorption column density of $N_\mathrm{H}\sim0.3~10^{22}\mathrm{cm^{-2}}$. However, significant X-ray temperature variations have been detected across the $3\degree$ angular extension of the SNR, and also the absorption column differs at different regions. Still, also when analyzing individual sub-regions, defined e.g. with a Voronoi binning analysis, most of the regions require more than one temperature for a satisfactory fit (see Tab.~\ref{TABIS1} and appendix~\ref{sec:append1},\ref{sec:append2}). Whether the plasma is in equilibrium or in non-equilibrium can however not be decided from the X-ray data alone.

However, significantly enhanced (above solar) abundances for O, Ne, and Mg seem to be required for an adequate description of the individual spectra and the total spectrum, when the simplest two-temperature models are adopted. No high-Z elements are observed.
This is noteworthy since only a small number of such O- Ne- dominated SNRs has been observed yet. The latter characteristics i.e., exhibiting clear O VII, O VIII Ly$\alpha$, Ne IX lines of ejecta origin as well as the Ne X Ly$\alpha$ line and He-like Mg XI unresolved triplet all across its surface, along with the detection of strong [OIII] lines in its optical spectra suggests that the SNR can be classified as O-rich. However, only a limited number of its optical filaments have been spectrally studied. Therefore, a more detailed analysis of its optical spectrum is essential to confirm its O-rich nature. If the SNR is confirmed to be O-rich, it would be the fourth known SNR of this class in our Galaxy, and the first evolved Galactic remnant which demonstrates such features. Two O-rich remnants have been observed in the Small Magellanic Clouds (SMC): 0102.2-272.9 \citep{2006ApJ...641..919F,Banovetz_2021}, 0103-72.6 \citep{2003ApJ...598L..95P}, another two in the Large Magellanic Clouds (LMC): N132D \citep{1987ApJ...314..103H}, 0540-69.3 \citep{1980ApJ...242L..73M,Park_2010}, and three  have been identified in the Milky Way: Cassiopeia A \citep{1976PASP...88..587K,2001AJ....122..297T,2006ApJ...645..283F,2008ApJS..179..195H}, Puppis A \citep{1985ApJ...299..981W}, and G292.0+1.8 \citep{1979MNRAS.189..501M,2009ApJ...692.1489W}. Even though the above remnants have the presence of O-enriched ejecta in common, their nature varies significantly. Such O-rich remnants, and in general SNRs exhibiting X-ray spectral line emission dominated by O, Ne and Mg, are considered to be the remnants of core-collapse supernovae of the stars of the highest mass. Such massive progenitor stars can reach masses of 20M$\odot$ or greater, and are objects of major importance for the study of SN nucleosynthesis. 

The SNR’s X-ray spatial morphology is in agreement with the radio continuum data as depicted in Fig.~\ref{Radioandoptical}. The radio continuum image of the remnant, using 4850 MHz data from the PMN \citep{1993AJ....106.1095C} Southern surveys,
yields a $~2.5-3\degree$ extension, almost identical to its X-ray size, and larger than what previously reported in \citet{1988MNRAS.234..971W,10.1111/j.1365-2966.2009.14476.x}. The remnant appears as an incomplete shell in both wavebands.  The regions of enhanced emission positioned to the North and South-East of the SNR are spatially identical in both radio and X-rays. Further X-ray spectral analysis revealed a strong X-ray absorption to the West, likely linked to the presence of enhanced IR emission originating from dust clouds (as seen in Fig~\ref{IRASS}). This is also supported by the obtained absorption column density parameter of the X-ray spectral fit (as seen in Fig.~\ref{ABSORB}). The results suggest that the SNR is characterized by a peculiar limb-brightened morphology in the X-ray energy band, since no enhanced absorption column density values were derived from its central region (region J). However, strong absorption column density values were derived from the Western part of the remnant accompanied by the absence or very weak X-ray diffuse emission, and thus we argue that the remnant is partially occluded to the West.

Fermi-LAT GeV data analysis from the location of the remnant using all available data yields consistent imaging results with earlier findings reported in \citet{2020MNRAS.492.5980A}. We confirm the detection of an extended GeV source spatially coincident with the remnant, named 4FGL J1000.0-5312e. In particular, three individual regions of enhanced $\gamma-\mathrm{ray}$ emission were detected with $4.6\sigma$, $5.5\sigma$, and $5.8\sigma$ significance, respectively, as shown in Fig.~\ref{GEVSPECONLY}. Modeling of the GeV excess favors the scenario that those three regions are part of the diffuse GeV emission from the SNR. A strong spatial anti-correlation is  observable between X-ray and GeV emission without any clear explanation. However, the spatial morphology of the extended GeV source (which is highly coincident with the location of the remnant), the high detection significance of three prominent regions, and its hard GeV spectral component (extending up to 0.5~TeV without any indications of spectral softening) strongly suggest that 4FGL J1000.0-5312e is likely the remnant's GeV counterpart.

The results obtained in this work cast some doubt on the so far prevailing interpretation that G279.0+1.1 is located at a distance of $\sim 2.7$\,kpc and has an age of $\sim 1$Myr. On the one hand, we could obtain a consistent picture for the state of the SNR under this assumption, specifically also considering an independent, consistent distance estimate using the derived X-ray absorption column and matching it to the latest GAIA-2MASS extinction maps \citep{2019A&A...625A.135L,2022A&A...661A.147L}. If taken face-value, the SNR has a 140~pc linear diameter and an age of $0.75-1.92$~Myrs, exploiting either the formula employed in \citet{2009A&A...507..841G} ($t\sim\frac{\tau}{n_e}$) or evolutionary models by \citet{Leahy_2017}. Then, the remnant would be categorized among the oldest Galactic SNRs, if not the oldest, and would as of now be the oldest SNR in which swept-up ISM and ejecta contributions are apparently detected. On the other hand, some of our observational results would be better consistent with a much smaller distance, e.g., $\sim$ 0.4\,kpc if one assumes that the SNR is associated with one of the potentially associated pulsars and their revised distance estimate from the YMW16 electron density model. The SNR would then have a $\sim$ 20\,pc linear diameter and a much younger age of $\sim 1\cdot10^4-5\cdot10^5$\,yrs, using again the same evolutionary models as above. The best-fit non-equilibrium X-ray models would be better in line with this younger age. Also, the association with a morphologically-associated GeV counterpart argues in favor of a younger age. Further studies are required to resolve this ambiguity regarding the SNR's estimated distance and age.\\

\noindent\textit{Acknowledgements}

M.M. and G.P. acknowledge support from the Deutsche Forschungsgemeinschaft through grant PU 308/2-1.

This work is based on data from eROSITA, the soft X-ray instrument aboard SRG, a joint Russian-German science mission supported by the Russian Space Agency (Roskosmos), in the interests of the Russian Academy of Sciences represented by its Space Research Institute (IKI), and the Deutsches Zentrum für Luft- und Raumfahrt (DLR). The SRG spacecraft was built by Lavochkin Association (NPOL) and its subcontractors, and is operated by NPOL with support from the Max Planck Institute for Extraterrestrial Physics (MPE).

The development and construction of the eROSITA X-ray instrument was led by MPE, with contributions from the Dr. Karl Remeis Observatory Bamberg $\&$ ECAP (FAU Erlangen-Nuernberg), the University of Hamburg Observatory, the Leibniz Institute for Astrophysics Potsdam (AIP), and the Institute for Astronomy and Astrophysics of the University of Tübingen, with the support of DLR and the Max Planck Society. The Argelander Institute for Astronomy of the University of Bonn and the Ludwig Maximilians Universität Munich also participated in the science preparation for eROSITA.
The eROSITA data shown here were processed using the eSASS/NRTA software system developed by the German eROSITA consortium.

We thank the EXPLORE team which provided us access to the G-TOMO tool of the EXPLORE platform \url{https://explore-platform.eu/} allowing us to exploit updated GAIA/2MASS data. We also thank Denys Malyshev, Victor Doroshenko and Lorenzo Ducci for fruitful discussions.

\bibliography{biblio}

\begin{thebibliography}{68}
\expandafter\ifx\csname natexlab\endcsname\relax\def\natexlab#1{#1}\fi

\bibitem[{Abdollahi {et~al.}(2020)Abdollahi, Acero, Ackermann, Ajello, Atwood,
  Axelsson, Baldini, Ballet, Barbiellini, Bastieri, Gonzalez, Bellazzini,
  Berretta, Bissaldi, Blandford, Bloom, Bonino, Bottacini, Brandt, Bregeon,
  Bruel, Buehler, Burnett, Buson, Cameron, Caputo, Caraveo, Casandjian, Castro,
  Cavazzuti, Charles, Chaty, Chen, Cheung, Chiaro, Ciprini, Cohen-Tanugi,
  Cominsky, Coronado-Blázquez, Costantin, Cuoco, Cutini, D’Ammando, DeKlotz,
  de~la Torre~Luque, de~Palma, Desai, Digel, Lalla, Mauro, Venere, Domínguez,
  Dumora, Dirirsa, Fegan, Ferrara, Franckowiak, Fukazawa, Funk, Fusco, Gargano,
  Gasparrini, Giglietto, Giommi, Giordano, Giroletti, Glanzman, Green, Grenier,
  Griffin, Grondin, Grove, Guiriec, Harding, Hayashi, Hays, Hewitt, Horan,
  Jóhannesson, Johnson, Kamae, Kerr, Kocevski, Kovac’evic’, Kuss, Landriu,
  Larsson, Latronico, Lemoine-Goumard, Li, Liodakis, Longo, Loparco, Lott,
  Lovellette, Lubrano, Madejski, Maldera, Malyshev, Manfreda, Marchesini,
  Marcotulli, Martí-Devesa, Martin, Massaro, Mazziotta, McEnery, Mereu, Meyer,
  Michelson, Mirabal, Mizuno, Monzani, Morselli, Moskalenko, Negro, Nuss, Ojha,
  Omodei, Orienti, Orlando, Ormes, Palatiello, Paliya, Paneque, Pei,
  Peña-Herazo, Perkins, Persic, Pesce-Rollins, Petrosian, Petrov, Piron, Poon,
  Porter, Principe, Rainò, Rando, Razzano, Razzaque, Reimer, Reimer, Remy,
  Reposeur, Romani, Parkinson, Schinzel, Serini, Sgrò, Siskind, Smith,
  Spandre, Spinelli, Strong, Suson, Tajima, Takahashi, Tak, Thayer, Thompson,
  Tibaldo, Torres, Torresi, Valverde, Klaveren, van Zyl, Wood, Yassine, \&
  Zaharijas}]{Abdollahi_2020}
Abdollahi, S., Acero, F., Ackermann, M., {et~al.} 2020, The Astrophysical
  Journal Supplement Series, 247, 33

\bibitem[{{Acero} {et~al.}(2016){Acero}, {Ackermann}, {Ajello}, {Baldini},
  {Ballet}, {Barbiellini}, {Bastieri}, {Bellazzini}, {Bissaldi}, {Blandford},
  {Bloom}, {Bonino}, {Bottacini}, {Brandt}, {Bregeon}, {Bruel}, {Buehler},
  {Buson}, {Caliandro}, {Cameron}, {Caputo}, {Caragiulo}, {Caraveo},
  {Casandjian}, {Cavazzuti}, {Cecchi}, {Chekhtman}, {Chiang}, {Chiaro},
  {Ciprini}, {Claus}, {Cohen}, {Cohen-Tanugi}, {Cominsky}, {Condon}, {Conrad},
  {Cutini}, {D'Ammando}, {de Angelis}, {de Palma}, {Desiante}, {Digel}, {Di
  Venere}, {Drell}, {Drlica-Wagner}, {Favuzzi}, {Ferrara}, {Franckowiak},
  {Fukazawa}, {Funk}, {Fusco}, {Gargano}, {Gasparrini}, {Giglietto}, {Giommi},
  {Giordano}, {Giroletti}, {Glanzman}, {Godfrey}, {Gomez-Vargas}, {Grenier},
  {Grondin}, {Guillemot}, {Guiriec}, {Gustafsson}, {Hadasch}, {Harding},
  {Hayashida}, {Hays}, {Hewitt}, {Hill}, {Horan}, {Hou}, {Iafrate}, {Jogler},
  {J{\'o}hannesson}, {Johnson}, {Kamae}, {Katagiri}, {Kataoka}, {Katsuta},
  {Kerr}, {Kn{\"o}dlseder}, {Kocevski}, {Kuss}, {Laffon}, {Lande}, {Larsson},
  {Latronico}, {Lemoine-Goumard}, {Li}, {Li}, {Longo}, {Loparco}, {Lovellette},
  {Lubrano}, {Magill}, {Maldera}, {Marelli}, {Mayer}, {Mazziotta}, {Michelson},
  {Mitthumsiri}, {Mizuno}, {Moiseev}, {Monzani}, {Moretti}, {Morselli},
  {Moskalenko}, {Murgia}, {Nemmen}, {Nuss}, {Ohsugi}, {Omodei}, {Orienti},
  {Orlando}, {Ormes}, {Paneque}, {Perkins}, {Pesce-Rollins}, {Petrosian},
  {Piron}, {Pivato}, {Porter}, {Rain{\`o}}, {Rando}, {Razzano}, {Razzaque},
  {Reimer}, {Reimer}, {Renaud}, {Reposeur}, {Rousseau}, {Saz Parkinson},
  {Schmid}, {Schulz}, {Sgr{\`o}}, {Siskind}, {Spada}, {Spandre}, {Spinelli},
  {Strong}, {Suson}, {Tajima}, {Takahashi}, {Tanaka}, {Thayer}, {Thompson},
  {Tibaldo}, {Tibolla}, {Torres}, {Tosti}, {Troja}, {Uchiyama}, {Vianello},
  {Wells}, {Wood}, {Wood}, {Yassine}, {den Hartog}, \&
  {Zimmer}}]{2016ApJS..224....8A}
{Acero}, F., {Ackermann}, M., {Ajello}, M., {et~al.} 2016, \apjs, 224, 8

\bibitem[{{Aharonian} {et~al.}(2004){Aharonian}, {Akhperjanian}, {Aye},
  {Bazer-Bachi}, {Beilicke}, {Benbow}, {Berge}, {Berghaus}, {Bernl{\"o}hr},
  {Bolz}, {Boisson}, {Borgmeier}, {Breitling}, {Brown}, {Bussons Gordo},
  {Chadwick}, {Chitnis}, {Chounet}, {Cornils}, {Costamante}, {Degrange},
  {Djannati-Ata{\"\i}}, {Drury}, {Ergin}, {Espigat}, {Feinstein}, {Fleury},
  {Fontaine}, {Funk}, {Gallant}, {Giebels}, {Gillessen}, {Goret}, {Guy},
  {Hadjichristidis}, {Hauser}, {Heinzelmann}, {Henri}, {Hermann}, {Hinton},
  {Hofmann}, {Holleran}, {Horns}, {de Jager}, {Jung}, {Kh{\'e}lifi}, {Komin},
  {Konopelko}, {Latham}, {Le Gallou}, {Lemoine}, {Lemi{\`e}re}, {Leroy},
  {Lohse}, {Marcowith}, {Masterson}, {McComb}, {de Naurois}, {Nolan},
  {Noutsos}, {Orford}, {Osborne}, {Ouchrif}, {Panter}, {Pelletier}, {Pita},
  {Pohl}, {P{\"u}hlhofer}, {Punch}, {Raubenheimer}, {Raue}, {Raux}, {Rayner},
  {Redondo}, {Reimer}, {Reimer}, {Ripken}, {Rivoal}, {Rob}, {Rolland},
  {Rowell}, {Sahakian}, {Saug{\'e}}, {Schlenker}, {Schlickeiser}, {Schuster},
  {Schwanke}, {Siewert}, {Sol}, {Steenkamp}, {Stegmann}, {Tavernet},
  {Th{\'e}oret}, {Tluczykont}, {van der Walt}, {Vasileiadis}, {Vincent},
  {Visser}, {V{\"o}lk}, \& {Wagner}}]{2004Natur.432...75A}
{Aharonian}, F.~A., {Akhperjanian}, A.~G., {Aye}, K.~M., {et~al.} 2004, \nat,
  432, 75

\bibitem[{{Anderson} {et~al.}(2014){Anderson}, {Bania}, {Balser}, {Cunningham},
  {Wenger}, {Johnstone}, \& {Armentrout}}]{2014ApJS..212....1A}
{Anderson}, L.~D., {Bania}, T.~M., {Balser}, D.~S., {et~al.} 2014, \apjs, 212,
  1

\bibitem[{{Araya}(2020)}]{2020MNRAS.492.5980A}
{Araya}, M. 2020, \mnras, 492, 5980

\bibitem[{Banovetz {et~al.}(2021)Banovetz, Milisavljevic, Sravan, Fesen,
  Patnaude, Plucinsky, Blair, Weil, Morse, Margutti, \& Drout}]{Banovetz_2021}
Banovetz, J., Milisavljevic, D., Sravan, N., {et~al.} 2021, The Astrophysical
  Journal, 912, 33

\bibitem[{Borkowski {et~al.}(2006)Borkowski, Hendrick, \&
  Reynolds}]{Borkowski_2006}
Borkowski, K.~J., Hendrick, S.~P., \& Reynolds, S.~P. 2006, The Astrophysical
  Journal, 652, 1259

\bibitem[{Borkowski {et~al.}(2001)Borkowski, Lyerly, \&
  Reynolds}]{Borkowski_2001}
Borkowski, K.~J., Lyerly, W.~J., \& Reynolds, S.~P. 2001, The Astrophysical
  Journal, 548, 820

\bibitem[{{Brunner} {et~al.}(2022){Brunner}, {Liu}, {Lamer}, {Georgakakis},
  {Merloni}, {Brusa}, {Bulbul}, {Dennerl}, {Friedrich}, {Liu}, {Maitra},
  {Nandra}, {Ramos-Ceja}, {Sanders}, {Stewart}, {Boller}, {Buchner}, {Clerc},
  {Comparat}, {Dwelly}, {Eckert}, {Finoguenov}, {Freyberg}, {Ghirardini},
  {Gueguen}, {Haberl}, {Kreykenbohm}, {Krumpe}, {Osterhage}, {Pacaud},
  {Predehl}, {Reiprich}, {Robrade}, {Salvato}, {Santangelo}, {Schrabback},
  {Schwope}, \& {Wilms}}]{2022A&A...661A...1B}
{Brunner}, H., {Liu}, T., {Lamer}, G., {et~al.} 2022, \aap, 661, A1

\bibitem[{{Cappellari} \& {Copin}(2003)}]{2003MNRAS.342..345C}
{Cappellari}, M. \& {Copin}, Y. 2003, \mnras, 342, 345

\bibitem[{{Cash}(1979)}]{1979ApJ...228..939C}
{Cash}, W. 1979, \apj, 228, 939

\bibitem[{{Collischon} {et~al.}(2021){Collischon}, {Sasaki}, {Mecke}, {Points},
  \& {Klatt}}]{2021A&A...653A..16C}
{Collischon}, C., {Sasaki}, M., {Mecke}, K., {Points}, S.~D., \& {Klatt}, M.~A.
  2021, \aap, 653, A16

\bibitem[{{Condon} {et~al.}(1991){Condon}, {Broderick}, \&
  {Seielstad}}]{1991AJ....102.2041C}
{Condon}, J.~J., {Broderick}, J.~J., \& {Seielstad}, G.~A. 1991, \aj, 102, 2041

\bibitem[{{Condon} {et~al.}(1994){Condon}, {Broderick}, {Seielstad}, {Douglas},
  \& {Gregory}}]{1994AJ....107.1829C}
{Condon}, J.~J., {Broderick}, J.~J., {Seielstad}, G.~A., {Douglas}, K., \&
  {Gregory}, P.~C. 1994, \aj, 107, 1829

\bibitem[{{Condon} {et~al.}(1993){Condon}, {Griffith}, \&
  {Wright}}]{1993AJ....106.1095C}
{Condon}, J.~J., {Griffith}, M.~R., \& {Wright}, A.~E. 1993, \aj, 106, 1095

\bibitem[{{Cordes} \& {Lazio}(2002)}]{2002astro.ph..7156C}
{Cordes}, J.~M. \& {Lazio}, T.~J.~W. 2002, arXiv e-prints, astro

\bibitem[{{Cram} {et~al.}(1998){Cram}, {Green}, \&
  {Bock}}]{1998PASA...15...64C}
{Cram}, L.~E., {Green}, A.~J., \& {Bock}, D.~C.~J. 1998, \pasa, 15, 64

\bibitem[{{Dopita} {et~al.}(1981){Dopita}, {Tuohy}, \&
  {Mathewson}}]{1981ApJ...248L.105D}
{Dopita}, M.~A., {Tuohy}, I.~R., \& {Mathewson}, D.~S. 1981, \apjl, 248, L105

\bibitem[{{Duncan} {et~al.}(1995){Duncan}, {Haynes}, {Stewart}, \&
  {Jones}}]{1995MNRAS.277..319D}
{Duncan}, A.~R., {Haynes}, R.~F., {Stewart}, R.~T., \& {Jones}, K.~L. 1995,
  \mnras, 277, 319

\bibitem[{{Fesen} {et~al.}(2006){Fesen}, {Hammell}, {Morse}, {Chevalier},
  {Borkowski}, {Dopita}, {Gerardy}, {Lawrence}, {Raymond}, \& {van den
  Bergh}}]{2006ApJ...645..283F}
{Fesen}, R.~A., {Hammell}, M.~C., {Morse}, J., {et~al.} 2006, \apj, 645, 283

\bibitem[{{Finkbeiner}(2003)}]{2003ApJS..146..407F}
{Finkbeiner}, D.~P. 2003, \apjs, 146, 407

\bibitem[{{Finkelstein} {et~al.}(2006){Finkelstein}, {Morse}, {Green},
  {Linsky}, {Shull}, {Snow}, {Stocke}, {Brownsberger}, {Ebbets}, {Wilkinson},
  {Heap}, {Leitherer}, {Savage}, {Siegmund}, \& {Stern}}]{2006ApJ...641..919F}
{Finkelstein}, S.~L., {Morse}, J.~A., {Green}, J.~C., {et~al.} 2006, \apj, 641,
  919

\bibitem[{{Foight} {et~al.}(2016){Foight}, {G{\"u}ver}, {{\"O}zel}, \&
  {Slane}}]{2016ApJ...826...66F}
{Foight}, D.~R., {G{\"u}ver}, T., {{\"O}zel}, F., \& {Slane}, P.~O. 2016, \apj,
  826, 66

\bibitem[{{Giacani} {et~al.}(2009){Giacani}, {Smith}, {Dubner}, {Loiseau},
  {Castelletti}, \& {Paron}}]{2009A&A...507..841G}
{Giacani}, E., {Smith}, M.~J.~S., {Dubner}, G., {et~al.} 2009, \aap, 507, 841

\bibitem[{{Goss} {et~al.}(1979){Goss}, {Shaver}, {Zealey}, {Murdin}, \&
  {Clark}}]{1979MNRAS.188..357G}
{Goss}, W.~M., {Shaver}, P.~A., {Zealey}, W.~J., {Murdin}, P., \& {Clark},
  D.~H. 1979, \mnras, 188, 357

\bibitem[{{H.~E.~S.~S. Collaboration} {et~al.}(2018{\natexlab{a}}){H.~E.~S.~S.
  Collaboration}, {Abdalla}, {Abramowski}, {Aharonian}, {Ait Benkhali},
  {Ang{\"u}ner}, {Arakawa}, {Arrieta}, {Aubert}, {Backes}, {Balzer}, {Barnard},
  {Becherini}, {Becker Tjus}, {Berge}, {Bernhard}, {Bernl{\"o}hr}, {Blackwell},
  {B{\"o}ttcher}, {Boisson}, {Bolmont}, {Bonnefoy}, {Bordas}, {Bregeon},
  {Brun}, {Brun}, {Bryan}, {B{\"u}chele}, {Bulik}, {Capasso}, {Caroff},
  {Carosi}, {Casanova}, {Cerruti}, {Chakraborty}, {Chaves}, {Chen},
  {Chevalier}, {Colafrancesco}, {Condon}, {Conrad}, {Davids}, {Decock}, {Deil},
  {Devin}, {deWilt}, {Dirson}, {Djannati-Ata{\"\i}}, {Donath}, {Drury},
  {Dutson}, {Dyks}, {Edwards}, {Egberts}, {Emery}, {Ernenwein}, {Eschbach},
  {Farnier}, {Fegan}, {Fernandes}, {Fernandez}, {Fiasson}, {Fontaine}, {Funk},
  {F{\"u}{\ss}ling}, {Gabici}, {Gallant}, {Garrigoux}, {Gat{\'e}}, {Giavitto},
  {Giebels}, {Glawion}, {Glicenstein}, {Gottschall}, {Grondin}, {Hahn},
  {Haupt}, {Hawkes}, {Heinzelmann}, {Henri}, {Hermann}, {Hinton}, {Hofmann},
  {Hoischen}, {Holch}, {Holler}, {Horns}, {Ivascenko}, {Iwasaki},
  {Jacholkowska}, {Jamrozy}, {Jankowsky}, {Jankowsky}, {Jingo}, {Jouvin},
  {Jung-Richardt}, {Kastendieck}, {Katarzy{\'n}ski}, {Katsuragawa}, {Katz},
  {Kerszberg}, {Khangulyan}, {Kh{\'e}lifi}, {King}, {Klepser}, {Klochkov},
  {Klu{\'z}niak}, {Komin}, {Kosack}, {Krakau}, {Kraus}, {Kr{\"u}ger}, {Laffon},
  {Lamanna}, {Lau}, {Lees}, {Lefaucheur}, {Lemi{\`e}re}, {Lemoine-Goumard},
  {Lenain}, {Leser}, {Lohse}, {Lorentz}, {Liu}, {L{\'o}pez-Coto}, {Lypova},
  {Malyshev}, {Marandon}, {Marcowith}, {Mariaud}, {Marx}, {Maurin}, {Maxted},
  {Mayer}, {Meintjes}, {Meyer}, {Mitchell}, {Moderski}, {Mohamed}, {Mohrmann},
  {Mor{\r{a}}}, {Moulin}, {Murach}, {Nakashima}, {de Naurois}, {Ndiyavala},
  {Niederwanger}, {Niemiec}, {Oakes}, {O'Brien}, {Odaka}, {Ohm}, {Ostrowski},
  {Oya}, {Padovani}, {Panter}, {Parsons}, {Pekeur}, {Pelletier}, {Perennes},
  {Petrucci}, {Peyaud}, {Piel}, {Pita}, {Poireau}, {Poon}, {Prokhorov},
  {Prokoph}, {P{\"u}hlhofer}, {Punch}, {Quirrenbach}, {Raab}, {Rauth},
  {Reimer}, {Reimer}, {Renaud}, {de los Reyes}, {Rieger}, {Rinchiuso},
  {Romoli}, {Rowell}, {Rudak}, {Rulten}, {Safi-Harb}, {Sahakian}, {Saito},
  {Sanchez}, {Santangelo}, {Sasaki}, {Schlickeiser}, {Sch{\"u}ssler}, {Schulz},
  {Schwanke}, {Schwemmer}, {Seglar-Arroyo}, {Settimo}, {Seyffert}, {Shafi},
  {Shilon}, {Shiningayamwe}, {Simoni}, {Sol}, {Spanier}, {Spir-Jacob},
  {Stawarz}, {Steenkamp}, {Stegmann}, {Steppa}, {Sushch}, {Takahashi},
  {Tavernet}, {Tavernier}, {Taylor}, {Terrier}, {Tibaldo}, {Tiziani},
  {Tluczykont}, {Trichard}, {Tsirou}, {Tsuji}, {Tuffs}, {Uchiyama}, {van der
  Walt}, {van Eldik}, {van Rensburg}, {van Soelen}, {Vasileiadis}, {Veh},
  {Venter}, {Viana}, {Vincent}, {Vink}, {Voisin}, {V{\"o}lk}, {Vuillaume},
  {Wadiasingh}, {Wagner}, {Wagner}, {Wagner}, {White}, {Wierzcholska},
  {Willmann}, {W{\"o}rnlein}, {Wouters}, {Yang}, {Zaborov}, {Zacharias},
  {Zanin}, {Zdziarski}, {Zech}, {Zefi}, {Ziegler}, {Zorn}, \&
  {{\.Z}ywucka}}]{2018A&A...612A...3H}
{H.~E.~S.~S. Collaboration}, {Abdalla}, H., {Abramowski}, A., {et~al.}
  2018{\natexlab{a}}, \aap, 612, A3

\bibitem[{{H.~E.~S.~S. Collaboration} {et~al.}(2018{\natexlab{b}}){H.~E.~S.~S.
  Collaboration}, {Abdalla}, {Abramowski}, {Aharonian}, {Ait Benkhali},
  {Ang{\"u}ner}, {Arakawa}, {Arrieta}, {Aubert}, {Backes}, {Balzer}, {Barnard},
  {Becherini}, {Becker Tjus}, {Berge}, {Bernhard}, {Bernl{\"o}hr}, {Blackwell},
  {B{\"o}ttcher}, {Boisson}, {Bolmont}, {Bonnefoy}, {Bordas}, {Bregeon},
  {Brun}, {Brun}, {Bryan}, {B{\"u}chele}, {Bulik}, {Capasso}, {Carrigan},
  {Caroff}, {Carosi}, {Casanova}, {Cerruti}, {Chakraborty}, {Chaves}, {Chen},
  {Chevalier}, {Colafrancesco}, {Condon}, {Conrad}, {Davids}, {Decock}, {Deil},
  {Devin}, {deWilt}, {Dirson}, {Djannati-Ata{\"\i}}, {Domainko}, {Donath},
  {Drury}, {Dutson}, {Dyks}, {Edwards}, {Egberts}, {Eger}, {Emery},
  {Ernenwein}, {Eschbach}, {Farnier}, {Fegan}, {Fernandes}, {Fiasson},
  {Fontaine}, {F{\"o}rster}, {Funk}, {F{\"u}{\ss}ling}, {Gabici}, {Gallant},
  {Garrigoux}, {Gast}, {Gat{\'e}}, {Giavitto}, {Giebels}, {Glawion},
  {Glicenstein}, {Gottschall}, {Grondin}, {Hahn}, {Haupt}, {Hawkes},
  {Heinzelmann}, {Henri}, {Hermann}, {Hinton}, {Hofmann}, {Hoischen}, {Holch},
  {Holler}, {Horns}, {Ivascenko}, {Iwasaki}, {Jacholkowska}, {Jamrozy},
  {Jankowsky}, {Jankowsky}, {Jingo}, {Jouvin}, {Jung-Richardt}, {Kastendieck},
  {Katarzy{\'n}ski}, {Katsuragawa}, {Katz}, {Kerszberg}, {Khangulyan},
  {Kh{\'e}lifi}, {King}, {Klepser}, {Klochkov}, {Klu{\'z}niak}, {Komin},
  {Kosack}, {Krakau}, {Kraus}, {Kr{\"u}ger}, {Laffon}, {Lamanna}, {Lau},
  {Lees}, {Lefaucheur}, {Lemi{\`e}re}, {Lemoine-Goumard}, {Lenain}, {Leser},
  {Lohse}, {Lorentz}, {Liu}, {L{\'o}pez-Coto}, {Lypova}, {Marandon},
  {Malyshev}, {Marcowith}, {Mariaud}, {Marx}, {Maurin}, {Maxted}, {Mayer},
  {Meintjes}, {Meyer}, {Mitchell}, {Moderski}, {Mohamed}, {Mohrmann},
  {Mor{\r{a}}}, {Moulin}, {Murach}, {Nakashima}, {de Naurois}, {Ndiyavala},
  {Niederwanger}, {Niemiec}, {Oakes}, {O'Brien}, {Odaka}, {Ohm}, {Ostrowski},
  {Oya}, {Padovani}, {Panter}, {Parsons}, {Paz Arribas}, {Pekeur}, {Pelletier},
  {Perennes}, {Petrucci}, {Peyaud}, {Piel}, {Pita}, {Poireau}, {Poon},
  {Prokhorov}, {Prokoph}, {P{\"u}hlhofer}, {Punch}, {Quirrenbach}, {Raab},
  {Rauth}, {Reimer}, {Reimer}, {Renaud}, {de los Reyes}, {Rieger}, {Rinchiuso},
  {Romoli}, {Rowell}, {Rudak}, {Rulten}, {Safi-Harb}, {Sahakian}, {Saito},
  {Sanchez}, {Santangelo}, {Sasaki}, {Schandri}, {Schlickeiser},
  {Sch{\"u}ssler}, {Schulz}, {Schwanke}, {Schwemmer}, {Seglar-Arroyo},
  {Settimo}, {Seyffert}, {Shafi}, {Shilon}, {Shiningayamwe}, {Simoni}, {Sol},
  {Spanier}, {Spir-Jacob}, {Stawarz}, {Steenkamp}, {Stegmann}, {Steppa},
  {Sushch}, {Takahashi}, {Tavernet}, {Tavernier}, {Taylor}, {Terrier},
  {Tibaldo}, {Tiziani}, {Tluczykont}, {Trichard}, {Tsirou}, {Tsuji}, {Tuffs},
  {Uchiyama}, {van der Walt}, {van Eldik}, {van Rensburg}, {van Soelen},
  {Vasileiadis}, {Veh}, {Venter}, {Viana}, {Vincent}, {Vink}, {Voisin},
  {V{\"o}lk}, {Vuillaume}, {Wadiasingh}, {Wagner}, {Wagner}, {Wagner}, {White},
  {Wierzcholska}, {Willmann}, {W{\"o}rnlein}, {Wouters}, {Yang}, {Zaborov},
  {Zacharias}, {Zanin}, {Zdziarski}, {Zech}, {Zefi}, {Ziegler}, {Zorn}, \&
  {{\.Z}ywucka}}]{2018A&A...612A...1H}
{H.~E.~S.~S. Collaboration}, {Abdalla}, H., {Abramowski}, A., {et~al.}
  2018{\natexlab{b}}, \aap, 612, A1

\bibitem[{{Hammell} \& {Fesen}(2008)}]{2008ApJS..179..195H}
{Hammell}, M.~C. \& {Fesen}, R.~A. 2008, \apjs, 179, 195

\bibitem[{{Hughes}(1987)}]{1987ApJ...314..103H}
{Hughes}, J.~P. 1987, \apj, 314, 103

\bibitem[{Hwang {et~al.}(2008)Hwang, Petre, \& Flanagan}]{Hwang_2008}
Hwang, U., Petre, R., \& Flanagan, K.~A. 2008, The Astrophysical Journal, 676,
  378

\bibitem[{Joye \& Mandel(2003)}]{joye2003new}
Joye, W.~A. \& Mandel, E. 2003, in Astronomical data analysis software and
  systems XII, Vol. 295, 489

\bibitem[{{Kamitsukasa} {et~al.}(2015){Kamitsukasa}, {Koyama}, {Uchida},
  {Nakajima}, {Hayashida}, {Mori}, {Katsuda}, \&
  {Tsunemi}}]{2015PASJ...67...16K}
{Kamitsukasa}, F., {Koyama}, K., {Uchida}, H., {et~al.} 2015, \pasj, 67, 16

\bibitem[{{Kamper} \& {van den Bergh}(1976)}]{1976PASP...88..587K}
{Kamper}, K. \& {van den Bergh}, S. 1976, \pasp, 88, 587

\bibitem[{{Kesteven} \& {Caswell}(1987)}]{1987A&A...183..118K}
{Kesteven}, M.~J. \& {Caswell}, J.~L. 1987, \aap, 183, 118

\bibitem[{{Khabibullin} {et~al.}(2024)}]{spaghettichugai}
{Khabibullin}, I. {et~al.} 2024, A\&A

\bibitem[{{Koyama} {et~al.}(1995){Koyama}, {Petre}, {Gotthelf}, {Hwang},
  {Matsuura}, {Ozaki}, \& {Holt}}]{1995Natur.378..255K}
{Koyama}, K., {Petre}, R., {Gotthelf}, E.~V., {et~al.} 1995, \nat, 378, 255

\bibitem[{{Lallement} {et~al.}(2019){Lallement}, {Babusiaux}, {Vergely},
  {Katz}, {Arenou}, {Valette}, {Hottier}, \& {Capitanio}}]{2019A&A...625A.135L}
{Lallement}, R., {Babusiaux}, C., {Vergely}, J.~L., {et~al.} 2019, \aap, 625,
  A135

\bibitem[{{Lallement} {et~al.}(2022){Lallement}, {Vergely}, {Babusiaux}, \&
  {Cox}}]{2022A&A...661A.147L}
{Lallement}, R., {Vergely}, J.~L., {Babusiaux}, C., \& {Cox}, N.~L.~J. 2022,
  \aap, 661, A147

\bibitem[{LASKER(1979)}]{4fbea282-ad61-35c8-8776-164566ece9b7}
LASKER, B.~M. 1979, Publications of the Astronomical Society of the Pacific,
  91, 153

\bibitem[{Leahy \& Williams(2017)}]{Leahy_2017}
Leahy, D.~A. \& Williams, J.~E. 2017, The Astronomical Journal, 153, 239

\bibitem[{{Lucke}(1978)}]{1978A&A....64..367L}
{Lucke}, P.~B. 1978, \aap, 64, 367

\bibitem[{{Manchester} {et~al.}(2005){Manchester}, {Hobbs}, {Teoh}, \&
  {Hobbs}}]{2005AJ....129.1993M}
{Manchester}, R.~N., {Hobbs}, G.~B., {Teoh}, A., \& {Hobbs}, M. 2005, \aj, 129,
  1993

\bibitem[{{Mathewson} {et~al.}(1980){Mathewson}, {Dopita}, {Tuohy}, \&
  {Ford}}]{1980ApJ...242L..73M}
{Mathewson}, D.~S., {Dopita}, M.~A., {Tuohy}, I.~R., \& {Ford}, V.~L. 1980,
  \apjl, 242, L73

\bibitem[{{McKee} \& {Cowie}(1975)}]{1975ApJ...195..715M}
{McKee}, C.~F. \& {Cowie}, L.~L. 1975, \apj, 195, 715

\bibitem[{{Merloni} {et~al.}(2012){Merloni}, {Predehl}, {Becker},
  {B{\"o}hringer}, {Boller}, {Brunner}, {Brusa}, {Dennerl}, {Freyberg},
  {Friedrich}, {Georgakakis}, {Haberl}, {Hasinger}, {Meidinger}, {Mohr},
  {Nandra}, {Rau}, {Reiprich}, {Robrade}, {Salvato}, {Santangelo}, {Sasaki},
  {Schwope}, {Wilms}, \& {German eROSITA Consortium}}]{2012arXiv1209.3114M}
{Merloni}, A., {Predehl}, P., {Becker}, W., {et~al.} 2012, arXiv e-prints,
  arXiv:1209.3114

\bibitem[{{Merloni} {et~al.}(2023)}]{Merloni2023}
{Merloni}, A. {et~al.} 2023, A\&A

\bibitem[{{Michailidis} {et~al.}(2024)}]{G279MILTOS}
{Michailidis}, M. {et~al.} 2024, A\&A

\bibitem[{{Murdin} \& {Clark}(1979)}]{1979MNRAS.189..501M}
{Murdin}, P. \& {Clark}, D.~H. 1979, \mnras, 189, 501

\bibitem[{{Park} {et~al.}(2003){Park}, {Hughes}, {Burrows}, {Slane}, {Nousek},
  \& {Garmire}}]{2003ApJ...598L..95P}
{Park}, S., {Hughes}, J.~P., {Burrows}, D.~N., {et~al.} 2003, \apjl, 598, L95

\bibitem[{Park {et~al.}(2010)Park, Hughes, Slane, Mori, \& Burrows}]{Park_2010}
Park, S., Hughes, J.~P., Slane, P.~O., Mori, K., \& Burrows, D.~N. 2010, The
  Astrophysical Journal, 710, 948

\bibitem[{{Predehl} {et~al.}(2021){Predehl}, {Andritschke}, {Arefiev},
  {Babyshkin}, {Batanov}, {Becker}, {B{\"o}hringer}, {Bogomolov}, {Boller},
  {Borm}, {Bornemann}, {Br{\"a}uninger}, {Br{\"u}ggen}, {Brunner}, {Brusa},
  {Bulbul}, {Buntov}, {Burwitz}, {Burkert}, {Clerc}, {Churazov}, {Coutinho},
  {Dauser}, {Dennerl}, {Doroshenko}, {Eder}, {Emberger}, {Eraerds},
  {Finoguenov}, {Freyberg}, {Friedrich}, {Friedrich}, {F{\"u}rmetz},
  {Georgakakis}, {Gilfanov}, {Granato}, {Grossberger}, {Gueguen}, {Gureev},
  {Haberl}, {H{\"a}lker}, {Hartner}, {Hasinger}, {Huber}, {Ji}, {Kienlin},
  {Kink}, {Korotkov}, {Kreykenbohm}, {Lamer}, {Lomakin}, {Lapshov}, {Liu},
  {Maitra}, {Meidinger}, {Menz}, {Merloni}, {Mernik}, {Mican}, {Mohr},
  {M{\"u}ller}, {Nandra}, {Nazarov}, {Pacaud}, {Pavlinsky}, {Perinati},
  {Pfeffermann}, {Pietschner}, {Ramos-Ceja}, {Rau}, {Reiffers}, {Reiprich},
  {Robrade}, {Salvato}, {Sanders}, {Santangelo}, {Sasaki}, {Scheuerle},
  {Schmid}, {Schmitt}, {Schwope}, {Shirshakov}, {Steinmetz}, {Stewart},
  {Str{\"u}der}, {Sunyaev}, {Tenzer}, {Tiedemann}, {Tr{\"u}mper}, {Voron},
  {Weber}, {Wilms}, \& {Yaroshenko}}]{2021A&A...647A...1P}
{Predehl}, P., {Andritschke}, R., {Arefiev}, V., {et~al.} 2021, \aap, 647, A1

\bibitem[{{Predehl} \& {Schmitt}(1995)}]{1995A&A...293..889P}
{Predehl}, P. \& {Schmitt}, J.~H.~M.~M. 1995, \aap, 293, 889

\bibitem[{{Shan} {et~al.}(2019){Shan}, {Zhu}, {Tian}, {Zhang}, {Yang}, \&
  {Zhang}}]{2019RAA....19...92S}
{Shan}, S.-S., {Zhu}, H., {Tian}, W.-W., {et~al.} 2019, Research in Astronomy
  and Astrophysics, 19, 092

\bibitem[{{Str{\"u}der} {et~al.}(2001){Str{\"u}der}, {Briel}, {Dennerl},
  {Hartmann}, {Kendziorra}, {Meidinger}, {Pfeffermann}, {Reppin}, {Aschenbach},
  {Bornemann}, {Br{\"a}uninger}, {Burkert}, {Elender}, {Freyberg}, {Haberl},
  {Hartner}, {Heuschmann}, {Hippmann}, {Kastelic}, {Kemmer}, {Kettenring},
  {Kink}, {Krause}, {M{\"u}ller}, {Oppitz}, {Pietsch}, {Popp}, {Predehl},
  {Read}, {Stephan}, {St{\"o}tter}, {Tr{\"u}mper}, {Holl}, {Kemmer}, {Soltau},
  {St{\"o}tter}, {Weber}, {Weichert}, {von Zanthier}, {Carathanassis}, {Lutz},
  {Richter}, {Solc}, {B{\"o}ttcher}, {Kuster}, {Staubert}, {Abbey}, {Holland},
  {Turner}, {Balasini}, {Bignami}, {La Palombara}, {Villa}, {Buttler},
  {Gianini}, {Lain{\'e}}, {Lumb}, \& {Dhez}}]{2001A&A...365L..18S}
{Str{\"u}der}, L., {Briel}, U., {Dennerl}, K., {et~al.} 2001, \aap, 365, L18

\bibitem[{Stupar \& Parker(2009)}]{10.1111/j.1365-2966.2009.14476.x}
Stupar, M. \& Parker, Q.~A. 2009, Monthly Notices of the Royal Astronomical
  Society, 394, 1791

\bibitem[{{Sunyaev} {et~al.}(2021){Sunyaev}, {Arefiev}, {Babyshkin},
  {Bogomolov}, {Borisov}, {Buntov}, {Brunner}, {Burenin}, {Churazov},
  {Coutinho}, {Eder}, {Eismont}, {Freyberg}, {Gilfanov}, {Gureyev}, {Hasinger},
  {Khabibullin}, {Kolmykov}, {Komovkin}, {Krivonos}, {Lapshov}, {Levin},
  {Lomakin}, {Lutovinov}, {Medvedev}, {Merloni}, {Mernik}, {Mikhailov},
  {Molodtsov}, {Mzhelsky}, {M{\"u}ller}, {Nandra}, {Nazarov}, {Pavlinsky},
  {Poghodin}, {Predehl}, {Robrade}, {Sazonov}, {Scheuerle}, {Shirshakov},
  {Tkachenko}, \& {Voron}}]{2021A&A...656A.132S}
{Sunyaev}, R., {Arefiev}, V., {Babyshkin}, V., {et~al.} 2021, \aap, 656, A132

\bibitem[{{Thorstensen} {et~al.}(2001){Thorstensen}, {Fesen}, \& {van den
  Bergh}}]{2001AJ....122..297T}
{Thorstensen}, J.~R., {Fesen}, R.~A., \& {van den Bergh}, S. 2001, \aj, 122,
  297

\bibitem[{{Turner} {et~al.}(2001){Turner}, {Abbey}, {Arnaud}, {Balasini},
  {Barbera}, {Belsole}, {Bennie}, {Bernard}, {Bignami}, {Boer}, {Briel},
  {Butler}, {Cara}, {Chabaud}, {Cole}, {Collura}, {Conte}, {Cros}, {Denby},
  {Dhez}, {Di Coco}, {Dowson}, {Ferrando}, {Ghizzardi}, {Gianotti}, {Goodall},
  {Gretton}, {Griffiths}, {Hainaut}, {Hochedez}, {Holland}, {Jourdain},
  {Kendziorra}, {Lagostina}, {Laine}, {La Palombara}, {Lortholary}, {Lumb},
  {Marty}, {Molendi}, {Pigot}, {Poindron}, {Pounds}, {Reeves}, {Reppin},
  {Rothenflug}, {Salvetat}, {Sauvageot}, {Schmitt}, {Sembay}, {Short},
  {Spragg}, {Stephen}, {Str{\"u}der}, {Tiengo}, {Trifoglio}, {Tr{\"u}mper},
  {Vercellone}, {Vigroux}, {Villa}, {Ward}, {Whitehead}, \&
  {Zonca}}]{2001A&A...365L..27T}
{Turner}, M.~J.~L., {Abbey}, A., {Arnaud}, M., {et~al.} 2001, \aap, 365, L27

\bibitem[{{Voges} {et~al.}(2000){Voges}, {Aschenbach}, {Boller}, {Brauninger},
  {Briel}, {Burkert}, {Dennerl}, {Englhauser}, {Gruber}, {Haberl}, {Hartner},
  {Hasinger}, {Pfeffermann}, {Pietsch}, {Predehl}, {Schmitt}, {Trumper}, \&
  {Zimmermann}}]{2000IAUC.7432....3V}
{Voges}, W., {Aschenbach}, B., {Boller}, T., {et~al.} 2000, \iaucirc, 7432, 3

\bibitem[{{Whiteoak} \& {Green}(1996)}]{1996A&AS..118..329W}
{Whiteoak}, J.~B.~Z. \& {Green}, A.~J. 1996, \aaps, 118, 329

\bibitem[{{Wilms} {et~al.}(2000){Wilms}, {Allen}, \&
  {McCray}}]{2000ApJ...542..914W}
{Wilms}, J., {Allen}, A., \& {McCray}, R. 2000, \apj, 542, 914

\bibitem[{{Winkler} \& {Kirshner}(1985)}]{1985ApJ...299..981W}
{Winkler}, P.~F. \& {Kirshner}, R.~P. 1985, \apj, 299, 981

\bibitem[{{Winkler} {et~al.}(2009){Winkler}, {Twelker}, {Reith}, \&
  {Long}}]{2009ApJ...692.1489W}
{Winkler}, P.~F., {Twelker}, K., {Reith}, C.~N., \& {Long}, K.~S. 2009, \apj,
  692, 1489

\bibitem[{{Woermann} \& {Jonas}(1988)}]{1988MNRAS.234..971W}
{Woermann}, B. \& {Jonas}, J.~L. 1988, \mnras, 234, 971

\bibitem[{Yamaguchi {et~al.}(2011)Yamaguchi, Koyama, \&
  Uchida}]{10.1093/pasj/63.sp3.S837}
Yamaguchi, H., Koyama, K., \& Uchida, H. 2011, Publications of the Astronomical
  Society of Japan, 63, S837

\bibitem[{{Yang} {et~al.}(2018){Yang}, {Kafexhiu}, \&
  {Aharonian}}]{2018A&A...615A.108Y}
{Yang}, R.-z., {Kafexhiu}, E., \& {Aharonian}, F. 2018, \aap, 615, A108

\bibitem[{{Yao} {et~al.}(2017){Yao}, {Manchester}, \&
  {Wang}}]{2017ApJ...835...29Y}
{Yao}, J.~M., {Manchester}, R.~N., \& {Wang}, N. 2017, \apj, 835, 29

\bibitem[{{Zeng} {et~al.}(2021){Zeng}, {Xin}, {Zhang}, \&
  {Liu}}]{2021ApJ...910...78Z}
{Zeng}, H., {Xin}, Y., {Zhang}, S., \& {Liu}, S. 2021, \apj, 910, 78

\end{thebibliography}
\bibliographystyle{aa}

\appendix
\section{Appendices}
\label{sec:append}

\subsection{XMM-Newton pointings specifics}
The European Photon Imaging Camera (EPIC) is a three-detector system (MOS1, MOS2 \citep{2001A&A...365L..27T}, and PN \citep{2001A&A...365L..18S}) mounted on the XMM-Newton telescope that operates in the energy range of 0.2-15 keV. EPIC detectors were operating in full frame mode in all three observations, allowing data analysis of diffuse X-ray emission. All detectors and thus all three observations were subjected to an exposure of a few tens of ks, resulting in good statistics for the analysis of images and spectra. In more detail, the profiles of the three observations are described as follows, ID:0823031001 : MOS1 full frame/MOS2 full frame/PN full frame, exposures: 17.7 ks/17.7 ks/15.8 ks, PI: Bettina Posselt,  ID:0823030401: MOS1 full frame/MOS2 full frame/PN full frame, exposures: 16.7 ks/16.7 ks/14.8 ks, PI: Bettina Posselt, and   ID:0823030301: MOS1 full frame/MOS2 full frame/PN full frame, exposures: 9.0 ks/9.0 ks/7.1 ks, PI: Bettina Posselt. 
The limited FoV of XMM-Newton, of $0.5\degree$ size, makes it impossible to derive information about the morphology of a remnant of size $3\degree$, as seen in Fig.~\ref{ROSATanalysis}. However, the improved statistical quality of the data in the respective limited areas, in comparison to ROSAT and/or eROSITA, allows us to examine the consistency of the findings among the three X-ray instruments and perform a detailed spectrum analysis verifying eROSITA best-fit spectral parameters.
Data reduction was performed using the eSAS software. In particular, eSAS tasks \texttt{emchain} and \texttt{epchain} were utilised to reprocess all observation data files. We furthermore employed the \texttt{mos-spectra} and \texttt{pn-spectra} commands to construct the images and extract the spectrum from the regions of interest. XMM-Newton sky maps, from all three observations, were produced by combining data from all three detectors but excluding CCD chips found in anomalous state. Moreover, point sources were filtered out, aiming at enhancing the visibility of the diffuse X-ray emission and avoiding possible contamination of the data. Vignetting corrections were applied, and adaptive smoothing using the default smoothing kernel of 50 counts was performed.

\subsection{\textit{XMM-Newton spectral analysis}}\label{XMMSPECI}

\begin{figure*}[h!]

    \includegraphics[width=0.5\textwidth,clip=true, trim= 0.9cm 0.1cm 1.2cm 0.5cm]{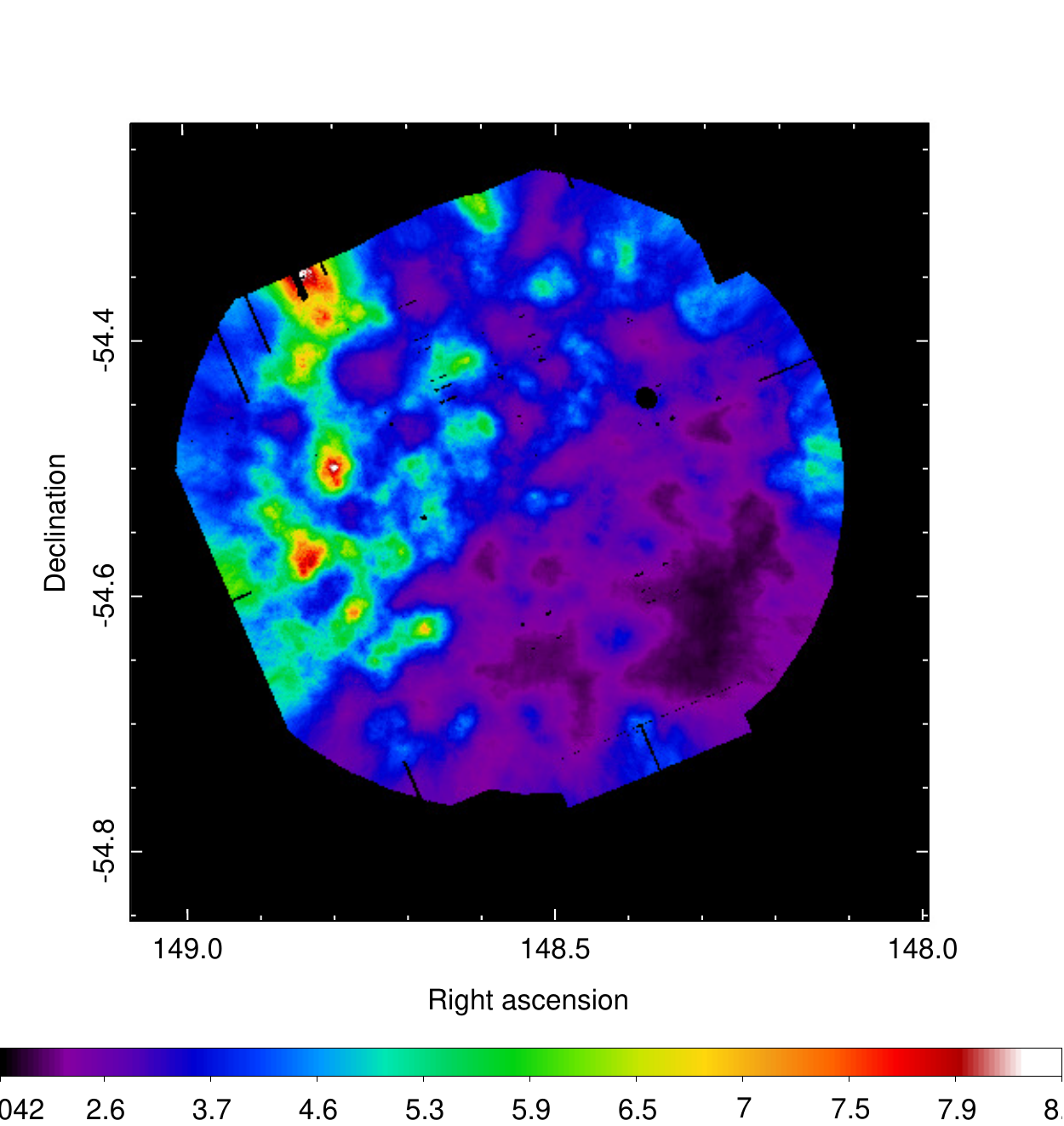}
    \includegraphics[width=0.5\textwidth,clip=true, trim= 0.9cm 0.1cm 0.9cm 0.5cm]{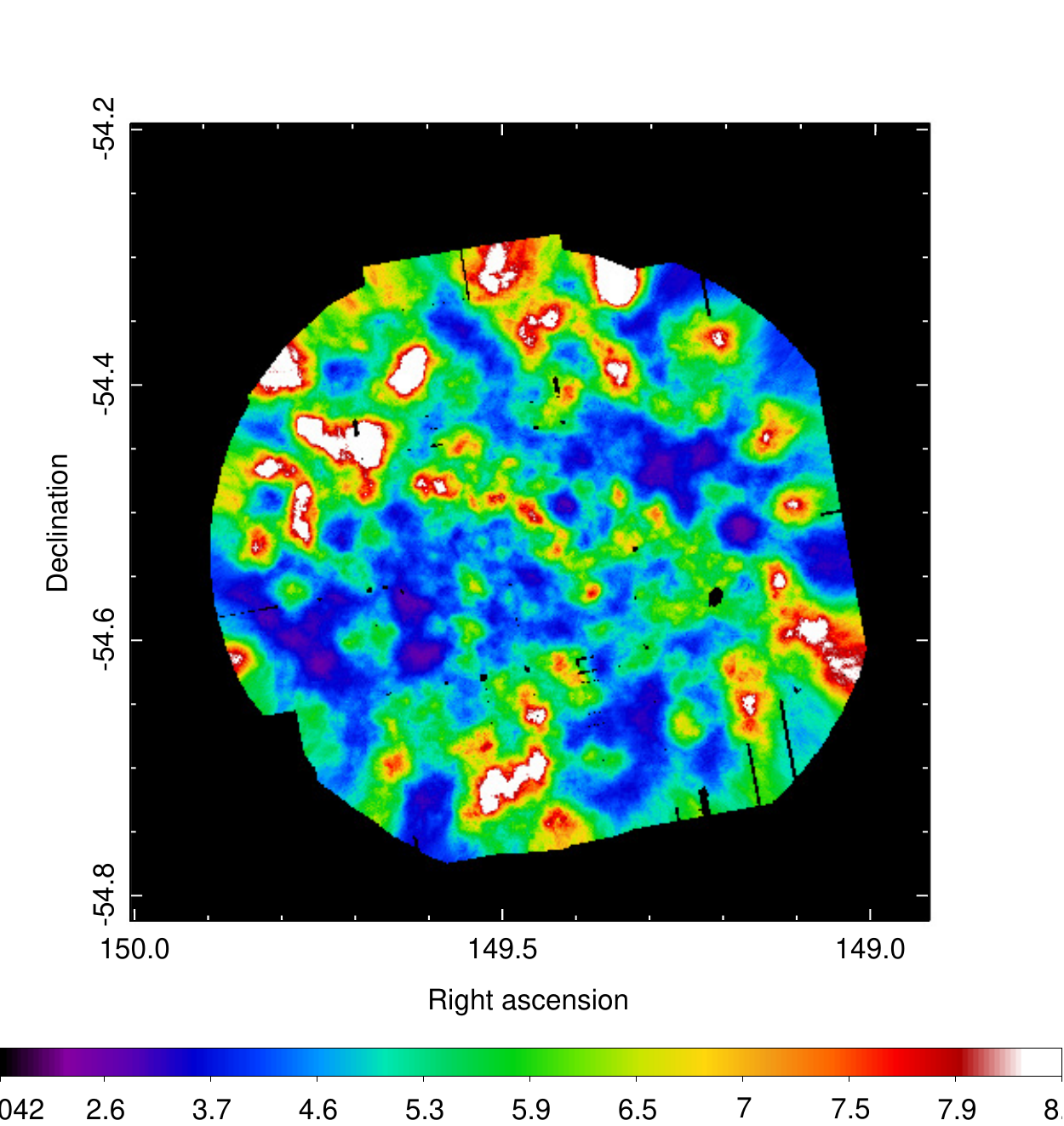}\\
    
    \centering\includegraphics[width=0.5\textwidth,clip=true, trim= 0.9cm 0.1cm 0.9cm 0.5cm]{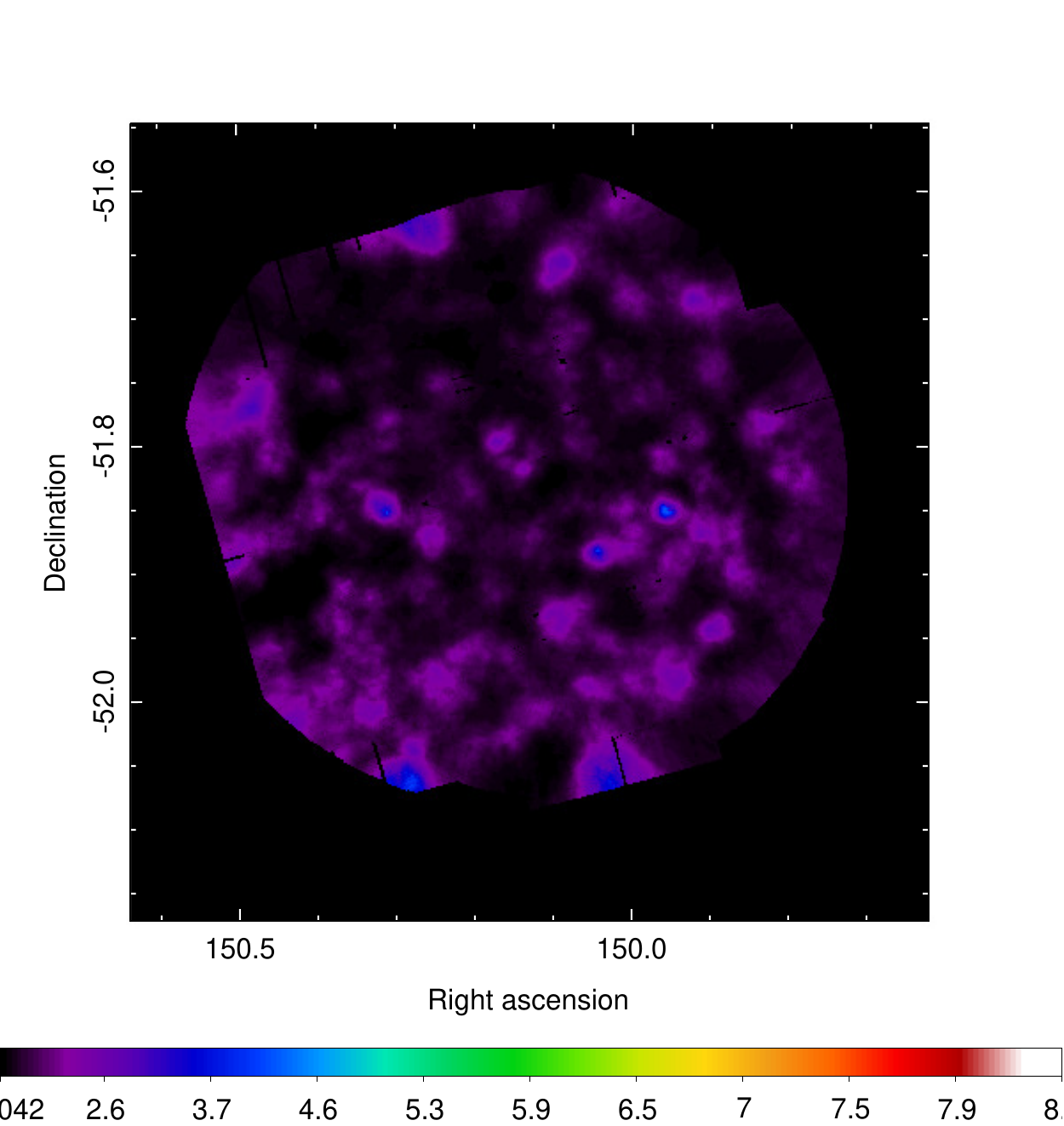}
    \caption{XMM-Newton surface brightness maps in the 0.3-1.1~keV energy band, in units of $\mathrm{counts/s/deg^2}$. Points sources are filtered out, and the maps, of $5''$ pixel size, are adaptively smoothed with a 50 counts kernel and vignetting-corrected. Upper left panel: 0823030401 XMM-Newton pointing. Upper right panel: 0823031001 XMM-Newton pointing. Lower left panel: 0823030301 XMM-Newton pointing. Within each circle one level contours are used, of identical scale for all three pointings, aiming at illustrating regions of enhanced X-ray emission. 
    The 0823031001 pointing is fully dominated by X-ray emission originating from the remnant, the 0823030401 pointing is only partially filled with X-ray emission, whereas the 0823030301 pointing is totally free of the remnant's X-ray emission.}
    \label{XMMIM}
\end{figure*}

\begin{figure*}[h]

    \includegraphics[width=0.54\textwidth]{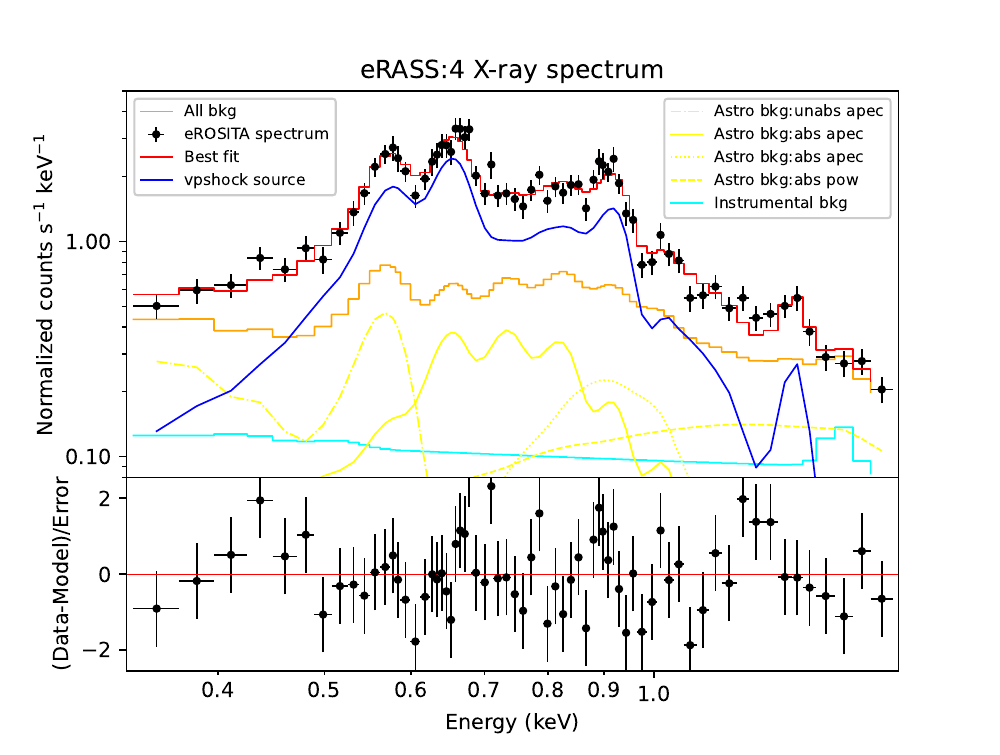}\includegraphics[width=0.54\textwidth]{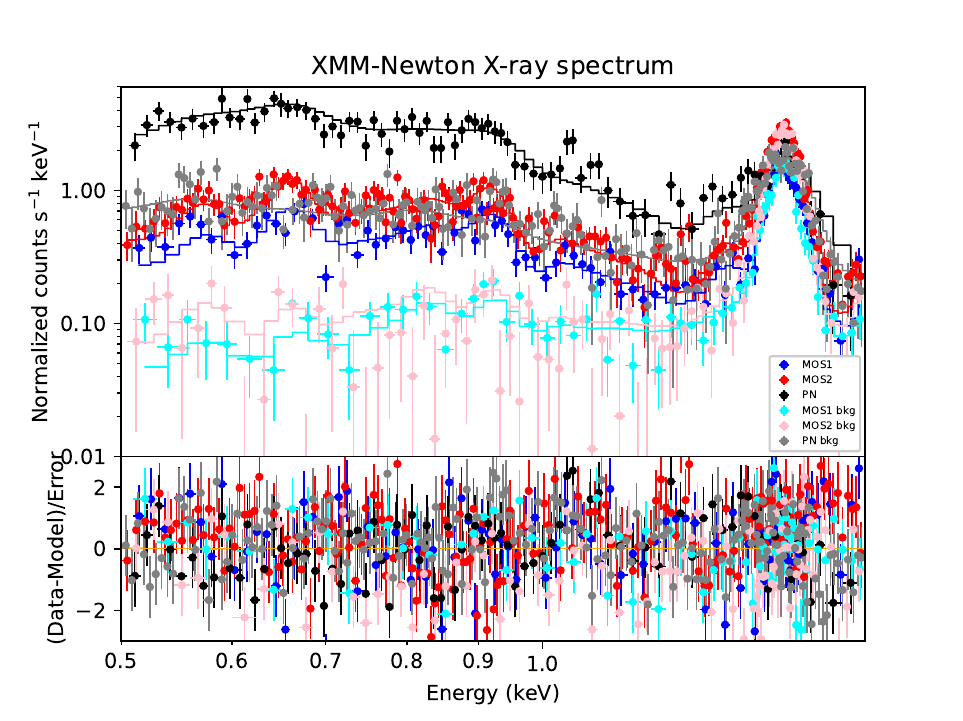}
    \caption{X-ray spectrum from a portion of the remnant which is spatially coincident with the FoV of 0823031001 XMM-Newton observation. Left panel: eRASS:4 data in the 0.3-1.7 keV energy band. To obtain a representative model for the background emission we made use of the black circlular nearby background control region, defined in Fig.~\ref{nett1}, and applied the obtained best-fit model in the simultaneous fitting of the source and background emission of the on-source region. Right panel: pn, mos1, and mos2 data in the 0.5-1.7 keV energy band. The corresponding XMM-Newton background spectrum was obtained from the nearby 0823031401 XMM-Newton observation, which is free of the remnant emission.}
    \label{eRASSvsXMM}
\end{figure*}

Spectral extraction and analysis was performed for all three available XMM-Newton observations. Apart from the \texttt{pn\_spectra} and \texttt{mos\_spectra} eSAS commands, we further made use of $\texttt{pn\_back}$ and $\texttt{mos\_back}$ eSAS tasks to estimate the quiescent particle background. XMM-Newton observation ID:0823031301 was used as a background control region since it is free of diffuse X-ray emission originating from the remnant. The whole XMM-Newton FoV was used for the spectral extraction of the observation ID:0823031001 since it is fully dominated by diffuse X-ray emission emanating from the remnant, as depicted on the upper right panel of Fig~\ref{XMMIM}. On the other hand, diffuse X-ray emission from the remnant covers only partially the observation ID:0823031401, and thus one can use approximately half of the XMM-Newton FoV as on-source region and half as background region. 
A simultaneous fitting of the source and background emission was performed in the 0.5-1.7~keV energy range (we restricted the XMM-Newton spectral analysis above 0.5 keV since some anomalous fluctuations were detected below 0.5 keV for PN and MOS2 instruments) in a similar manner to the eROSITA spectral analysis. Three distinct models were used to describe the X-ray emission from portions of the remnant as seen with XMM-Newton. Those three model components can be discriminated to source emission: \texttt{tbabs$\times$vpshock}, X-ray background: \texttt{const$\times$ const$\times$(apec+tbabs$\times$(apec+apec+pow))}, and soft proton events/instrumental background: \texttt{unabsorbed~power~law+gauss+gauss}. The entire FoV of the XMM-Newton ID:0823031001 observation exhibits spectral features which can be well described by either an absorbed CIE model of kT$\sim0.16$~keV or an absorbed NEI model of kT$\sim0.8-0.9$~keV. The latter finding is in excellent agreement with the eROSITA spectral results from the South-Western parts of the remnant. However, among the selected models for spectral analysis, we show spectral fit results for the \texttt{tbabs(vpshock)} model (as shown on the right panel of Fig~\ref{eRASSvsXMM}). We suggest that the latter model describes this region best since it provides the highest fit quality and absorption column density values that are well aligned with the expected $\mathrm{N_H}$ values based on optical extinction measurements at the distance of the remnant, as discussed in sec.~\ref{dist}. The latter assertion is also supported by the fact that the corresponding XMM observation does not coincide with regions of enhanced IR emission which is likely related to X-ray absorption (dust clouds), see sec.~\ref{dist} for a detailed discussion.
In addition, aiming at performing a direct spectral consistency check between the two X-ray instruments, we extracted the eROSITA spectrum from a region which is identical to the region of the 0823031001 XMM-Newton observation. The obtained spectral fitting results are shown in the two side-by-side panels of Fig.~\ref{eRASSvsXMM}. 
The main parameters of the spectral fit of ObsID: 0823031001 and eROSITA from the exact same portion of the remnant are summarized in Tab~\ref{TABIS2}, and are found to be in excellent agreement. One noticeable difference, still not significant, is that while letting O and Ne vary, on top of Mg, moderately improves the fit for XMM-Newton data, it does not improve the one for eRASS:4 data (which can be likely attributed to the lower statistics of the eRASS data compared to XMM-Newton). For completeness, we state that, even though we report on Tab~\ref{TABIS2} the best-fit results, if one employs the exact same absorbed CIE model for XMM-Newton data as for eRASS:4 (i.e., only Mg elemental abundance is let vary) almost identical fit results are obtained, kT=$0.17_{-0.01}^{+0.01}$~keV, $\mathrm{N_{H}}=0.52_{-0.02}^{+0.02}\mathrm{10^{22}cm^{-2}}$, and a Mg abundance of $1.95_{-0.33}^{+0.34}$. A single temperature plasma component appears sufficient to describe the region's spectral features. The latter finding does not come as a surprise since the XMM-Newton 0823031001 observation partially overlaps with the C region, as defined in sec.~\ref{ROSITAspec}, which exhibits similar characteristics. It is noteworthy that when allowing the plasma temperature to vary freely (i.e., no constraints on the acceptable temperature range) both instruments favor a somewhat questionably high plasma temperature when employing NEI models (of the scale of 5-10 keV). However, when forcing the plasma temperature to values below 1 keV and then letting it vary; a best fit of $\sim0.8$~keV temperature is obtained when utilizing eRASS:4 data. On the contrary, when exploiting XMM-Newton data, and applying an identical strategy, the same high plasma temperature persists (i.e., the plasma temperature hits the forced high-temperature boundary); thus we fix the latest parameter (i.e., plasma temperature) to the obtained best-fit values from eROSITA (no significant change in the fit quality is found and thus the fit is considered reasonable). Finally, if one tries to use multiple component models (e.g., \texttt{tbabs(vapec+vapec)}) no improvement is obtained for eRASS:4 data whereas only an insignificant improvement is obtained for XMM-Newton data. The obtained parameters are common for both data sets, $\mathrm{N_{H}}\sim0.44~\mathrm{10^{22}cm^{-2}}$, kT$\sim0.54-0.64$~keV, and kT$\sim0.16-0.18$~keV. Once again, multiple component models which include NEI models favor questionably high temperatures and the obtained best-fit model is generally unstable.

\begin{table}
\centering
\caption{Best-fit parameters, with $1\sigma$ errors, of the remnant's X-ray spectrum when utilizing both XMM-Newton and eRASS:4 data from the exact same portion of the remnant (i.e., location of 0823031001 XMM-Newton observation). Where not defined; elemental abundances are set to solar values.}
\renewcommand{\arraystretch}{1.7}
\begin{tabular}
{p{3.5cm} p{10.5cm} p{10.5cm} p{10.5cm} p{10.5cm}}
\hline
Instrument & \multicolumn{2}{c}{eROSITA}& \multicolumn{2}{c}{XMM-Newton}  \\ \hline
Area ($10^{6}~\mathrm{arcs^2}$) & \multicolumn{2}{c}{2.55}& \multicolumn{2}{c}{2.25} \\ \hline
$\mathrm{Surf\_bri}$ ($10^{-3}~\mathrm{c/arcs^2}$) & \multicolumn{2}{c}{2.66}& \multicolumn{2}{c}{212.92} \\ \hline
Model & \multicolumn{4}{c}{vapec}  \\ \hline
kT{\ssmall (keV)}& \multicolumn{2}{c}{$0.17_{-0.01}^{+0.01}$}& \multicolumn{2}{c}{$0.16_{-0.01}^{+0.01}$}  \\ 
$\mathrm{N_{H}}${\ssmall ($\mathrm{10^{22}cm^{-2}}$)}&\multicolumn{2}{c}{$0.51_{-0.03}^{+0.03}$}& \multicolumn{2}{c}{$0.63_{-0.02}^{+0.02}$}   \\
O&\multicolumn{2}{c}{-}& \multicolumn{2}{c}{$0.78_{-0.12}^{+0.17}$}  \\
Ne&\multicolumn{2}{c}{-}& \multicolumn{2}{c}{$0.44_{-0.07}^{+0.09}$}\\ 
Mg&\multicolumn{2}{c}{$2.02_{-0.56}^{+0.61}$}& \multicolumn{2}{c}{$0.93_{-0.22}^{+0.26}$}\\
 \hline
 $\mathrm{\chi^2/dof}$ &\multicolumn{2}{c}{1.17}& \multicolumn{2}{c}{-} \\ 
\hline
Model & \multicolumn{4}{c}{vnei}  \\ \hline
kT{\ssmall (keV)}& \multicolumn{2}{c}{$0.84_{-0.21}^{+0.34}$}& \multicolumn{2}{c}{0.84}  \\ 
$\mathrm{N_{H}}${\ssmall ($\mathrm{10^{22}cm^{-2}}$)}&\multicolumn{2}{c}{$0.11_{-0.03}^{+0.03}$}& \multicolumn{2}{c}{0.11}   \\
O&\multicolumn{2}{c}{$1.48_{-0.18}^{+0.22}$}& \multicolumn{2}{c}{$1.23_{-0.11}^{+0.12}$}  \\
Ne&\multicolumn{2}{c}{$2.84_{-0.45}^{+0.51}$}& \multicolumn{2}{c}{$2.33_{-0.22}^{+0.24}$}\\ 
Mg&\multicolumn{2}{c}{$3.03_{-0.75}^{+0.90}$}& \multicolumn{2}{c}{$2.86_{-0.34}^{+0.36}$}\\
Ionization time (\ssmall{$\mathrm{10^{10}s/cm^{3}}$)} &\multicolumn{2}{c}{$1.83_{-0.56}^{+0.76}$}& \multicolumn{2}{c}{$1.77_{-0.12}^{+0.12}$}\\
\hline
$\mathrm{\chi^2/dof}$ &\multicolumn{2}{c}{1.14}& \multicolumn{2}{c}{-} \\ 
 \hline
Model & \multicolumn{4}{c}{vpshock}  \\ \hline
kT{\ssmall (keV)}& \multicolumn{2}{c}{$0.88_{-0.21}^{+0.67}$}& \multicolumn{2}{c}{0.88}  \\ 
$\mathrm{N_{H}}${\ssmall ($\mathrm{10^{22}cm^{-2}}$)}&\multicolumn{2}{c}{$0.14_{-0.04}^{+0.04}$}& \multicolumn{2}{c}{0.14}   \\
O&\multicolumn{2}{c}{$1.43_{-0.18}^{+0.21}$}& \multicolumn{2}{c}{$1.32_{-0.16}^{+0.19}$}  \\
Ne&\multicolumn{2}{c}{$2.25_{-0.57}^{+0.59}$}& \multicolumn{2}{c}{$1.94_{-0.22}^{+0.25}$}\\ 
Mg&\multicolumn{2}{c}{$2.32_{-0.82}^{+0.92}$}& \multicolumn{2}{c}{$2.25_{-0.29}^{+0.31}$}\\ 
Ionization time (\ssmall{$\mathrm{10^{10}s/cm^{3}}$)} &\multicolumn{2}{c}{$4.36_{-2.13}^{+3.11}$}& \multicolumn{2}{c}{$4.81_{-0.63}^{+0.80}$}\\
 \hline
$\mathrm{\chi^2/dof}$ &\multicolumn{2}{c}{1.06}& \multicolumn{2}{c}{-} \\ 
\hline
\label{TABIS2}
\end{tabular}

\end{table}

\subsection{Single versus multiple component spectral fitting result for the representative regions A, B, C, and the entire remnant}\label{sec:append1}

For each individual region, we started the fitting procedure attempting to fit the data with both a single VAPEC model (CIE) and a single VNEI (or VPSHOCK) (NEI) model.
In more detail, it appears that a simple absorption model of collisionally ionised diffuse gas, \texttt{tbabs(apec)} in Xspec notation, is not sufficient to describe the spectral data, even when O, Ne, and Mg elemental abundances are let to vary ($\chi^2/dof = 1.78$ when fixing the elemental abundances to solar values, and $\chi^2/dof = 1.69$ when letting O, Ne and Mg vary) for region A (North-West part of the remnant), while it provides relatively good fits, nonetheless not the finest possible, only when the aforementioned elemental abundances are free, for region B (central part of the remnant, $\chi^2/dof = 1.27$) and C (South-West part of the remnant, $\chi^2/dof = 1.22$). It is noteworthy that when letting Fe vary (on top of the elemental abundances mentioned above) a $\chi^2/dof=1.51$ value is obtained for region A, however, the elemental abundances for both Mg and Fe are found to be questionably high: around 10 and 30 solar values, respectively, making the derived fit questionable. For regions B and C the improvement is more modest ($\chi^2/dof=1.26, 1.19$ for regions B and C respectively) and is not characterized by the same extreme elemental abundance values of Ne and Mg as those obtained for region A (i.e., 2.5 and 3.3 solar values were derived for Ne and Mg in the case of region C). A significantly improved fit in all the above cases can also be obtained when letting N free (on top of O, Ne, Mg, and Fe) which results in extremely low N values, and thus one can fix it to zero. In spite of this, the fits obtained are still not ideal (e.g., $\chi^2/dof/=1.24$ for region A), whereas even if it improves significantly the fits for region B and C ($\chi^2/dof/=1.07$ and $\chi^2/dof/=1.08$ respectively) strong residuals $\sim1.0-1.05$~ keV still persist. When switching to non-equilibrium models (NEI), with free O and Ne elemental abundances (also Mg for region C), a \texttt{tbabs(vnei)} model provides significant improvement in the spectral fitting results for region A ($\chi^2/dof = 1.21$), it suggests that a NEI model can provide a fit of equal goodness for region C ($\chi^2/dof = 1.24$, but a 15~keV temperature is derived which is highly questionable), while it results to a somewhat worse fit for region B ($\chi^2/dof = 1.5$). It is noteworthy that the Mg contribution (Mg bump at $\sim1.3-1.4$~keV) to the total spectrum is well modeled when employing NEI models by keeping the corresponding elemental abundance fixed to solar value (except for region C which exhibits the strongest Mg peak among all three regions and thus letting the latest elemental abundance varying improves the fit), whereas Fe does not change dramatically the quality of the fit as in the case of the CIE model (the same applies for N elemental abundance). Finally, it is also worth mentioning that NEI models favor elemental abundances values which are in the range of 0.9-2 solar values (i.e., no extreme values are derived). In all of the above cases, a much higher temperature plasma and a significantly lower absorption column density are obtained in comparison to CIE models, as shown in Tab.~\ref{TABIS1}. Therefore, NEI models are considered preferable since the derived absorption column density based on the known distance of the remnant is well-aligned with the spectral fit results of NEI models whereas it falls short of the CIE model (see sec.~\ref{dist} for a detailed discussion). Similarly, a \texttt{tbabs(vpshock)} (NEI) model provides highly consistent results, except for region C, in terms of plasma temperature and absorption column density with a \texttt{tbabs(vnei)} model, when applying the same conventions, and the following fitting $\chi^2/dof$ values: 1.17 (region A), 1.42 (region B), and 1.16 (region C with a reasonable temperature of 0.99~keV in comparison to 15~keV obtained when employing a \texttt{tbabs(vnei)} model).

Multiple component models were also employed, and in particular, two temperature plasmas, aiming at improving the fitting results of the three representative spectra extraction regions, even though a \texttt{tbabs(vpshock)} model for regions A and C and a \texttt{tbabs(vapec)} model for region B seem to describe the remnant's spectral data relatively well. When using a \texttt{tbabs(vapec+vapec)} model we obtained significantly improved fitting results for regions A and B in comparison to single component models, as follows: $\chi^2/dof=1.03$ for region A, $\chi^2/dof=1.01$ for region B (letting O of the cooler component and Ne of the hotter one vary for both regions as shown in Tab.~\ref{TABIS1}, whereas both O and Ne for the cooler component and Ne and Mg for the hotter one can be let vary obtaining almost identical results with only slightly modified elemental abundance values), whereas for region C it becomes clear that a single component model is sufficient for the fitting process, as no improvement and an extremely low normalization value is obtained for the second plasma model component (a result which does not come as a surprise as it was already indicated by the eRASS:4 RGB image of Fig.~\ref{RGB}). By using a \texttt{tbabs(vnei+vnei)} model a $\chi^2/dof=0.88$ is obtained for region A with elemental abundances fixed to solar values (one obtains similar results when letting O, Ne, and Mg of the two models vary since they are found to be relatively close to solar values), a $\chi^2/dof=1.06$ is obtained for region B but letting varying O and Ne elemental abundances for both plasma components, whereas for region C a $\chi^2/dof=1.06$ can be obtained when letting the O elemental abundance of the cooler component vary but fixing the higher temperature component to 2.0 keV since when letting kT parameter of that hottest component vary, a questionable high temperature of 10 keV scale is obtained, however, it still remains unconstrained. In this case, a kT=$0.81_{-0.09}^{+0.15}$~keV is obtained for the cooler component with an $\mathrm{N_{h}}=0.29_{-0.04}^{+0.03}\mathrm{10^{22}cm^{-2}}$. Finally, we considered a \texttt{tbabs(vpshock+vpshock)} model, which results to a $\chi^2/dof=0.87$ for region A with elemental abundances fixed to solar values (once again when letting 0 and Ne free almost identical results are obtained), a $\chi^2/dof=1.06$ for region B (0 and Ne elemental abundance free for both model components whereas the temperature of the hotter component is fixed to the best fit before running the Xspec error task, since it remains unconstrained when free), whereas for region C it provides a fit of moderately improved quality ($\chi^2/dof=1.08$ instead of a $\chi^2/dof=1.16$ of a single \texttt{vpshock} model), however, inspecting the corresponding fit and residuals one realizes that the two obtained results (derived from the two distinct models) are almost identical indicating once again that a single component model can sufficiently describe the spectral features of that part of the remnant. In addition, the \texttt{tbabs(vpshock+vpshock)} model is highly unstable for that region, since error computation of the individual parameters of the model result to constant re-fitting. For completeness, we report the obtained best-fit parameters of the \texttt{tbabs(vpshock+vpshock)} model for region C as follows: kT$=3.57$~keV, $\tau=3.87_{-0.73}^{+1.15}\mathrm{10^{9}~s/cm^{3}}$, and kT$=0.78_{-0.04}^{+0.07}$~keV, $\tau=1.17_{-0.18}^{+0.56}\mathrm{10^{11}~s/cm^{3}}$, and $\mathrm{N_{h}}=0.27_{-0.05}^{+0.06}~\mathrm{10^{22}cm^{-2}}$. In what above, we have fixed the hotter plasma temperature to the best-fit value since it remains unconstrained.
We note that for region B when employing a \texttt{tbabs(vnei+vnei)} or a \texttt{tbabs(vpshock+vpshock)} model since both NEI models contribute to the Mg bump the latest is well modeled by keeping the Mg elemental abundance fixed to solar value. However, one can let Mg elemental abundance of the hotter component vary and obtain a fit of equal goodness which favors a somewhat lower temperature for the cooler component $~\sim0.23$~keV and higher elemental abundances, twice as high as shown in Tab.~\ref{TABIS1}, of O, Ne for the hotter component.
We additionally considered mixed two temperature plasma model components, i.e., one plasma is found in equilibrium while the other one is in non-equilibrium ionization (\texttt{tbabs(vapec+vnei)} or \texttt{tbabs(vapec+vpshock)}), and we tried to fit the data. When employing the \texttt{tbabs(vapec+vnei)} model a $\chi^2/dof=1.03$ is obtained for region A with abundances fixed to solar,
a $\chi^2/dof=0.99$ is derived for region B by letting vary O and Ne of the cooler component and Ne and Mg of the hotter one, however, the ionization time of the NEI model remains unconstrained - which suggests uncertainty on the nature of the hotter plasma component (NEI or CIE) -  thus we fix it to the best-fit value before running the Xspec error task. For region C, the fit is of equal goodness with single component models with one component being substantially weaker (at the level of the astrophysical background components) than the other. Similarly, $\chi^2/dof=0.96$, $\chi^2/dof=1.02$, are obtained for regions A and B, respectively, when using a \texttt{tbabs(vapec+vpshock)} model, while for region C the single model component remains the preferred option. To sum up, a two-temperature plasma model describes best the spectral features of regions A and B (without a clear preference when it comes to selection between CIE, NEI, or mixed morphology models), whereas region C can be best described by a single plasma in non-equilibrium (\texttt{tbabs(vpshock)}). However, for the latter a \texttt{tbabs(vpshock+vpshock)} temperature component in equilibrium cannot be excluded. It is noteworthy, that when employing the magenta or blue background regions as control background regions, the obtained $\chi^2/dof$ value for all single-component models and for all three regions are worse (significantly higher by a factor of $~\sim1.2-1.3$), whereas the two temperature plasma component provide fits of equal goodness regardless of the selected background and of consistent best fit spectral parameters. 
On the other hand, when fitting the spectrum of the entire remnant, two temperature plasma components provide by far better-fit quality compared to single-temperature models (either CIE or NEI) which can only model the spectrum poorly even when letting N, O, Ne, Mg, and Fe elemental abundances vary. Among all models mentioned above, a two-temperature plasma component in non-equilibrium (\texttt{tbabs(vphock+vpshock)}), letting O and Ne to vary, provides a fit of $\chi^2/dof=1.3$. A best-fit of $\chi^2/dof=1.19$ quality can be achieved as shown on the last column of Tab.~\ref{TABIS1}. One can try to let Fe of the cooler component vary and obtain a $\chi^2/dof=1.15$ fit quality, however, the latest action further increases the value of the rest of elemental abundances which are left vary and results in a 7 solar value Fe elemental abundance.

\subsection{Spectral analysis of the remnant's sub-regions defined by Voronoi binning analysis}\label{sec:append2}

In what below, we consider a \texttt{vpshock} model as the representative NEI model for the fitting process, as a \texttt{vnei} model results in unconstrained plasma temperatures in many cases. However, we note that in most of the cases, a \texttt{vnei} provides fitting results of only marginally worse fit quality but of the same best-fit values as the \texttt{vpshock}. In addition, we suggest that where applicable NEI models are favored over a CIE model since the distance of the remnant (either $\sim2.5$~kpc or $0.4$~kpc) supports an absorption column density which is well-aligned with non-equilibrium plasma models whereas it falls short of single plasma temperature models in equilibrium. However, in a number of sub-regions in the remnant's surroundings (i.e., sub-regions that do not exhibit high diffuse X-ray excess compared to the astrophysical background) CIE models provide an excellent fit to the data. In Fig.~\ref{ALLSUBSPEC} we summarize the spectral best-fit results from all sub-regions of the remnant. In Tab.~\ref{TABISOO} we report on the best-fit spectral parameters of each sub-region.

\begin{table*}
\centering
\caption{Best-fit parameters, with $1\sigma$ errors, of the sub-regions (defined via the Voronoi analysis process). Where not defined; elemental abundances are set to solar values.}
\renewcommand{\arraystretch}{1.7}
\setlength{\tabcolsep}{9pt}
\begin{tabular}
{p{2.5cm} p{5.5cm} p{5.5cm} p{5.5cm} p{5.5cm} p{5.5cm} p{5.5cm} p{5.5cm} p{5.5cm} p{5.5cm} p{5.5cm} p{5.5cm} p{5.5cm}}
\hline
Region & \multicolumn{3}{c}{A' region}& \multicolumn{3}{c}{B' region} &\multicolumn{3}{c}{C' region} &\multicolumn{3}{c}{D region}\\ \hline
Model &\multicolumn{3}{c}{vpshock} &\multicolumn{3}{c}{vpshock}&\multicolumn{3}{c}{vpshock}&\multicolumn{3}{c}{vpshock}  \\ \hline
kT{\ssmall (keV)}& \multicolumn{3}{c}{$0.61_{-0.17}^{+0.38}$}& \multicolumn{3}{c}{$0.61$} & \multicolumn{3}{c}{$2.0$}& \multicolumn{3}{c}{$2.0$} \\ 
$\mathrm{N_{H}}${\ssmall ($\mathrm{10^{22}cm^{-2}}$)}&\multicolumn{3}{c}{$0.42_{-0.06}^{+0.06}$}& \multicolumn{3}{c}{$0.38_{-0.03}^{+0.03}$}  & \multicolumn{3}{c}{$0.23_{-0.04}^{+0.03}$}&\multicolumn{3}{c}{$0.42_{-0.07}^{+0.06}$}\\
O&\multicolumn{3}{c}{-}& \multicolumn{3}{c}{-} & \multicolumn{3}{c}{$0.88_{-0.11}^{+0.15}$}&\multicolumn{3}{c}{-} \\
Ne&\multicolumn{3}{c}{-}& \multicolumn{3}{c}{-}& \multicolumn{3}{c}{$1.12_{-0.21}^{+0.30}$}&\multicolumn{3}{c}{-}\\ 
Mg&\multicolumn{3}{c}{-}& \multicolumn{3}{c}{-}& \multicolumn{3}{c}{$2.46_{-0.67}^{+0.92}$}&\multicolumn{3}{c}{-}\\
 Ionization time (\ssmall{$\mathrm{10^{9}s/cm^{3}}$)} &\multicolumn{3}{c}{$0.90_{-0.10}^{+0.12}$}& \multicolumn{3}{c}{$8.88_{-1.47}^{+1.60}$} & \multicolumn{3}{c}{$6.31_{-0.86}^{+1.17}$}&\multicolumn{3}{c}{$2.13_{-0.42}^{+0.67}$}\\
\hline
 $\mathrm{\chi^2/dof}$ &\multicolumn{3}{c}{1.2}& \multicolumn{3}{c}{1.33} &  \multicolumn{3}{c}{1.02}&\multicolumn{3}{c}{1.02} \\ 
\hline
Region & \multicolumn{3}{c}{E region}& \multicolumn{3}{c}{F region} &\multicolumn{3}{c}{G region} &\multicolumn{3}{c}{H region}\\ \hline
Model & \multicolumn{3}{c}{vpshock+vpshock} &\multicolumn{3}{c}{vpshock+vpshock}&\multicolumn{3}{c}{vpshock}&\multicolumn{3}{c}{vpshock}  \\ \hline
kT{\ssmall (keV)}& \multicolumn{3}{c}{$0.75_{-0.10}^{+0.14}$ $1.68_{-0.93}^{+2.24}$}& \multicolumn{3}{c}{$0.8$ $1.37_{-0.28}^{+0.86}$} & \multicolumn{3}{c}{$0.8$}&\multicolumn{3}{c}{$0.43_{-0.17}^{+0.33}$}\\ 
$\mathrm{N_{H}}${\ssmall ($\mathrm{10^{22}cm^{-2}}$)}&\multicolumn{3}{c}{$0.27_{-0.04}^{+0.04}$}& \multicolumn{3}{c}{$0.21_{-0.03}^{+0.03}$} & \multicolumn{3}{c}{$0.40_{-0.03}^{+0.03}$} &\multicolumn{3}{c}{$0.54_{-0.11}^{+0.15}$} \\
O&\multicolumn{3}{c}{-}& \multicolumn{3}{c}{-} & \multicolumn{3}{c}{-} &\multicolumn{3}{c}{$0.15_{-0.03}^{+0.04}$}\\
Ne&\multicolumn{3}{c}{-}& \multicolumn{3}{c}{-}& \multicolumn{3}{c}{-}&\multicolumn{3}{c}{ $0.15_{-0.06}^{+0.06}$}\\ 
Mg&\multicolumn{3}{c}{-}& \multicolumn{3}{c}{-}& \multicolumn{3}{c}{-}&\multicolumn{3}{c}{-}\\
Ionization time (\ssmall{$\mathrm{10^{9}s/cm^{3}}$)} &\multicolumn{3}{c}{$167_{-75}^{+116}$ $1.15_{-0.18}^{+0.20}$}& \multicolumn{3}{c}{$1.84_{-0.34}^{+0.67}$ $62.0_{-24.7}^{+34.3}$} & \multicolumn{3}{c}{$9.75_{-1.46}^{+1.86}$}&\multicolumn{3}{c}{$9.30_{-3.11}^{+9.75}$}\\
\hline
$\mathrm{\chi^2/dof}$ &\multicolumn{3}{c}{1.2}& \multicolumn{3}{c}{1.01} & \multicolumn{3}{c}{1.01}&\multicolumn{3}{c}{1.3}\\ 
 \hline
 Region & \multicolumn{3}{c}{I region}& \multicolumn{3}{c}{J region} &\multicolumn{3}{c}{K region} &\multicolumn{3}{c}{L region}\\ \hline
Model &  \multicolumn{3}{c}{vpshock+vpshock} &\multicolumn{3}{c}{vpshock}&\multicolumn{3}{c}{vpshock+vpshock}&\multicolumn{3}{c}{vpshock+vpshock}  \\ \hline
kT{\ssmall (keV)}& \multicolumn{3}{c}{$0.96_{-0.53}^{+0.59}$ $1.15_{-0.26}^{+0.82}$}& \multicolumn{3}{c}{$2.0$} & \multicolumn{3}{c}{$0.88_{-0.12}^{+0.23}$ $1.56_{-0.67}^{+2.74}$} & \multicolumn{3}{c}{$0.14_{-0.01}^{+0.04}$ $1.26_{-0.40}^{+1.50}$} \\ 
$\mathrm{N_{H}}${\ssmall ($\mathrm{10^{22}cm^{-2}}$)}&\multicolumn{3}{c}{$0.18_{-0.02}^{+0.02}$}& \multicolumn{3}{c}{$0.17_{-0.03}^{+0.03}$} & \multicolumn{3}{c}{$0.28_{-0.05}^{+0.05}$} & \multicolumn{3}{c}{$0.53_{-0.13}^{+0.10}$} \\
O&\multicolumn{3}{c}{-}& \multicolumn{3}{c}{-} & \multicolumn{3}{c}{-} & \multicolumn{3}{c}{$0.57_{-0.18}^{+0.17}$ -} \\
Ne&\multicolumn{3}{c}{-}& \multicolumn{3}{c}{-} & \multicolumn{3}{c}{-} & \multicolumn{3}{c}{$0.21_{-0.21}^{+0.30}$ $0.82_{-0.60}^{+0.67}$}\\ 
Mg&\multicolumn{3}{c}{-}& \multicolumn{3}{c}{-}& \multicolumn{3}{c}{-} & \multicolumn{3}{c}{-} \\ 
Ionization time (\ssmall{$\mathrm{10^{9}s/cm^{3}}$)} &\multicolumn{3}{c}{$6.11_{-1.89}^{+3.31}$ $68.5_{-35.1}^{+70.2}$}& \multicolumn{3}{c}{$6.26_{-0.63}^{+0.79}$}& \multicolumn{3}{c}{$147_{-61}^{+95}$ $1.02_{-0.16}^{+0.19}$} & \multicolumn{3}{c}{$23400$ $35.1_{-19.7}^{+16.21}$}\\
 \hline
$\mathrm{\chi^2/dof}$ &\multicolumn{3}{c}{1.33}& \multicolumn{3}{c}{1.03} & \multicolumn{3}{c}{0.99} &\multicolumn{3}{c}{1.08} \\ 
\hline
Region & \multicolumn{3}{c}{M region}& \multicolumn{3}{c}{N region} &\multicolumn{3}{c}{O region} &\multicolumn{2}{c}{P region}&\multicolumn{1}{c}{Q region}\\ \hline
Model&  \multicolumn{3}{c}{vpshock+vpshock} &\multicolumn{3}{c}{vpshock}&\multicolumn{3}{c}{vpshock}&\multicolumn{2}{c}{vpshock} &\multicolumn{1}{c}{vpshock}\\ \hline
kT{\ssmall (keV)}& \multicolumn{3}{c}{$0.79_{-0.08}^{+0.11}$ $1.20_{-0.35}^{+0.52}$}& \multicolumn{3}{c}{$2.43_{-0.75}^{+1.22}$} &\multicolumn{3}{c}{$0.53_{-0.11}^{+0.13}$} &\multicolumn{2}{c}{$0.8$}&\multicolumn{1}{c}{$0.8$}\\ 
$\mathrm{N_{H}}${\ssmall ($\mathrm{10^{22}cm^{-2}}$)}&\multicolumn{3}{c}{$0.18_{-0.02}^{+0.03}$}& \multicolumn{3}{c}{$0.24_{-0.03}^{+0.03}$}& \multicolumn{3}{c}{$0.24_{-0.04}^{+0.05}$} &\multicolumn{2}{c}{$0.51_{-0.04}^{+0.03}$} & \multicolumn{1}{c}{$0.52_{-0.04}^{+0.04}$} \\
O&\multicolumn{3}{c}{$0.63_{-0.07}^{+0.07}$ -}& \multicolumn{3}{c}{$0.78_{-0.07}^{+0.08}$} &\multicolumn{3}{c}{$0.66_{-0.10}^{+0.12}$}&\multicolumn{2}{c}{-} & \multicolumn{1}{c}{$0.27_{-0.06}^{+0.08}$} \\
Ne&\multicolumn{3}{c}{$0.56_{-0.12}^{+0.13}$ -}& \multicolumn{3}{c}{$1.08_{-0.13}^{+0.15}$}&\multicolumn{3}{c}{$1.07_{-0.16}^{+0.19}$}&\multicolumn{2}{c}{-} &\multicolumn{1}{c}{$0.19_{-0.06}^{+0.07}$}\\ 
Mg&\multicolumn{3}{c}{-}& \multicolumn{3}{c}{-}&\multicolumn{3}{c}{-}& \multicolumn{2}{c}{-} &\multicolumn{1}{c}{$0.76_{-0.23}^{+0.29}$}\\
Fe&\multicolumn{3}{c}{-}& \multicolumn{3}{c}{-}& \multicolumn{3}{c}{ $0.28_{-0.06}^{+0.08}$}&\multicolumn{2}{c}{-}&\multicolumn{1}{c}{-} \\
Ionization time (\ssmall{$\mathrm{10^{9}s/cm^{3}}$)} &\multicolumn{3}{c}{$157_{-56}^{+80}$ $3.50_{-0.87}^{+1.18}$}& \multicolumn{3}{c}{$8.91_{-0.91}^{+1.20}$}&\multicolumn{3}{c}{$236_{-142}^{+360}$}& \multicolumn{2}{c}{$3.75_{-0.43}^{+0.67}$}& \multicolumn{1}{c}{$7.6_{-0.27}^{+0.35}$}\\
 \hline
$\mathrm{\chi^2/dof}$ &\multicolumn{3}{c}{1.1}& \multicolumn{3}{c}{1.04} &\multicolumn{3}{c}{1.19}&\multicolumn{2}{c}{1.25}& \multicolumn{1}{c}{1.2}\\ 
\hline
\label{TABISOO}
\end{tabular}

\end{table*}

Overall, the detailed spectral analysis of sub-regions defined by the Voronoi binning algorithm favors an increased absorption column density in the remnant's surroundings, especially to the South and West of the remnant, supporting previous indications of IRAS data (see Fig.~\ref{IRASS}) for X-ray absorption in those regions due to the prevalence of dust. In addition, the current analysis concludes that the remnant is of incomplete shell-type (or fragmented annulus) morphology, since its central region, namely region J in this work, is both free of X-ray emission and is not characterized by enhanced absorption column density, and its Western part is not observable in X-rays (likely due to X-ray absorption) as resulted from the X-ray spectral fit. The obtained absorption column density map is shown on the left panel of Fig.~\ref{ABSORB}.

\begin{figure*}[]
    \centering
    
    \includegraphics[width=0.496\textwidth]{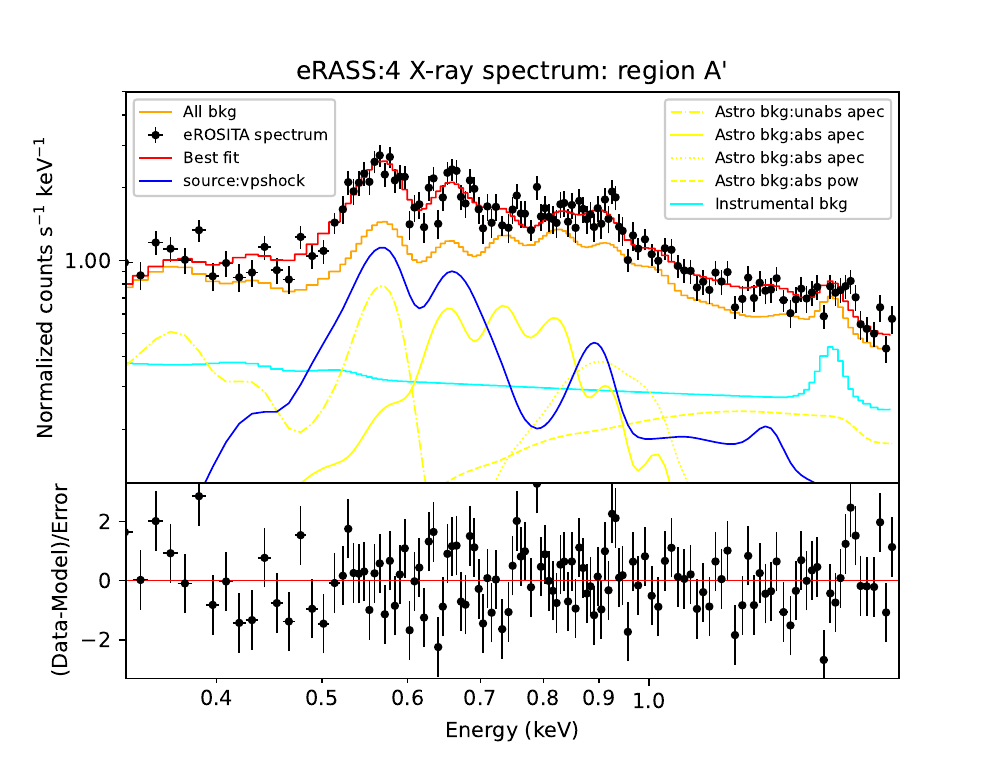}
    \includegraphics[width=0.498\textwidth]{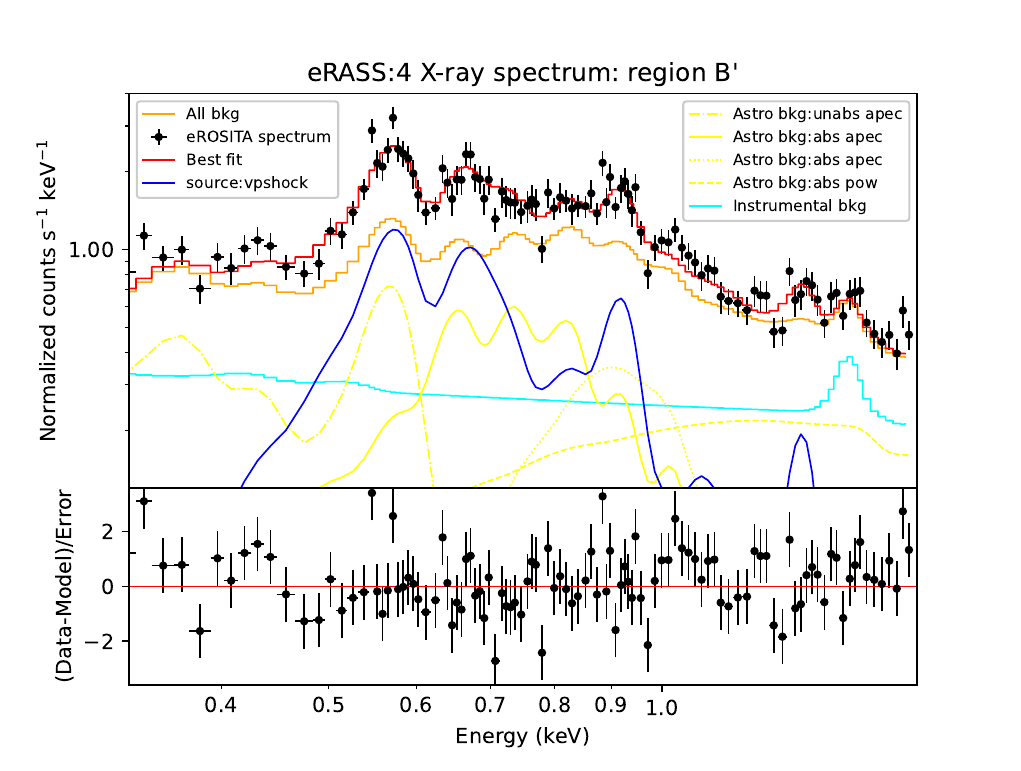}
    \includegraphics[width=0.495\textwidth]{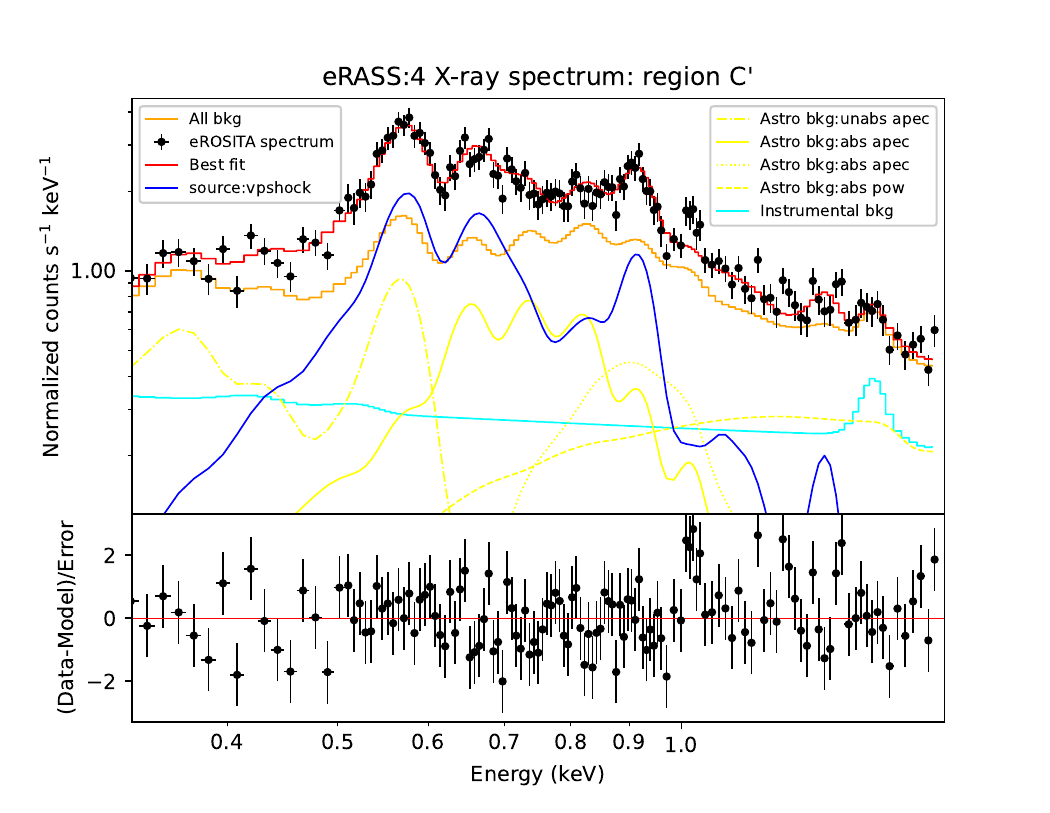}
    \includegraphics[width=0.495\textwidth]{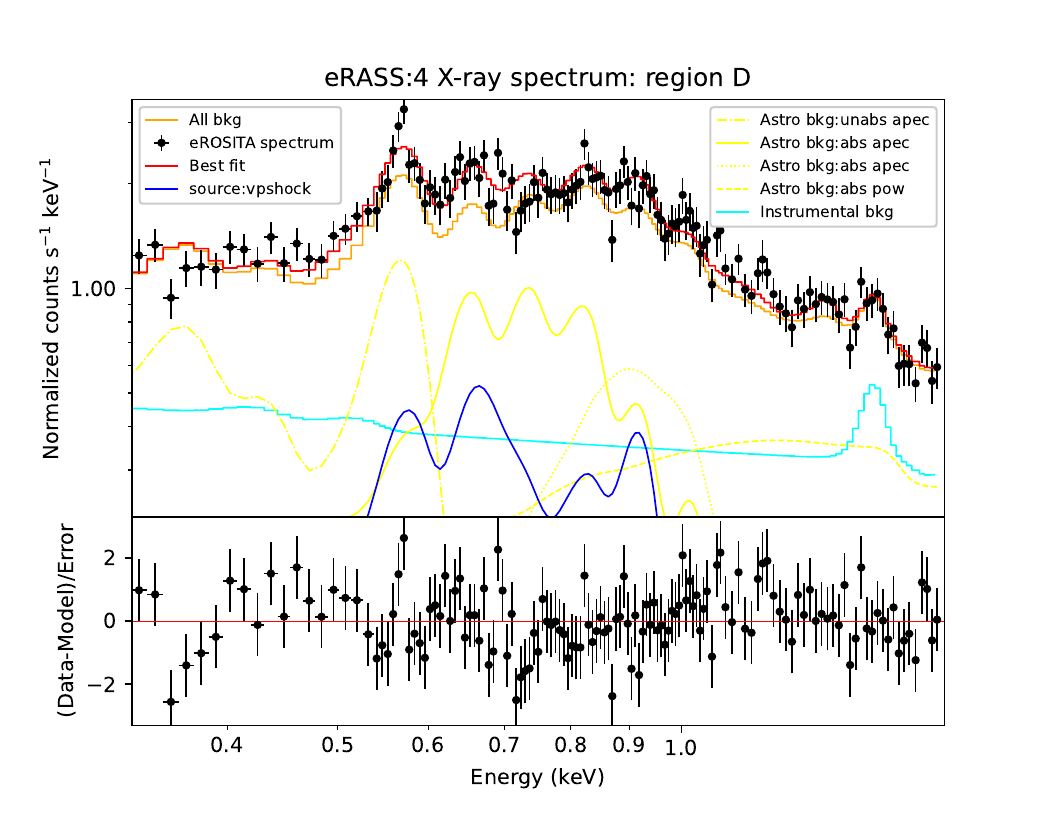}

    \includegraphics[width=0.488\textwidth]{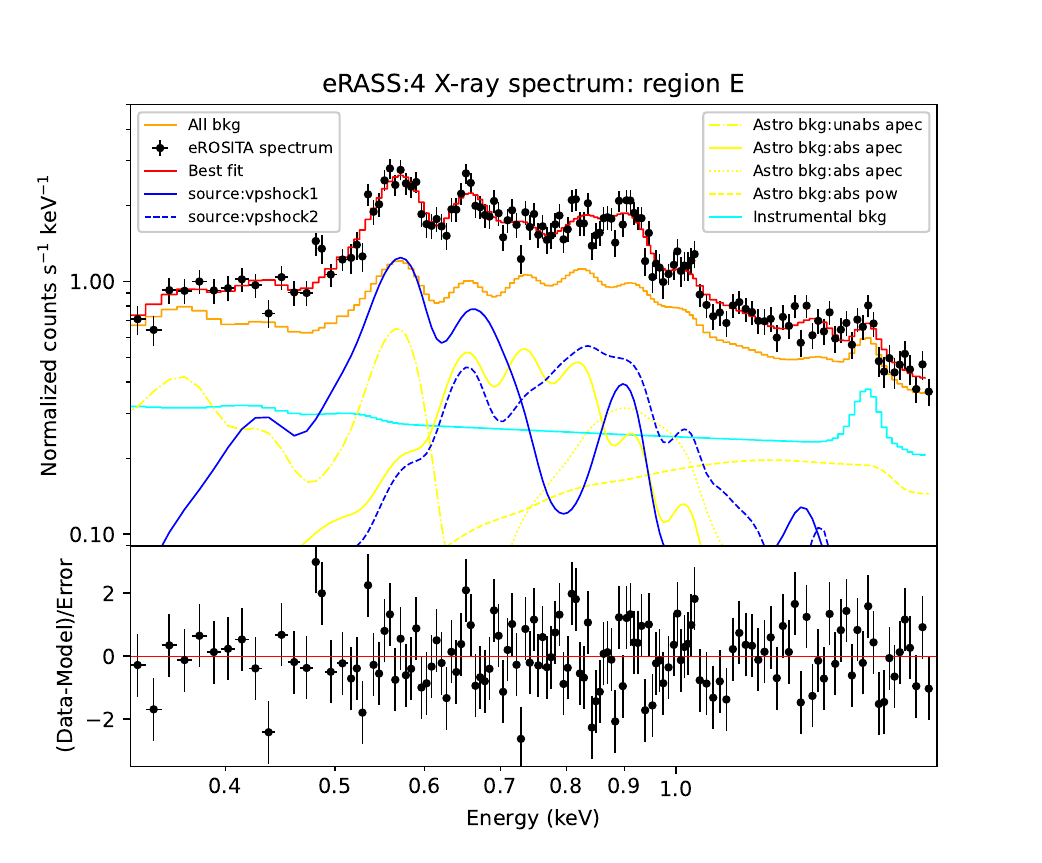}
    \includegraphics[width=0.502\textwidth]{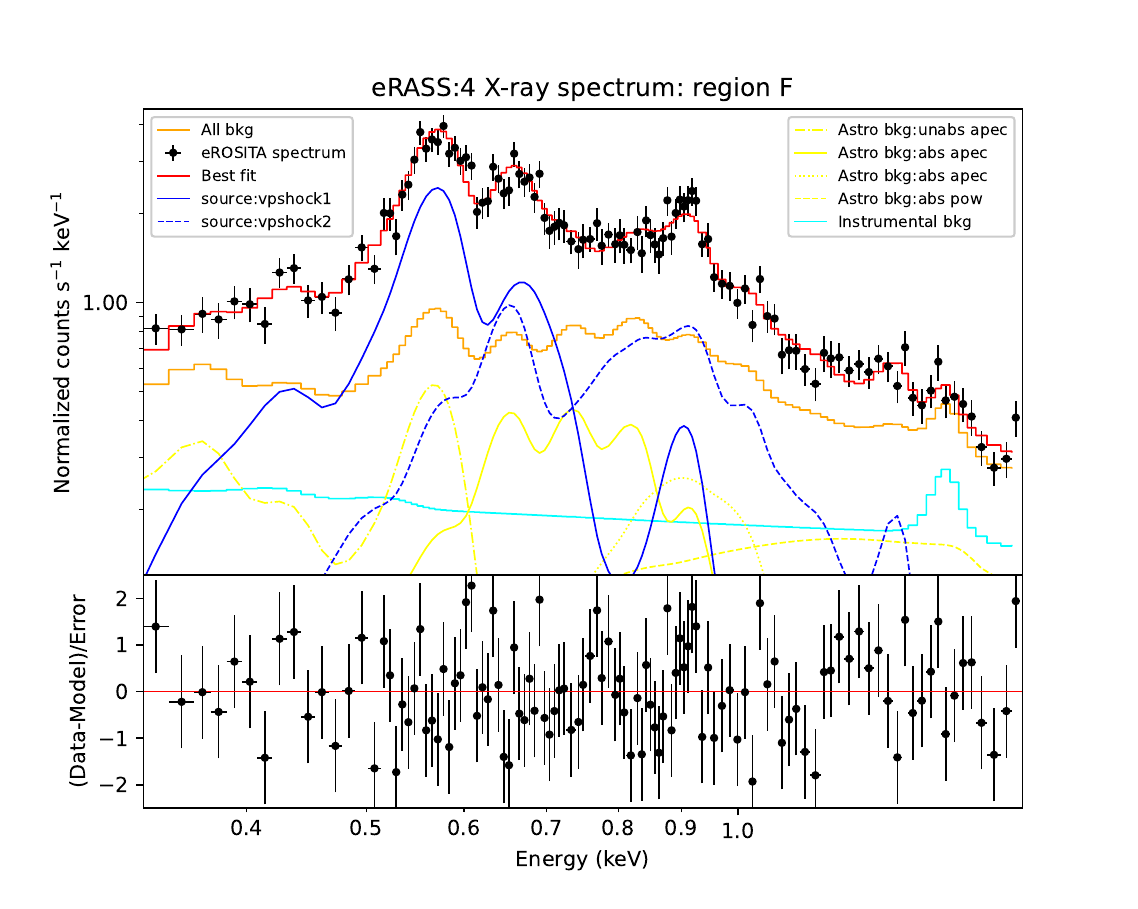}

    \phantomcaption
\end{figure*}
\begin{figure*}
\ContinuedFloat   
    \centering

     \includegraphics[width=0.495\textwidth]{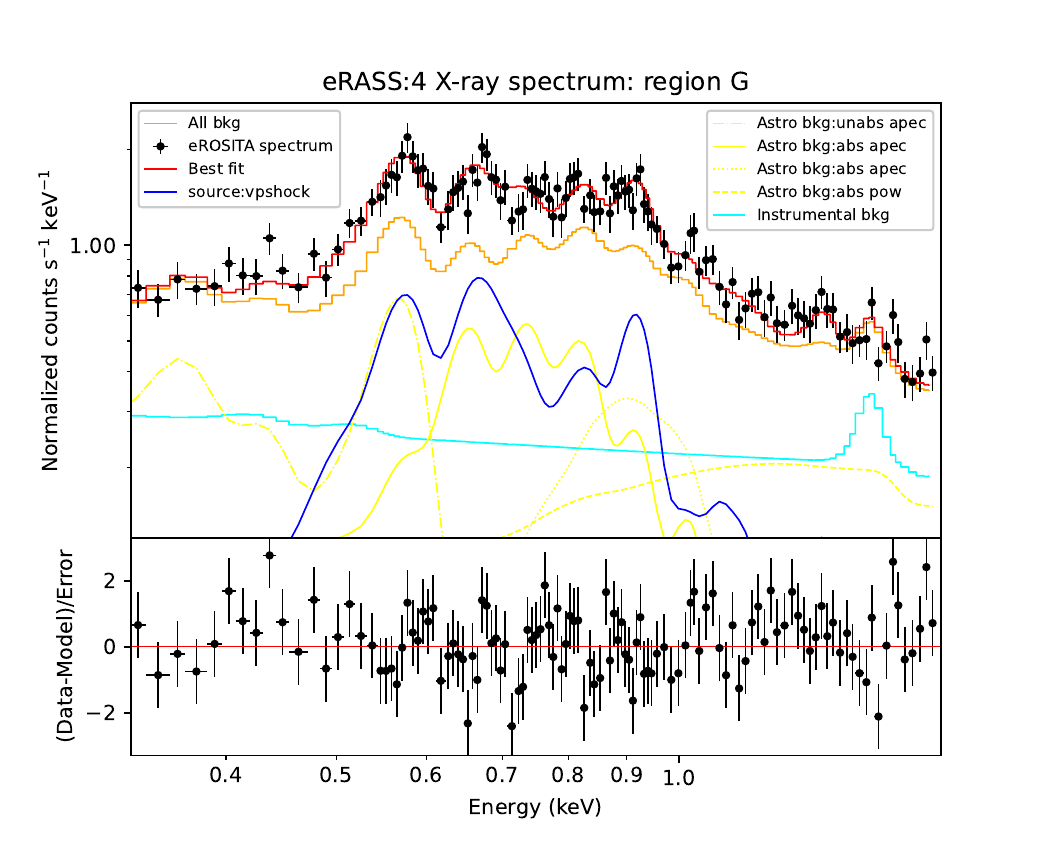}
    \includegraphics[width=0.495\textwidth]{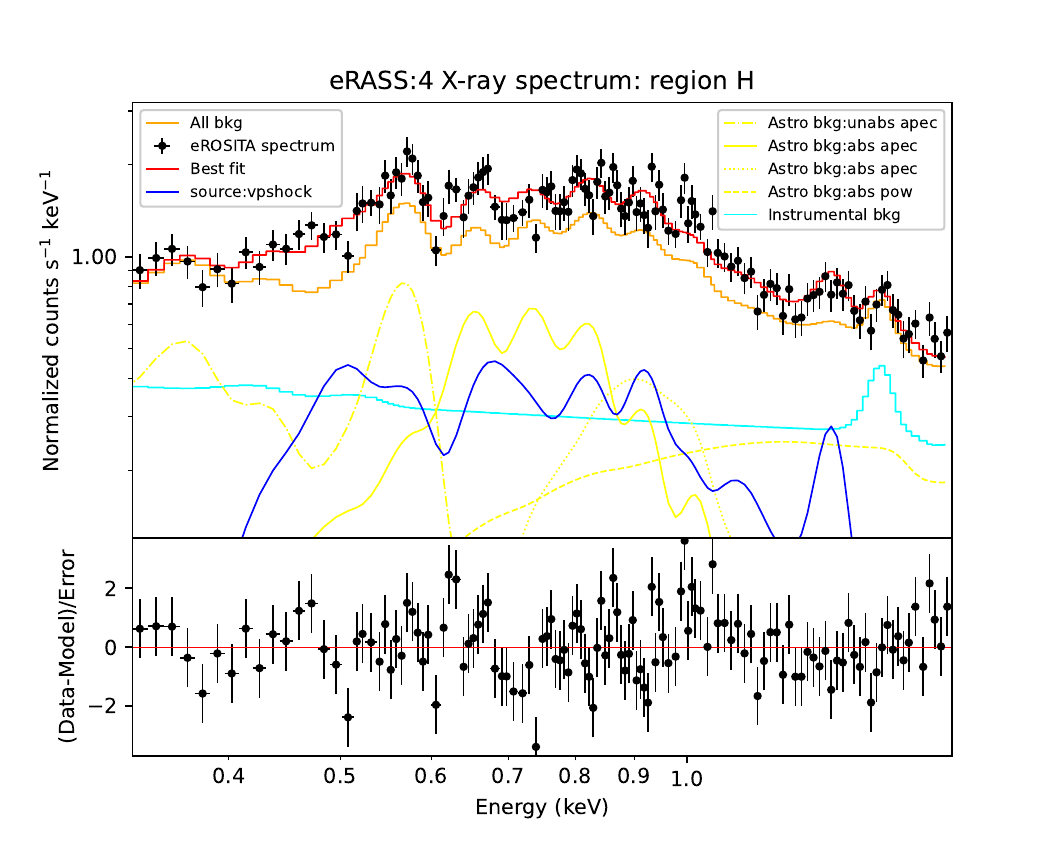}   
    \includegraphics[width=0.501\textwidth]{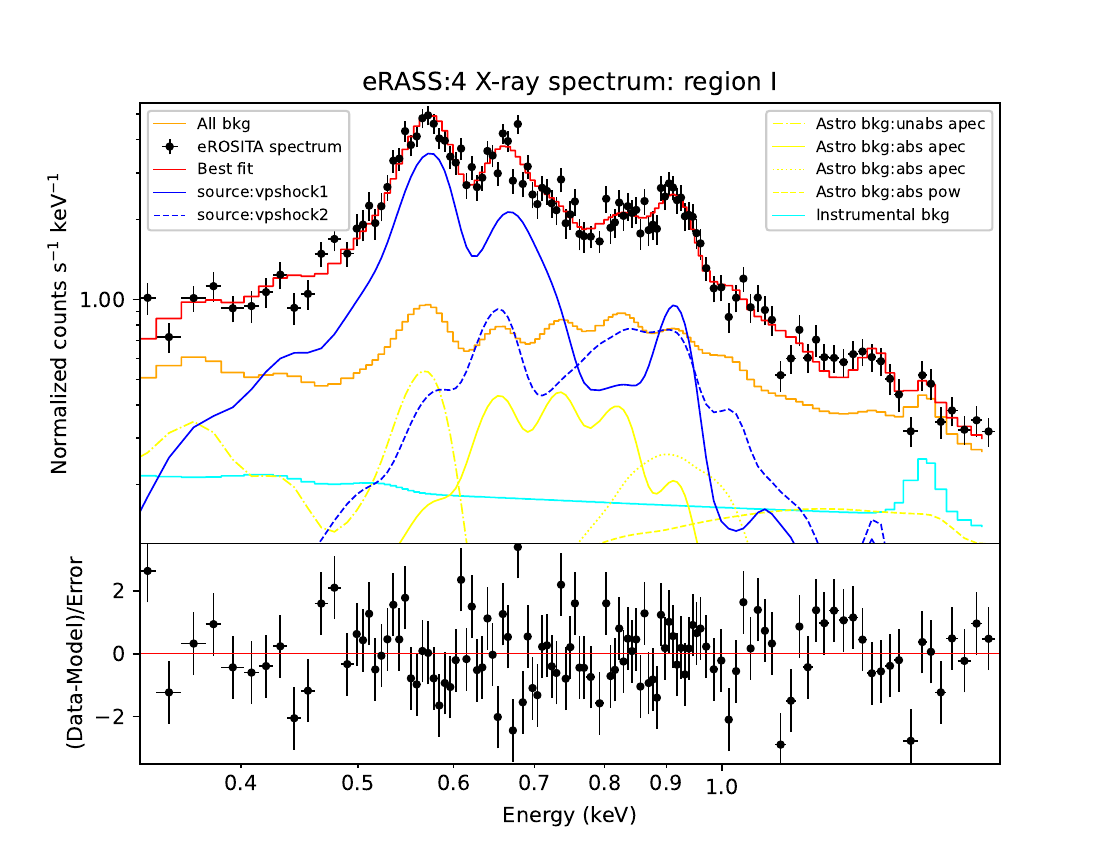}
    \includegraphics[width=0.49\textwidth]{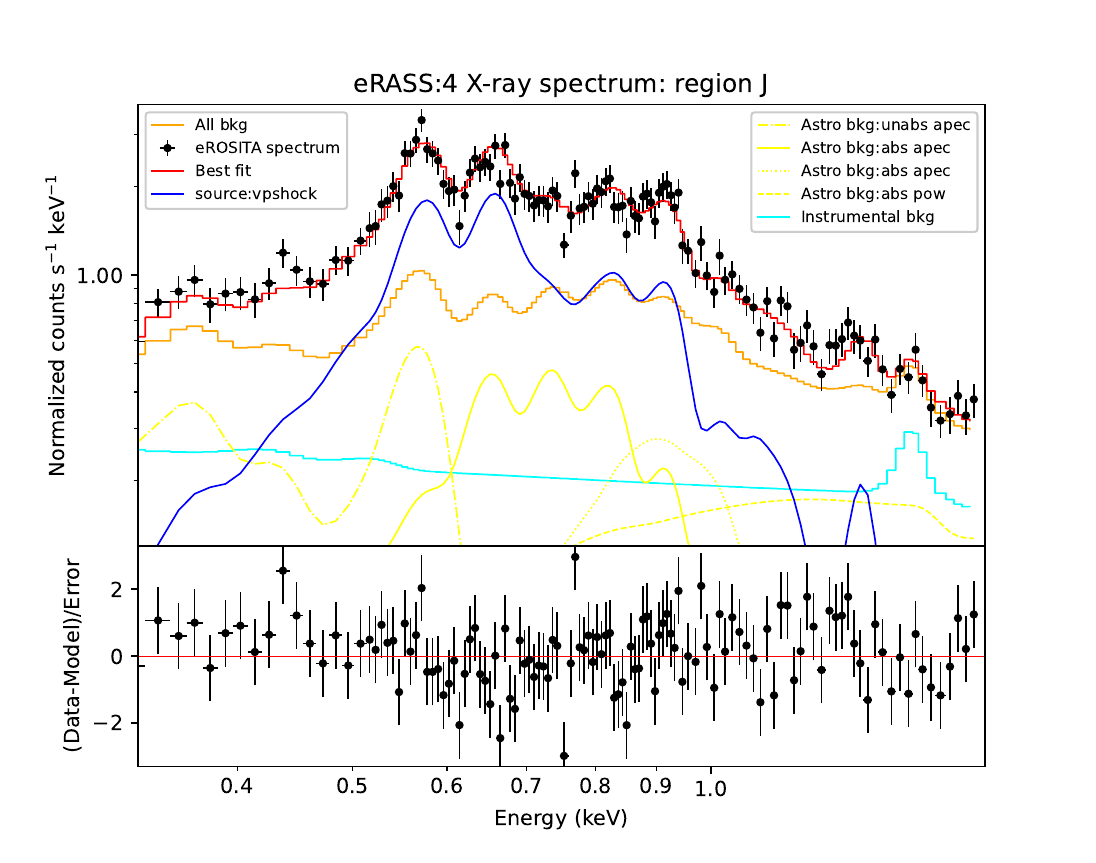}
    \includegraphics[width=0.497\textwidth]{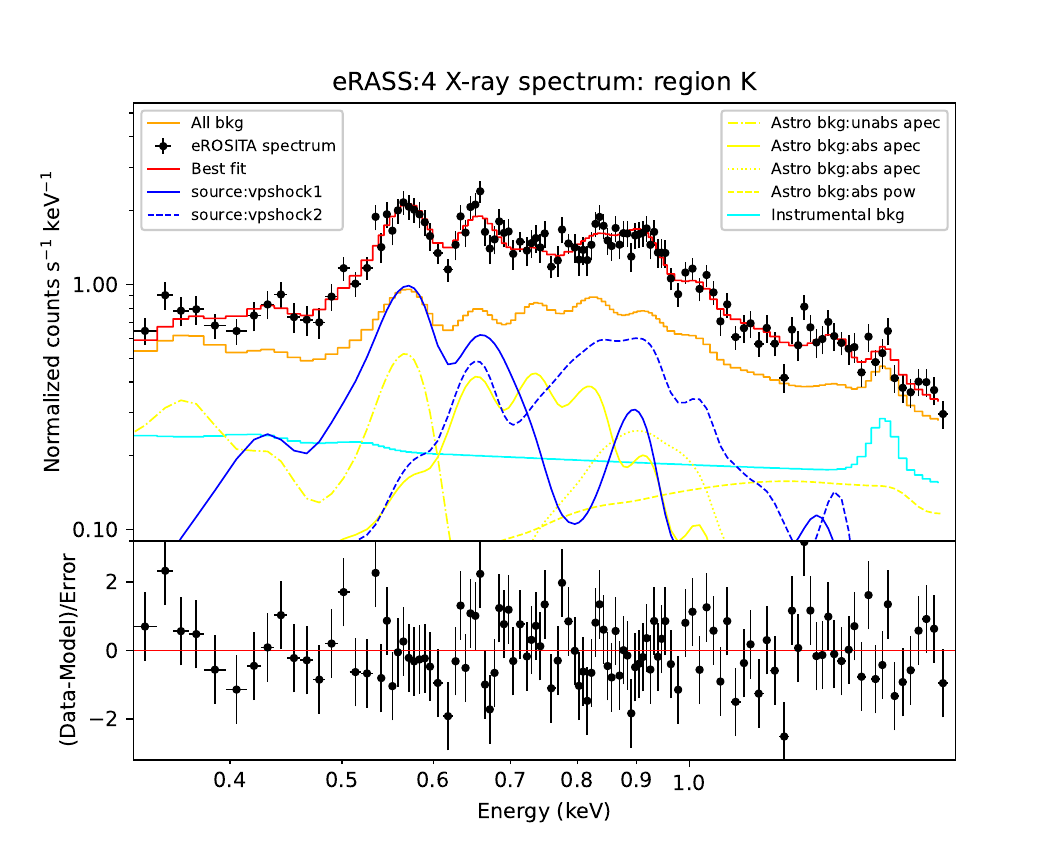}
    \includegraphics[width=0.495\textwidth]{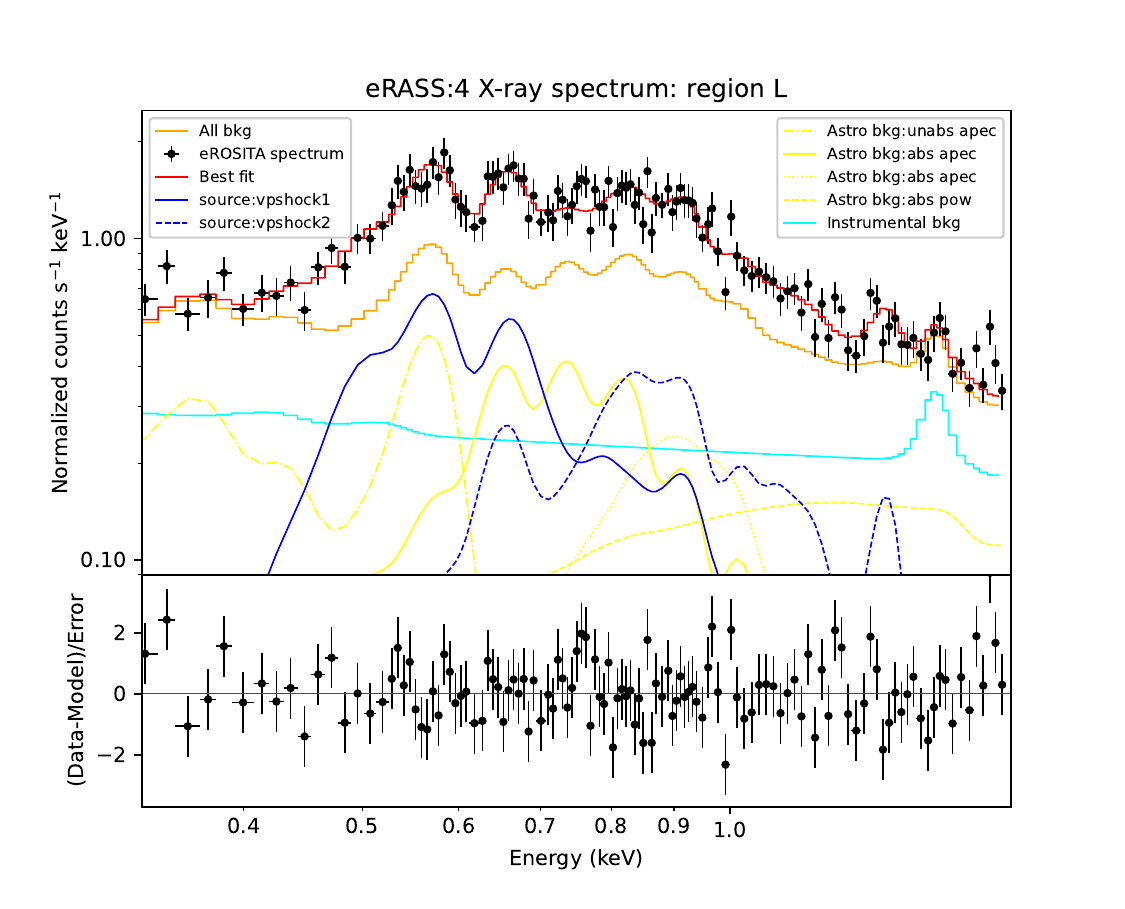}

\end{figure*}
\begin{figure*}
\ContinuedFloat   
    \centering
    \includegraphics[width=0.496\textwidth]{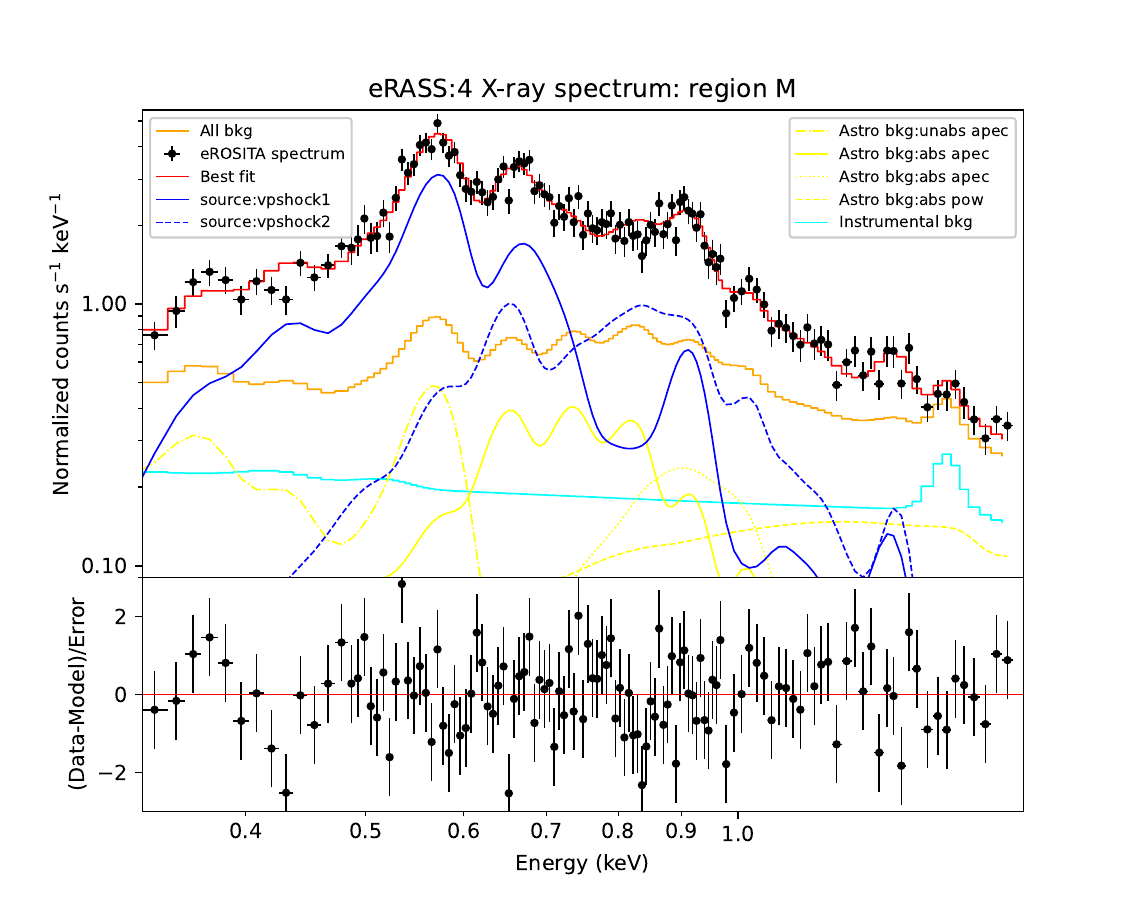}
    \includegraphics[width=0.498\textwidth]{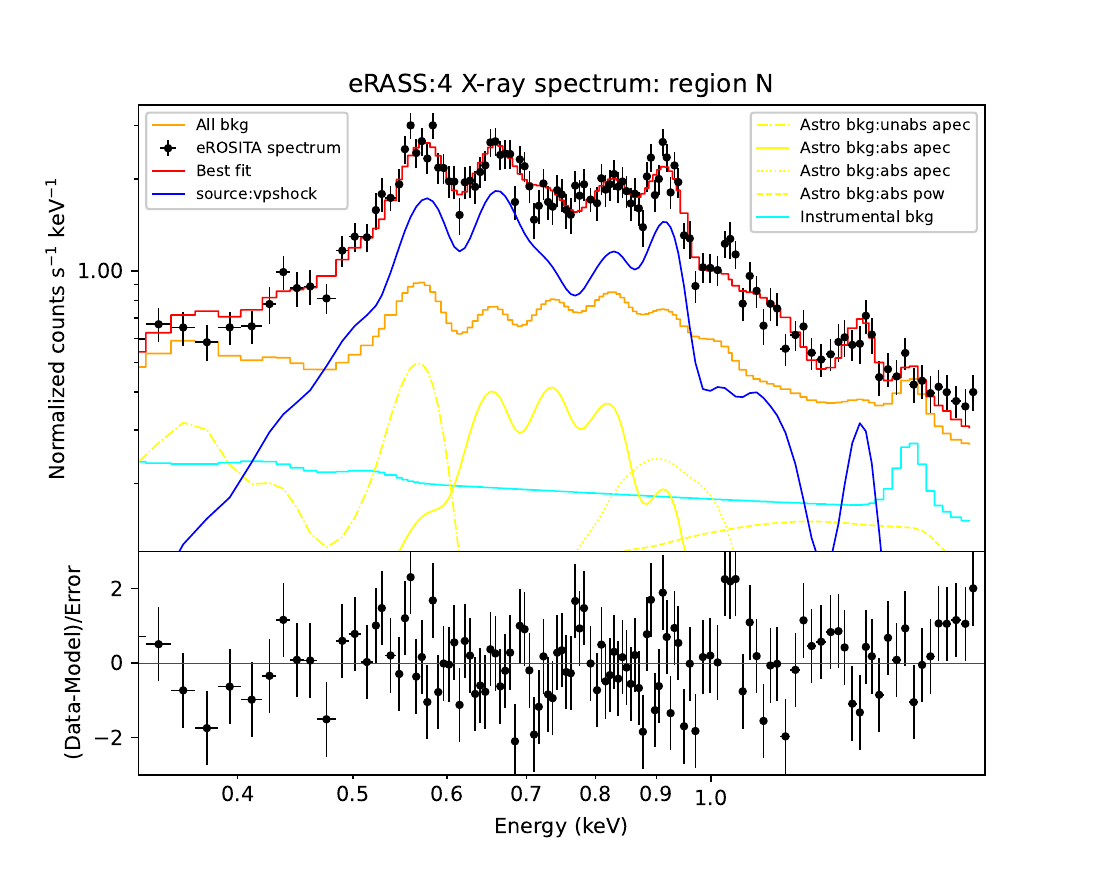}
    \includegraphics[width=0.502\textwidth]{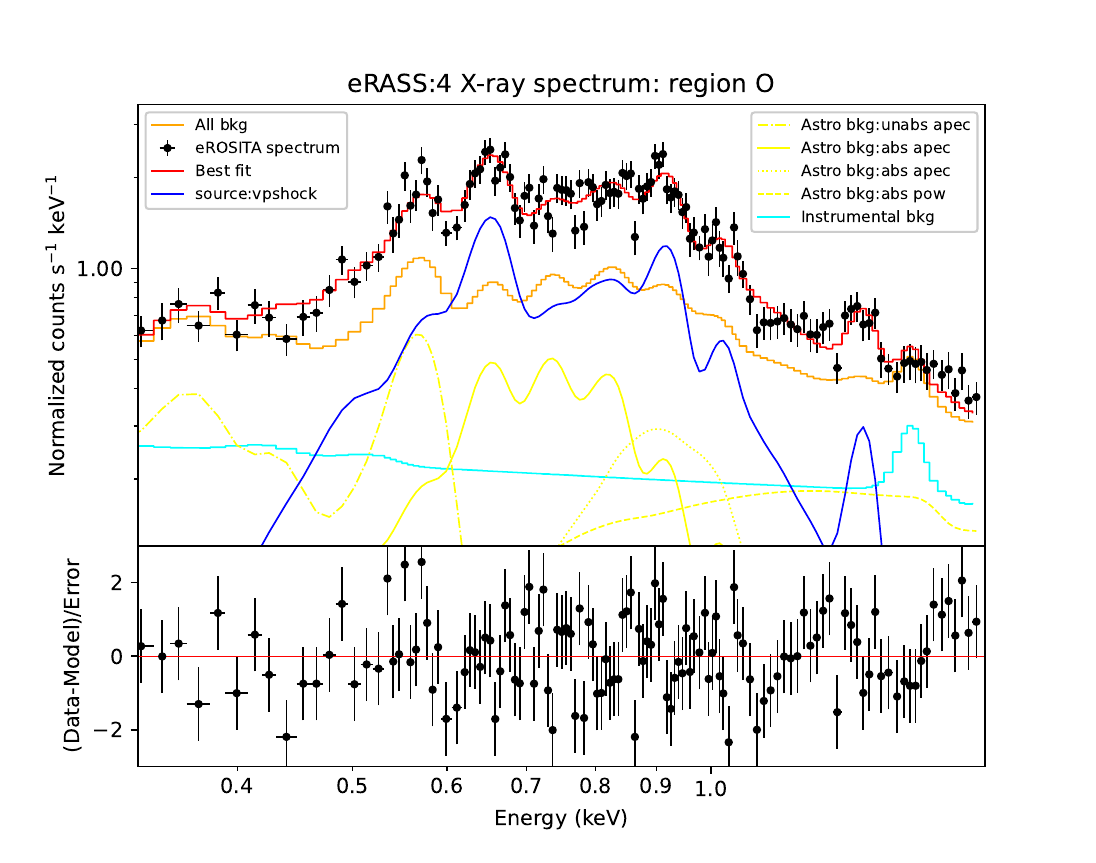}
    \includegraphics[width=0.493\textwidth]{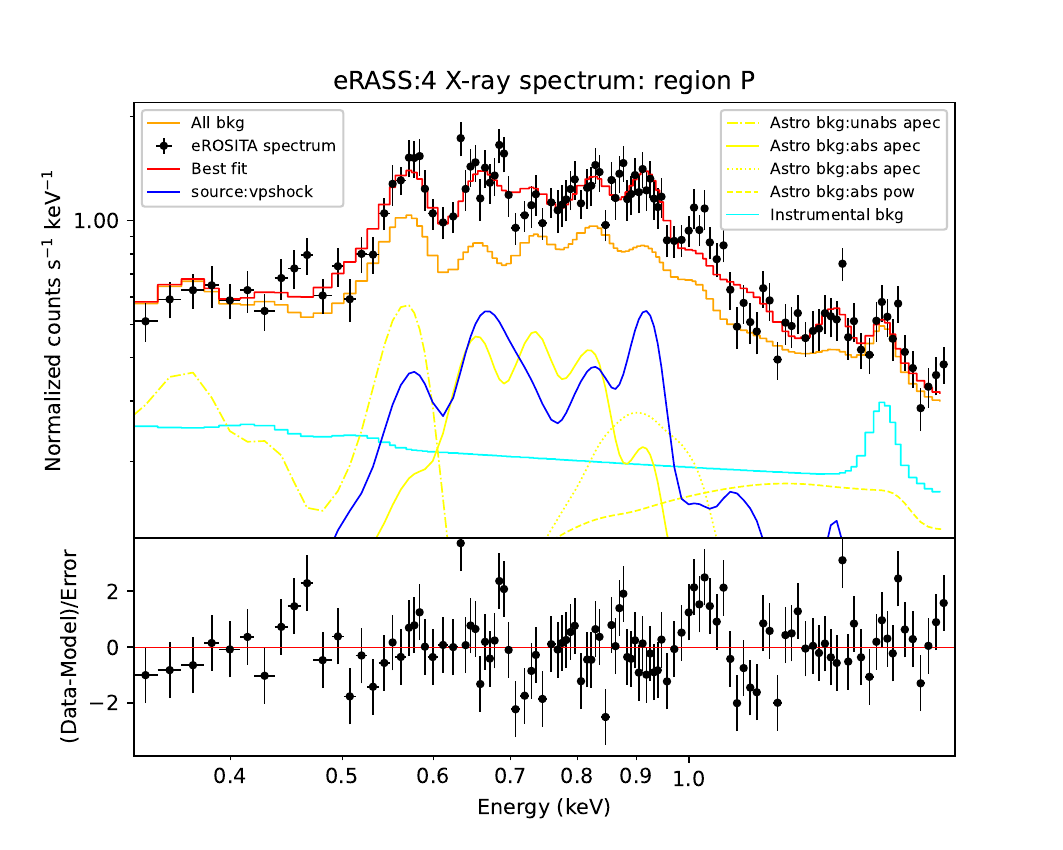}
    \includegraphics[width=0.5\textwidth]{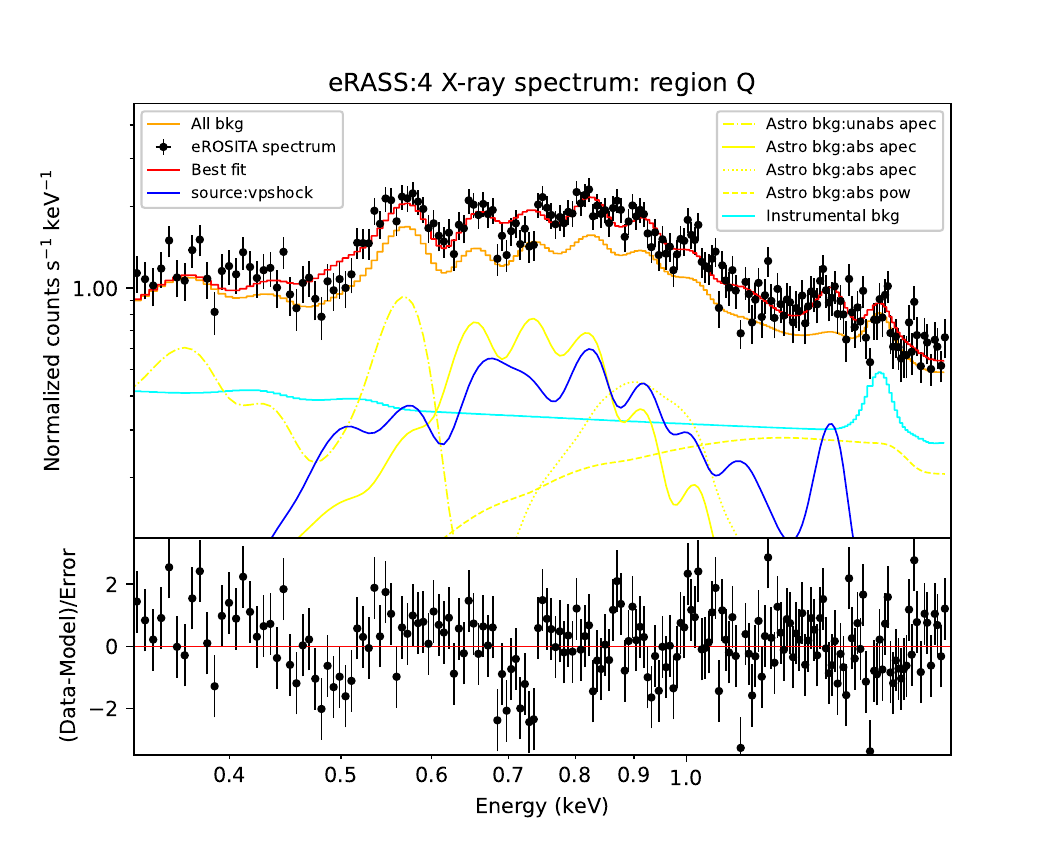}

    \caption{eRASS:4 X-ray spectrum in the 0.3-1.7 keV energy band, from the selected sub-regions of the remnant defined using Voronoi binning analysis, which clearly illustrates the spectral shape change within the remnant.}
    \label{ALLSUBSPEC}
\end{figure*}


\end{document}